\documentclass[12pt]{article}

\usepackage{authblk}
\usepackage{parskip}
\usepackage[normalem]{ulem} 
\usepackage{fullpage}
\usepackage{url}
\usepackage{verbatim}
\usepackage{amstext}
\usepackage{amsmath,amssymb}
\usepackage{amsfonts}
\usepackage{mathtools}
\usepackage{float}
\usepackage{tikz}
\usepackage{subcaption}
\usetikzlibrary {arrows.meta} 
\usepackage{color}
\usepackage{amsthm}
\usepackage{dsfont}
\usepackage[continuous]{causets}

\newtheorem*{definition}{Definition}

\newcommand{\av}[1]{\langle #1\rangle}

\newcommand{\mink}{\mathbb M} 
\newcommand{\ccL}{\mathcal L}
\newcommand{\tcL}{\tilde{\mathcal L}}
\newcommand{ \bC}{\mathbf C}
\newcommand{ \bN}{\mathbf N}

\newcommand{ \nN}{\overline{\mathbf N}}
\newcommand{\re}{\mathbb R}
\newcommand{\bZ}{\mathbb Z} 

\newcommand{\achain}{\mathbf a}
\newcommand{\chain}{\mathbf c} 
\newcommand{\vol}{\mathrm{vol}}

\newcommand{\oR}{\overline {\re}^{n-1,+}}
\newcommand{\oRone}{\overline {\re}^{1,+}}
\newcommand{\oRtwo}{\overline {\re}^{2,+}}

\newcommand{\dist}{\mathcal D}
\newcommand{\nnd}{\dist}
\newcommand{\nd}{\bar \dist}
\newcommand{\nndr}{{\dist}^{(r)}}

\newcommand{\diam}{\mathbb D}

\newcommand{\spec}{\mathcal S} 

\newcommand{\PnCmgone}{P_n(C|M_1,g_1)}
\newcommand{\PnCmgtwo}{P_n(C|M_2,g_2)}

\newcommand{\bomd}{\mathcal D_{B}}

\newcommand{\Mg}{(M,g)}
\newcommand{\Mgprime}{(M',g')}
\newcommand{\Mgone} {(M_1,g_1)}
\newcommand{\Mgtwo} {(M_2,g_2)}
\newcommand{\Mgi} {(M_i,g_i)}
\newcommand{\cR}{\mathcal R}
\newcommand{\bm}{\mathbf m}
\newcommand{\cN}{\mathcal N}
\newcommand{\bNmn}{\mathbf N_m^{(n)}}
\newcommand{\nbNmn}{\overline{\mathbf N}_m^{(n)}}
\newcommand{\Nmn}{N_m^{(n)}}
\newcommand{\bZplus}{{\overline{\mathbb Z}}^{n-1,+}}
\newcommand{\bbF}{\mathbf F}
\newcommand{\dS}{\mathbf {dS}}
\def\bit{\begin{itemize}}
\def\eit{\end{itemize}}
\usepackage{enumerate}
\usepackage[continuous]{causets}
\date{\today}
\title{A Closeness Function on \\ Coarse Grained Lorentzian Geometries} 
\author{Sumati Surya, \\ Raman Research Institute, CV Raman Ave, \\ Sadashivanagar, Bangalore 560080,  India.}

\begin{document}
\maketitle
\begin{abstract} {We construct a family of  closeness functions on the space
    of finite volume Lorentzian geometries using the
    abundance of discrete intervals in the underlying random causal
    sets.  Although strictly weaker than a Lorentzian Gromov
    Hausdorff distance function, it  has the advantage of being
    numerically calculable for large causal sets. It thus provides a 
    concrete and quantitative measure of continuumlike
    behaviour in causal set theory and can be used to define a weak
    convergence condition for Lorentzian geometries.}
 \end{abstract}

 \section{Introduction} 

 Quantifying the closeness of two spacetimes remains an important open
 problem in Lorentzian geometry.  While an exact isometry is well
 defined, there is considerable ambiguity in what it means for
 non-isometric spacetimes to be close, or ``approximately
 isometric''. From a physicist's perspective,  such a characterisation
 can help quantify the smallness of a perturbation, which in turn can
 be used to define a convergence condition.

A  quantitative
characterisation of closeness is also essential for formulations of quantum
gravity in which the continuum spacetime is  replaced by  a more
fundamental,   proto-geometric  structure,  which does not satisfy the
regularity requirements associated with the classical
theory, as in the causal set approach \cite{blms}.  Continuum
spacetimes thus  arise as  approximations which are close, but not   identical,  to the
underlying proto-geometry.

The question of closeness of Riemannian spaces is of interest
to geometrodynamics and  the  initial
value formulation of general relativity  \cite{wheeler,fischer,edwards}. The 
superspace $S(M)$ of Riemannian spaces modulo
diffeomorphisms  corresponds to possible initial spatial
geometries. Having a quantitative measure of when two initial
spatial geometries are close, therefore helps decide if one can be treated as a
perturbation of the other.     
The  distance function  in  a Riemannian space obeys the triangle
inequality  and can be used to 
define the so-called Hausdorff distance between any two of its subspaces.
Roughly speaking  this distance  is obtained by maximising over the minimum
distance between a point in one  subspace and the points in the other subspace. The Gromov-Hausdorff distance between two spaces is
then obtained by minimising this Hausdorff distance over all possible
embedding of the two spaces, or
equivalently by varying over all possible distances functions over the union of the
two spaces \cite{gromov,petersen}.

It is not  straightforward to apply  this construction  to  Lorentzian
spacetimes since the Lorentzian distance function obeys
a reverse triangle inequality \cite{beemehrlich}. Despite this,  there
has  been considerable progress in constructing 
Lorentzian versions of the  Gromov-Hausdorff
distance  between Lorentzian spacetimes. In \cite{bomclose}  
 Bombelli used the underlying random causal set associated with a
causal spacetime to define a closeness function on the space
$\ccL$ of finite volume spacetimes from which he extracted a distance
function, subject to an assumption which was recently proved in
\cite{braun}.  Subsequently, Noldus \cite{nolone,noltwo} and Bombelli and Noldus
\cite{bomnol} used the  Lorentzian distance function\footnote{ The
  Lorentzian distance function satisfies the condition: $d(p,q)>0$ if
  $ p$ is causally to the past of $q$ and $d(p,q)=0$  otherwise.}  to define a strong metric on a
spacetime,  from which a Lorentzian Gromov Hausdorff notion of
convergence was constructed on $\ccL$.  There has
been significant progress more recently  in rigorously defining convergence
conditions,  both for Lorentzian geometries,   as well as their generalisation
to the Lorentzian length spaces of Kunzinger and S\"amann \cite{mueller,minsuhr,kunstein,monsam,braunsam,kunsam}.  
In  \cite{sorveg} Sormani and Vega  used an extended version of the
strong metric  constructed from cosmological time functions
to define a ``null-distance function''. This was then used to obtain convergence conditions for globally hyperbolic cosmological
spacetimes \cite{allenburt,sorsak,sorsaktwo} as well as  Lorentzian length
spaces \cite{kunstein, kunsam}.

In this work we follow Bombelli's lead and focus on this question from the perspective of causal
set theory (CST) which is a discrete approach to quantum gravity
\cite{blms}. Central to CST   is the idea  that the causal
structure poset $(M,\prec)$ associated with a causal spacetime $(M,g)$
contains all the relevant information about the spacetime,
upto a discreteness scale. The approach is inspired by theorems in Lorentzian
geometry which state that the causal structure of a distinguishing
spacetime determines its  conformal class 
\cite{hkm,mal,kp}.

In CST the spacetime continuum is replaced by a locally finite
partially ordered set, a causal set,  where the order
relation $\prec$ is taken to be irreflexive \cite{blms,lr,book}.  Local
finiteness encapsulates  
the  assumption of a fundamental spacetime discreteness,  which is a
cornerstone of CST. This in turn implies that the spacetime continuum
must be an  emergent entity, arising as an approximation to the 
underlying proto-geometric structure. In CST this is the 
random causal set generated by a uniform Poisson
point process or sprinkling  into the spacetime at a density $\rho=1/V_c$, where $V_c$ is
the $d$-dimensional discreteness scale.  An important aspect of CST is
the reconstruction of  geometric and topological properties of the
continuum spacetime from its underlying  random causal set. A fundamental conjecture  of CST (its
``Hauptvermutung'') 
 states that if a causal set is approximated  at a given $V_c$ by two different
spacetimes,   then they must be ``approximately
isometric'', which roughly means that they are identical down to the
discreteness scale 
$V_c$ \cite{braun,bommeyer,muellerH}.  Quantifying approximate
isometry is however non-trivial, since it requires a notion of closeness on the
space of Lorentzian geometries. 

The  Poisson point process into a finite volume spacetime
$(M,g)$ induces a probability distribution $P_n(M,g)$ over the space  $\Omega_n$ of  $n$-element
causal sets. 
In \cite{bomclose} Bombelli defined a closeness function on 
the space of  finite volume causal Lorentzian spacetimes $\ccL$ using
the $P_n(M,g)$.  At finite $n$ this is not a distance
function on $\ccL$ since for every $n$ there exist pairs of non-isometric
spacetimes over which the function vanishes.  A distance
function on $\ccL$ can however be constructed in the large $n$ limit,  subject
to an assumption that has recently been proved in \cite{braun}.

While $ P_n(M,g)$ is   mathematically well
defined, it  is highly non-trivial  to compute.  $\Omega_n$  grows as
$\sim 2^{n^2/4+o(n^2)}$ \cite{kr},  making it impossible in real time to
 find $ P_n(M,g)$,  even for relatively small
 values of $n$ and simple spacetimes. $\Omega_n$ itself has been
 explicitly enumerated only upto $n=16$, which is insufficient
 to recover any continuum information.   Instead of considering the entirety of
 $P_n(M,g)$,  one could  use a coarser  set of  {\sl geometric order invariants} to define 
weaker families of  closeness functions. A geometric order invariant in a  random
 causal set $\bC$ obtained from $\Mg$, is a labelling invariant
 quantity in $\bC$ (an order invariant)  whose expectation value corresponds to a geometric or topological
 invariant of the spacetime.  Many such invariants 
have been constructed \cite{lr,book} and lend support to the
fundamental conjecture of CST. However,  a single invariant like a dimension estimator
\cite{myrheim,meyer} is, in an obvious sense, too coarse grained to be
very helpful.  One must therefore find a family of order invariants
which is numerically calculable, but a  better approximation of $P_n(M,g)$.   

 In this work we consider the family of order-invariants in a causal
 set obtained by
 counting  the number, or abundance  $N_m(C)$ of {\sl $m$-element order
   intervals} with $m=0,\ldots n-1$. An order interval is the discrete
 analogue of an Alexandrov interval and $m$ is an estimate of its
 volume in discrete units.  Let $\bC_n$ denote the random $n$-element
 causal set associated with the finite volume region $\Mg$
 of a spacetime. We  define the {\sl $n$-interval spectrum} of 
$\Mg$ as  $\spec_{n}{\Mg} = \{ \av{\bNmn}\}$ for $m=0,\ldots n-2$, 
 where $\bNmn$ denotes the random number associated with the abundance
 $N_m(C_n)$. While there is
 no direct geometric or topological analogue for each of the  $\bNmn$,
 they form the 
 essential ingredients for the discrete Einstein-Hilbert or
 Benincasa-Dowker-Glaser action  for small
 values of $m$ \cite{bd,dg,glaser}.   

As pointed out in \cite{ls},  $\spec_{n}{\Mg}$ contains  a
surprisingly large amount of information about the spacetime, thus
providing a coarse grained but robust characterisation of  $(M,g)$, upto the
discreteness scale $V_c=\vol(M,g)/n$. This  is roughly analogous to
the spectrum of a discretised elliptic
operator in a Riemannian space.  Simulations  of $\spec_{n}{\Mg}$ in
different  spacetimes suggest that it is a  scale-dependent geometric
``fingerprint'' of $\Mg$ (see Figs. \ref{Fig1.fig}-\ref{Fig6.fig}).  Though it  contains strictly  less
information than $P_n\Mg$,  it  defines  a new class of $L^r$  closeness functions $\nndr_n(.,.)$
both over  $\Omega_n \times \Omega_n$ 
as well as over $\ccL \times \ccL$. While it satisfies both symmetry
and the triangle inequality, $\nndr_n(.,.)$ vanishes between ``interval
isospectral'' Lorentzian geometries and hence is not a distance
function on $\ccL$.   This spectral degeneracy can in some cases be lifted by including
additional  order invariants to the spectrum for certain sequences of
spacetimes, which gives a  (weak) convergence criterion in $\ccL$.  

In Section \ref{intabund.sec} we review the results of \cite{ls} on
the abundance of intervals and the spectrum $\spec_{n}{\Mg}$ of  the  random $n$-element causal set $\bC_n$ associated with a
 finite volume spacetime $(M,g)$.  The analytic results of \cite{ls}
for $\spec_n(\diam^d)$ for a causal diamond $\diam^d$ in Minkowski
spacetime $\mink^d$ provides a concrete example with which to compare
simulations.  In 
Section \ref{distance.sec} we define a class of $L^r$ closeness functions
$\nnd_n^{(r)}(.,.)$ and an  associated notion of convergence. We then use
the recent results of  Braun \cite{braun} to define an approximate isometry
between non-isometric spacetimes.  Numerical simulations are used
throughout to  illustrate these ideas and guide our
intuition. 
We end with a summary and discussion  in Section \ref{conclusion.sec} .

\section{Interval Abundances in Causal Sets}\label{intabund.sec} 

Let $C_n$ be  an $n$-element causal set, or locally finite partially
ordered set,  with the order relation between elements denoted by
$\prec$.  A {\sl labelling}  of $C_n$ is a bijective map  $L: C_n \rightarrow
\tilde C_n$ which is an assignment of a distinct 
label $L(e)=e_i$, $i \in \{1,\ldots,n \}$ to each element $e \in C_n$.  An {\sl order invariant}  of $C_n$
is a function $f: C_n \rightarrow \cR$, for some choice of
range $\cR$, which is 
independent of the labelling of $C_n$. This means that 
the induced labelled order invariant $\tilde f$, defined by  $f = \tilde f \circ L$
satisfies  $\tilde f_1 \circ L_1 =\tilde f_2 \circ L_2$ for any pair
of  labelings $L_{1,2}$ of $C_n$.     An  example of an order
invariant is the number of related pairs  in $C_n$. 

For any $e\prec e' $ in $ C_n$ the discrete Alexandrov or {\sl   order interval}  is
\begin{equation}
  (e,e')= \{e''| e \prec e'' \prec e  \}.
\end{equation}
$(e,e')$ is called  an
{\sl $m$-element order interval}  if its cardinality $|(e,e')|=m$. 
An order interval is itself  not label-invariant since  the {\it
  labelled}  interval  $(e_i,e_j)$ in $\tilde C_n$  need not
 remain invariant. On the other hand, the
number or {\sl abundance of $m$-element intervals} $N_m(C_n)$ in $C_n$  {is}   an  order
invariant, even in a labelled causal set.   For example, $N_0(C_n)$  is the number of  links or
nearest neighbour relations in $C_n$  and is independent of labelling. 

In CST we are not only interested in abstract causal sets,  but also
those that arise from the continuum approximation defined as follows. A
random causal set $\bC$ at a  discreteness scale $V_c=\rho^{-1}$ is associated with every causal  spacetime
$\Mg$, which is generated via a Poisson Point Process (PPP) $\Phi:(M,g) \rightarrow_\rho
\bC$, where the ordering of elements is induced by the causal relation
$\prec_M$  in $\Mg$.  For a PPP, the probability of
finding $m$-elements in a spacetime region of volume $V$ is given by  
\begin{equation}
P_m(V) = \frac{(\rho V)^m}{m!} e^{-\rho V}, 
\end{equation}
which implies that  $\av{\bm} = \rho V$.  In CST  a causal set is said to be approximated by a spacetime, i.e.,
is continuumlike,  if it
is generated from a PPP in this manner. It is important to note that
most causal sets cannot  be
obtained in this way and belong to the
Kleitman-Rothschild (KR) class of $3$-layer posets which are not
continuumlike \cite{kr}.

An important feature of
continuumlike causal sets is that they are non-local in a very
specific sense -- the set of nearest neighbours or links to  an
element can span the entire height of a causal set.  Thus, the $\bNmn$
cannot be estimated from a local counting, as one might do in a fixed
valency graph, but is globally determined. For the random causal set $\bC_n$ associated with a finite volume
spacetime $(M,g)$,  the expectation value 
\begin{equation}
\av{\bNmn\Mg} = \rho^2 \int_{M} dV_x \int_{I^+(x) \cap M} dV_y
\frac{(\rho \,\,  \vol(x,y))^m}{m!} e^{-\rho
  \vol(x,y)}, \label{bNmn.eq} 
\end{equation}
where $\vol(x,y)$ is the volume of the Alexandrov interval $I^+(x)
\cap I^-(y)$ in $(M,g)$.     
We can therefore associate  an    {\sl $n$-interval-spectrum} $\spec_{n}{\Mg} 
\equiv \{ \av{\bNmn\Mg} \}$ with the
spacetime $\Mg$,  where $m=0,\ldots n-2$. We can  similarly
define $\spec_{n}(C_n) \equiv \{ N_m(C_n)\}$ for any abstract causal
set $C_n$.   In MCMC 
simulations of causal set dynamics,   $\spec_{n}(\bC)$ 
has been  used as an observable to 
distinguish between a continuum phase and a non-continuum phase in 
causal set quantum gravity \cite{2dqg,hh,2dcyl}. Extensive
simulations carried out in 
\cite{ls} show that  the  $\spec_{n}{\Mg}$ carries
a significant amount of global information about $\Mg$. Moreover,  
there is recognisable ``continuumlike'' features in 
$\spec_n\Mg$ which differs from $\spec_n(C_n)$ for 
non-continuumlike $C_n$.  Very generally, 
continuumlike behaviour manifests itself  with $N_m(C_n)$  being a
monotonically decreasing
function of $m$.  By numerically comparing $\spec_n\Mg$ for
a variety of spacetimes it is clear that it  has more detailed
information than this,  and  carries enough geometric and topological information
to distinguish between  different  
continuum spacetimes, even for  relatively small $n$ (see Figs. \ref{Fig3.fig}-
\ref{Fig6.fig}) \cite{ls}.   

In general the expression Eqn. \eqref{bNmn.eq} is not easy to
calculate analytically mainly because 
\begin{equation}
  \vol(x,y) = \int_{I^+(x) \cap I^-(y)} \sqrt{-g(z)} d^dz 
\end{equation} 
itself can be a complicated function with  no known closed
form expression for  a generic  spacetime.     
The simplest calculable case is the  $d$-dimensional
Minkowski causal diamond  $\diam^d \subset \mink^d$. In \cite{ls} 
$\av{\bNmn(\diam^d)}$ was calculated to be 
 \begin{eqnarray} 
\av{\bNmn(\diam^d)} = \frac{\Gamma(d)^2}{m!} n^m \sum_{j=0}^\infty
   \frac{(-n)^{j+2}}{j!(j+m+1)(j+m+2)}\frac{\Gamma(\frac{d}{2}(j+m)+1)\Gamma(\frac{d}{2}(j+m+1)+1)}{\Gamma(\frac{d}{2}(j+m+2))
   \Gamma(\frac{d}{2}(j+m+3))},  \label{nmid.eq} 
 \end{eqnarray}
 where $n=\rho \vol\Mg$, and  is  convergent in the large $n$ limit.

 For $m \ll n$,  the  leading order term for $d>2$ was shown to simplify  to 
\begin{equation}
\av{\bNmn(\diam^d) } = n^{2-2/d} \frac{\Gamma(2/d+m)}{m!}
\frac{\Gamma(d)}{(d/2-1)(d/2+1)_{d-2}} + o(n^{2-2/d}).   \label{asympnmid.eq} 
  \end{equation} 
  On the other hand, it is easy to see that for $m$ comparable to $n\gg1 $,  Eqn. \eqref{nmid.eq}
  goes like $e^{-n}$, approximately independent of
  dimension.  For $m \ll n $, the leading order dependence on $n$ is $m$
independent, and hence  for $d>2$ one can define 
\begin{equation}
\nbNmn(d,n) \equiv \frac{\av{\bN_m(d,n)}}{\av{\bN_0(d,n)}} = \frac{\Gamma(2/d+m)}{m! \Gamma(2/d)}
+ o(n^{-2+2/d}),   \label{normNs.eq}
  \end{equation} 
which  is independent of $n$.  For $m \sim n$ on the other hand,
the exponential decay dominates and to leading order
\begin{equation}
  \nN_m(d,n) \sim o(1).
\end{equation}

As discussed in \cite{ls} the $n$-interval spectrum can also used to pick out local
continuumlike  neighbourhoods $\cN(e)$ in a causal set as an intrinsic, order
theoretic version of a Riemann Normal Neighbourhood (RNN) where the
curvature is approximately constant.

\subsubsection{Simulations} 

\begin{figure}[H]
\begin{subfigure}{.3\textwidth}
  \includegraphics[width=.8\linewidth]{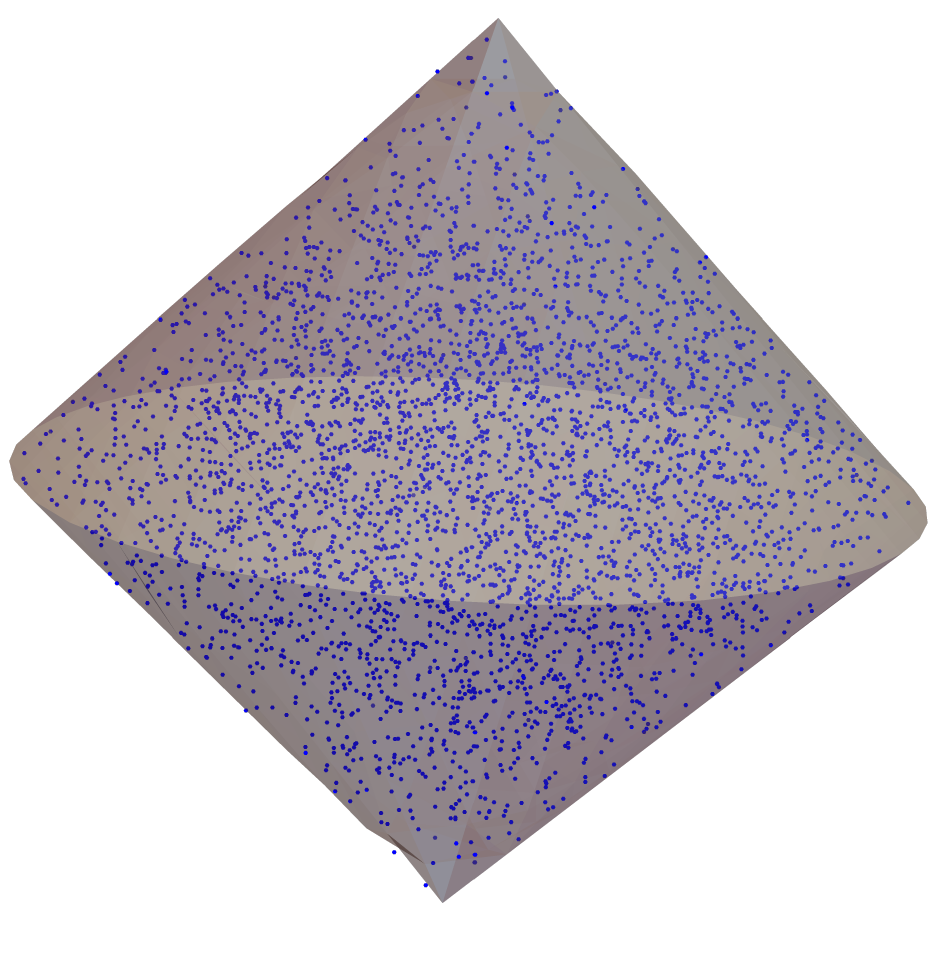} \caption{$\diam^3$} \end{subfigure}
\begin{subfigure}{.3\textwidth}
  \includegraphics[width=.8\linewidth]{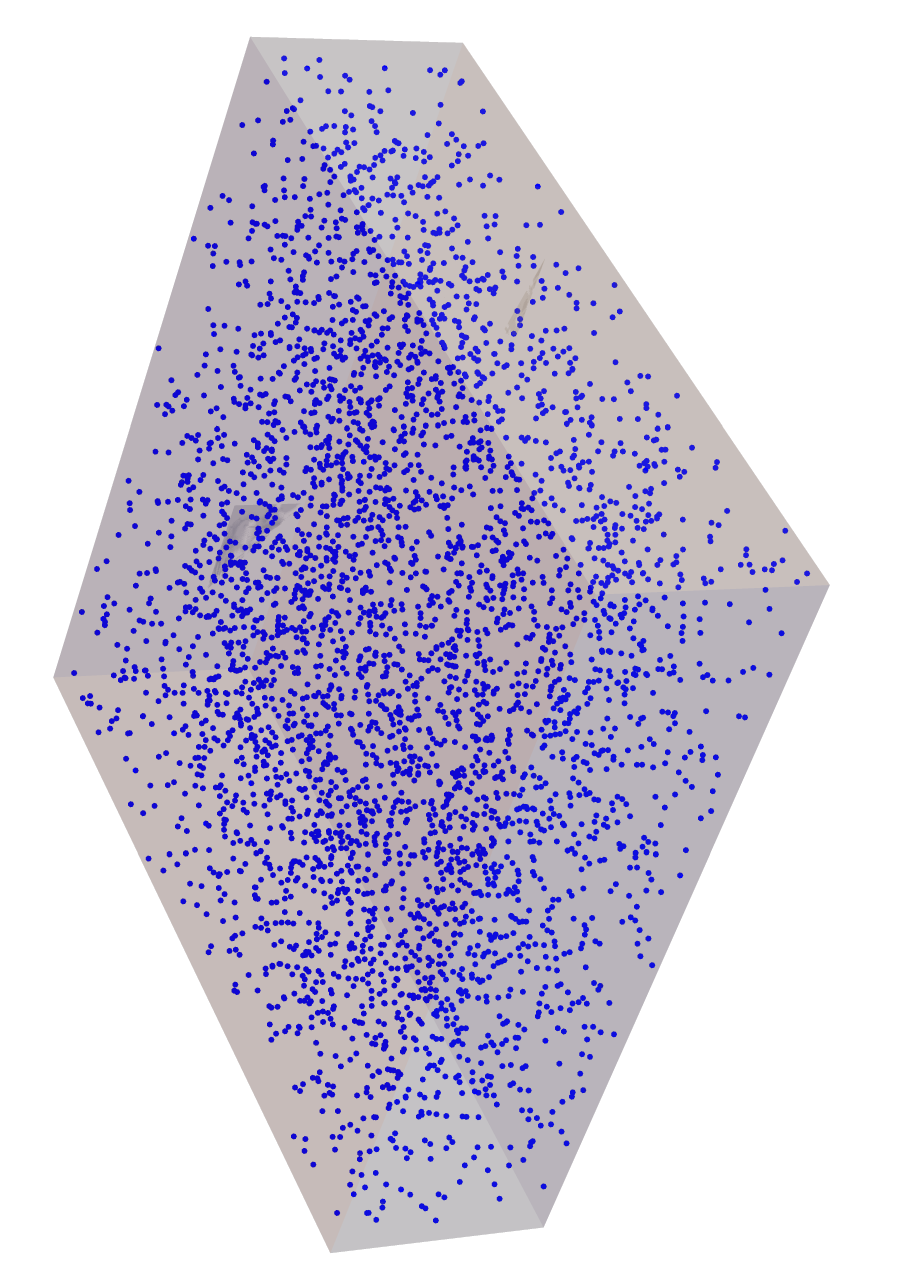} \caption{
    $\diam^2 \times \mathbb I_t$}\end{subfigure}
\begin{subfigure}{.3\textwidth}
  \includegraphics[width=.8\linewidth]{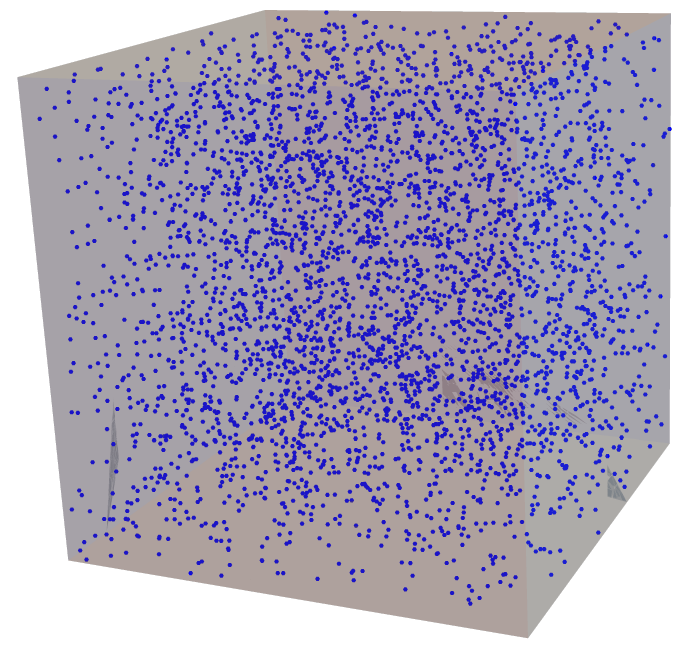} \caption{
    $\mathbb I^3$}\end{subfigure}
\begin{subfigure}{.5\textwidth}
  \includegraphics[width=.8\linewidth]{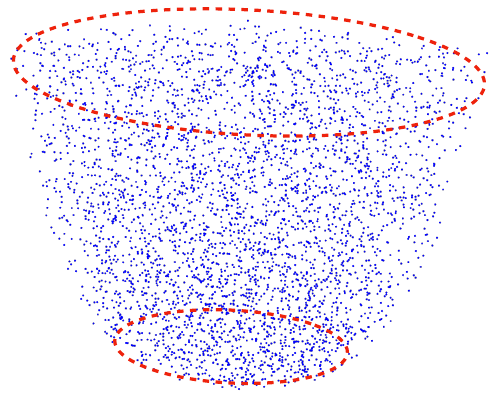} \caption{
    $\bbF_3(0.5)$}\end{subfigure}
\begin{subfigure}{.5\textwidth}
  \includegraphics[width=.8\linewidth]{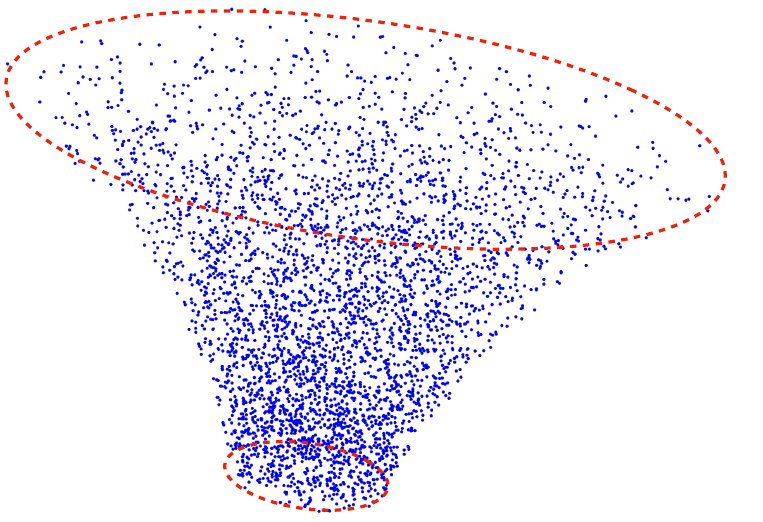} \caption{
    $\dS_3(0.3)$}\end{subfigure}
                \caption{Causal sets generated by Poisson sprinklings
                  into various regions of spacetime.} \label{regions.fig}
\end{figure}

We now show examples of $\spec_n\Mg$ generated via Poisson
sprinklings into various spacetime regions.  In this work we consider
(a) the causal 
diamond or Alexandrov interval  $\diam^d$  in Minkowski
spacetime  for  $d=2,\ldots 8$, (b) the thickened causal
diamond $\diam^d \times \mathbb I_t$ in Minkowski
spacetime, with  $d=2,\ldots 8$ and where $\mathbb I_t \equiv (0,t)$ 
with $t= 0.05, 0.1, 0.15, 0.2, 0.25, 0.3$ (c) the
Minkowski hypercube $\mathbb I^d$ for $d=2, \ldots 8$ with two spacelike
boundaries (the rest being timelike) and finally (d) FRW spacetime
regions  with line element 
\begin{equation}
  ds_d^2= -dt^2 +a^2(t)\biggl( \sum_{i=1}^{d-1} dx_i^2 \biggr) 
\end{equation}
for $d=3,\ldots 9$  and scale factors (i) $\bbF_d(p)$ with $a_p(t)=t^{p}$, $p=0.33,0.5,0.67$ and
(ii) $\dS_d(c)$ with $a_c(t)=e^{c
  \, t}$, and $c=0.025,0.05,0.1,0.15,0.2$. Fig. \ref{regions.fig} shows some examples of
these regions in $d=3$.

\begin{figure}[!htbp]
  \begin{subfigure}{.5\textwidth}
    \includegraphics[height=5cm]{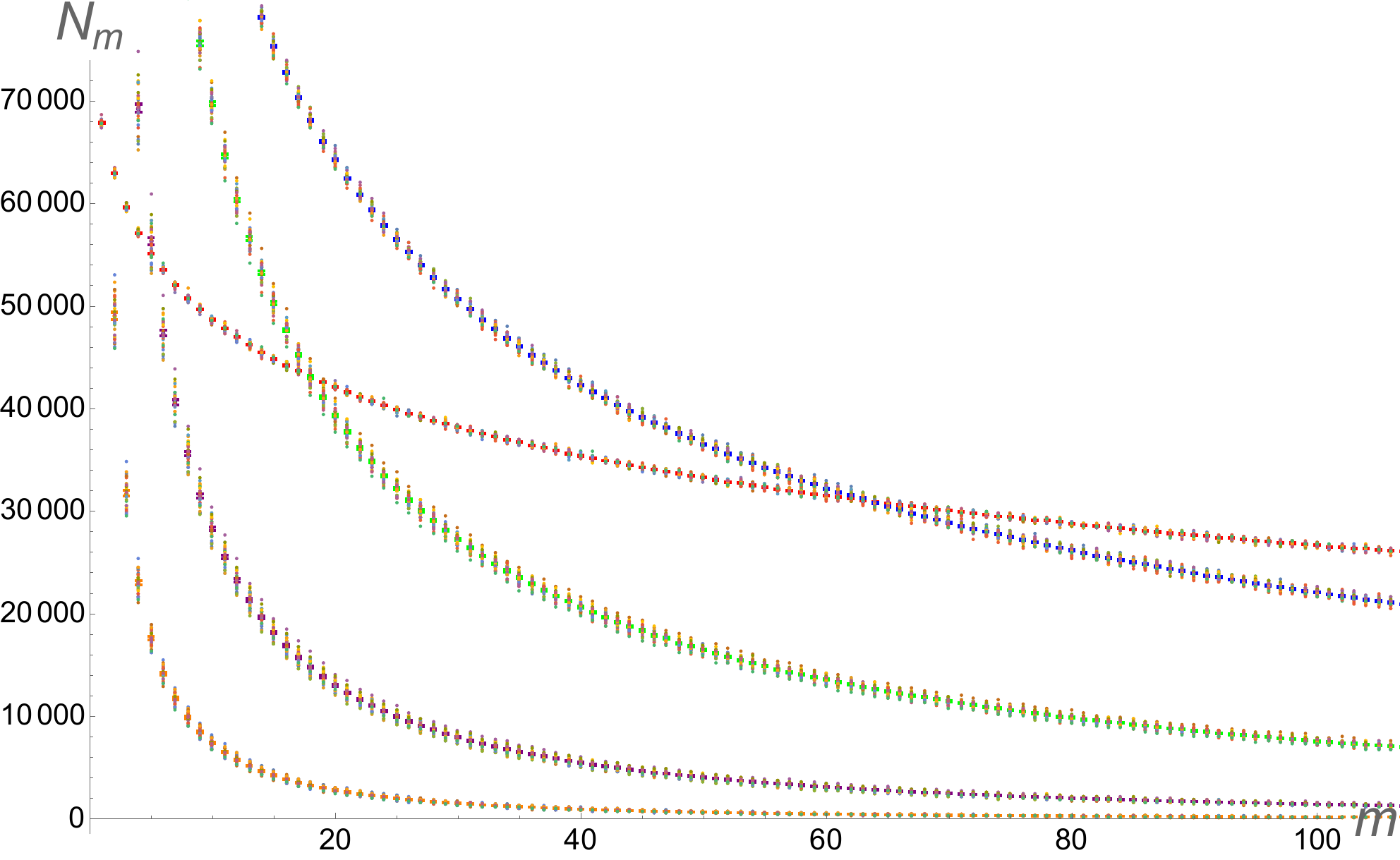} \caption{$\spec_n(\diam^d)$
  as a function of $d$}\end{subfigure} \begin{subfigure}{.5\textwidth}
\includegraphics[height=5cm]{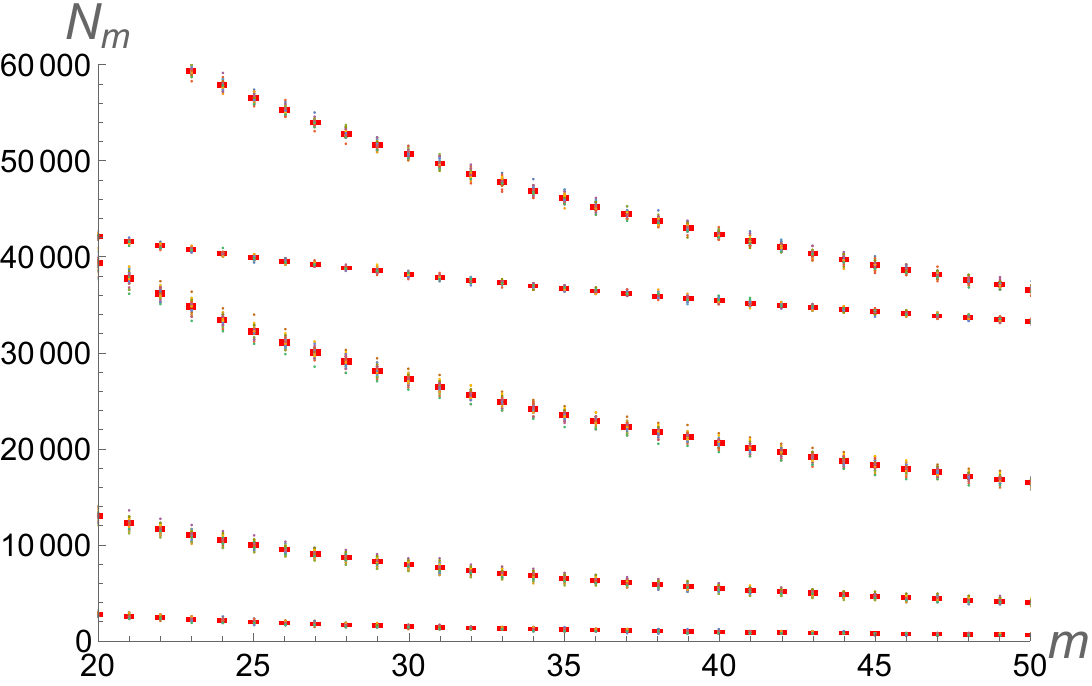}
 \caption{Zoomed in view} \end{subfigure} 
\caption{
  We show 30 different realisations of 
  $\spec_n(\diam^d)$ for  
  $n=10,000$, $d=2,\ldots 6$ plotted with the mean and standard
  error. One can make
  out the spread, which is large for any given $m$, but is relatively small
  as far as the entire spectrum is concerned. Compare with
  Fig. \ref{Fig1.fig} below.} \label{Fig3.fig}
                 \end{figure}

\begin{figure}[!htbp]
\begin{subfigure}{.5\textwidth}
  \includegraphics[height=5.2cm]{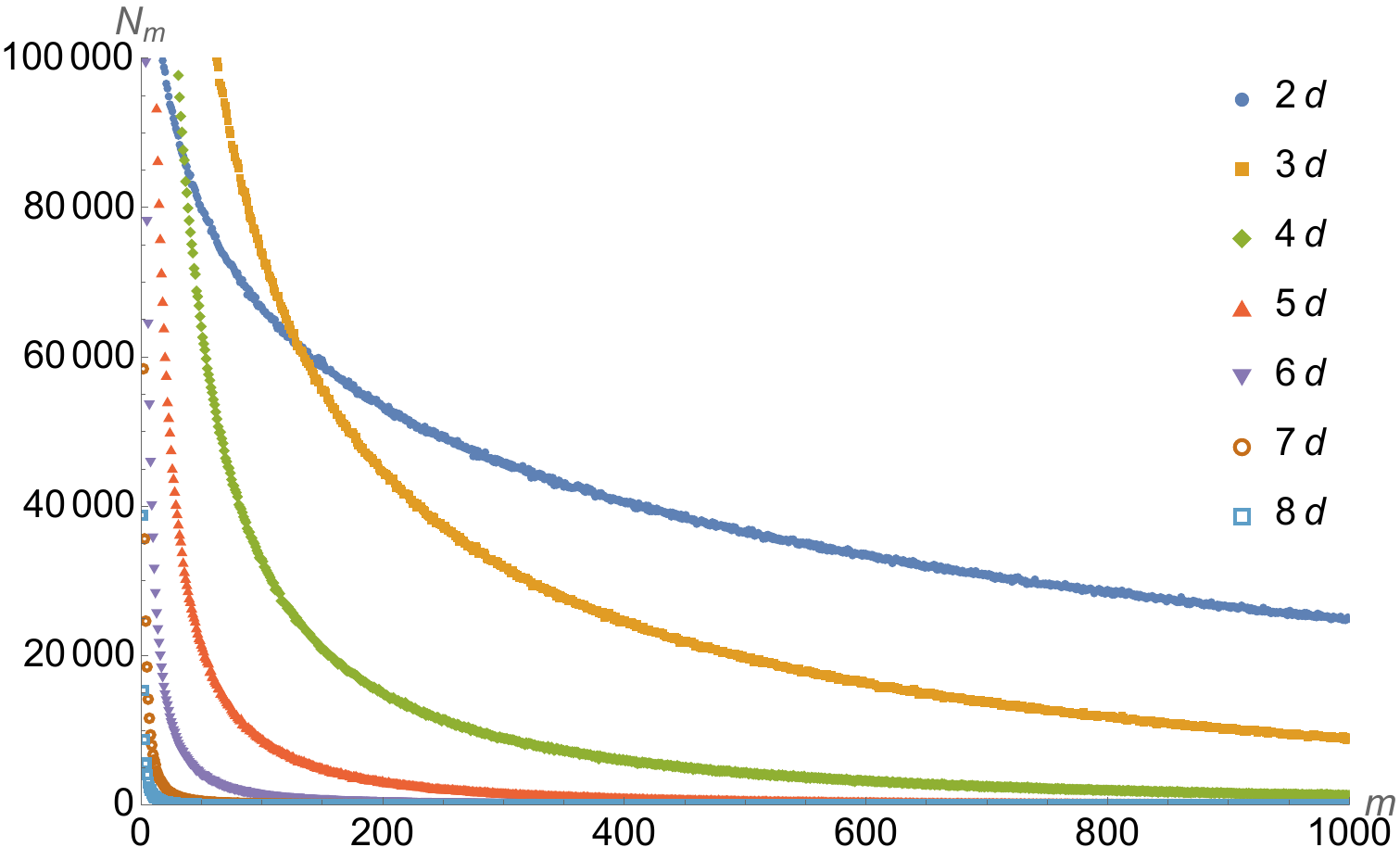} \caption{$\spec_n(\diam^d)$
  as a function of $d$} \end{subfigure}
\begin{subfigure}{.5\textwidth}
  \includegraphics[height=5.2cm]{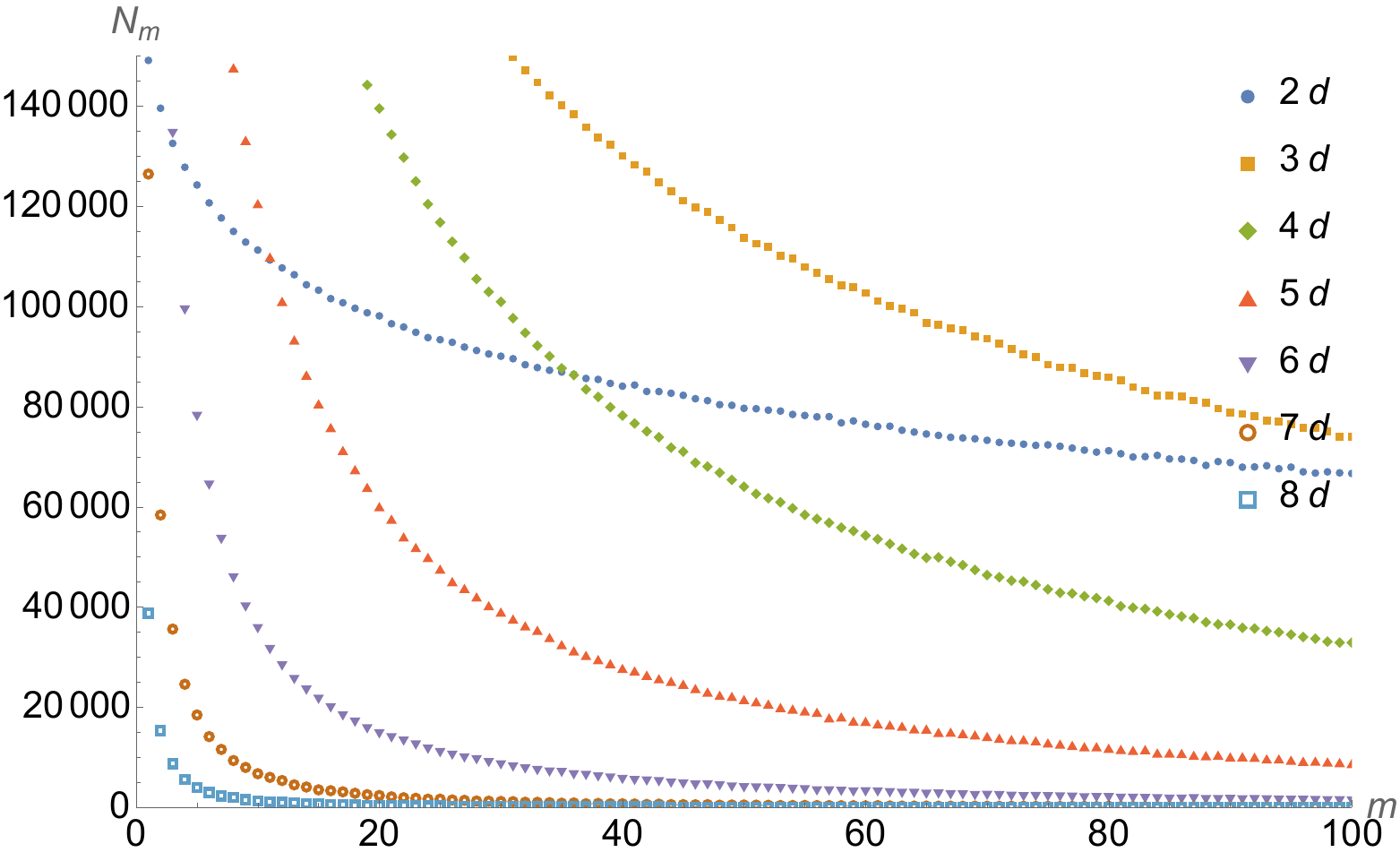} \caption{Zoomed
    in view}\end{subfigure}
                \caption{  $\spec_n(\diam^d)$ for  $d=2, \ldots 8$ for
                  a single  realisation with  $n=20,000$.} \label{Fig1.fig}
              \end{figure}

\begin{figure}[!htbp]
\begin{subfigure}{.5\textwidth}
    \includegraphics[height=5.2cm]{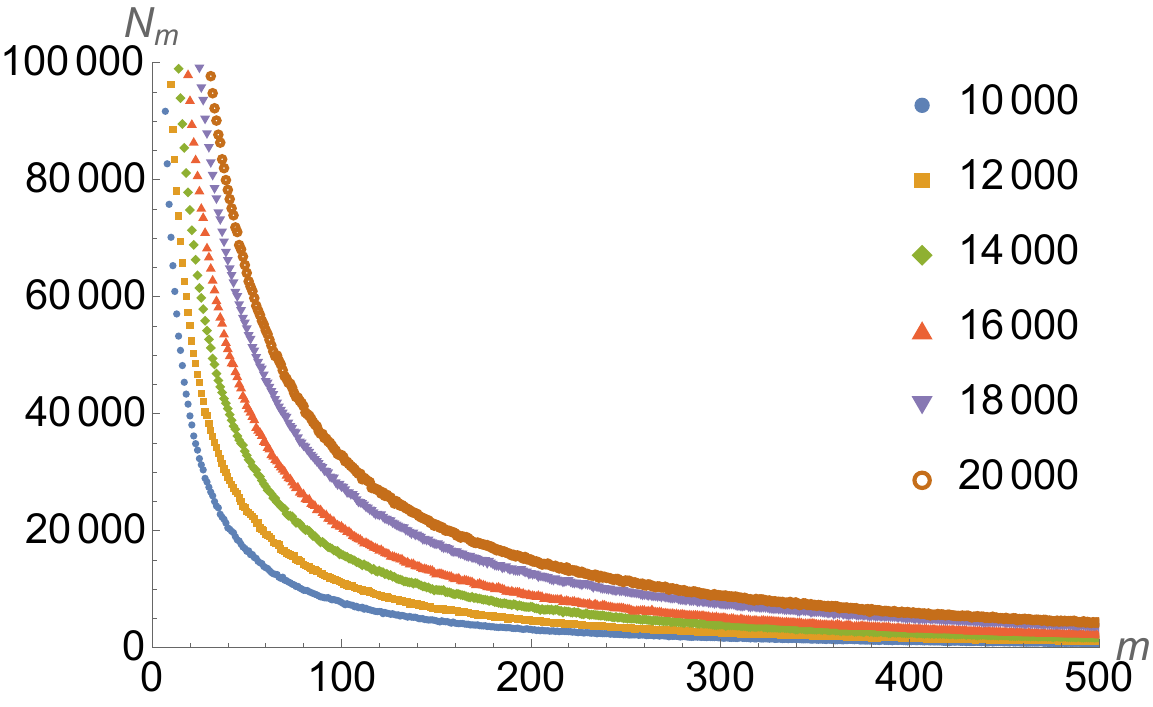} \caption{$\spec_n(\diam^d)$
  as a function of $n$}\end{subfigure} \begin{subfigure}{.5\textwidth}
\includegraphics[height=5.2cm]{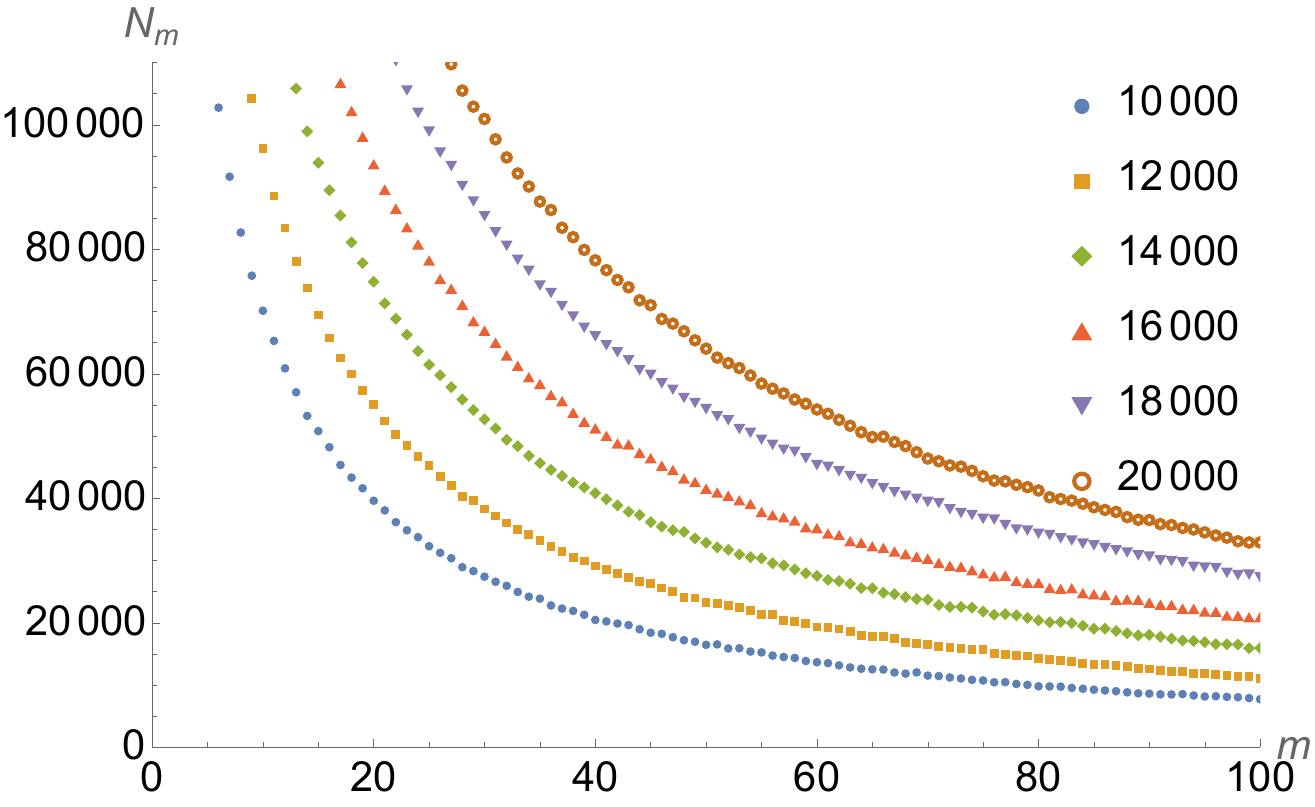} \caption{Zoomed
    in view} \end{subfigure}               \caption{ $\spec_n(\diam^4)$ for  
                  a range of $n$ values. While the number of intervals increases 
                  with the size of the causal set, the overall
                  shape of the function does not change. 
                  } \label{Fig2.fig}
    \end{figure}

\begin{figure}[!htbp]
 \begin{subfigure}{.5\textwidth}
   \includegraphics[height=5.2cm]{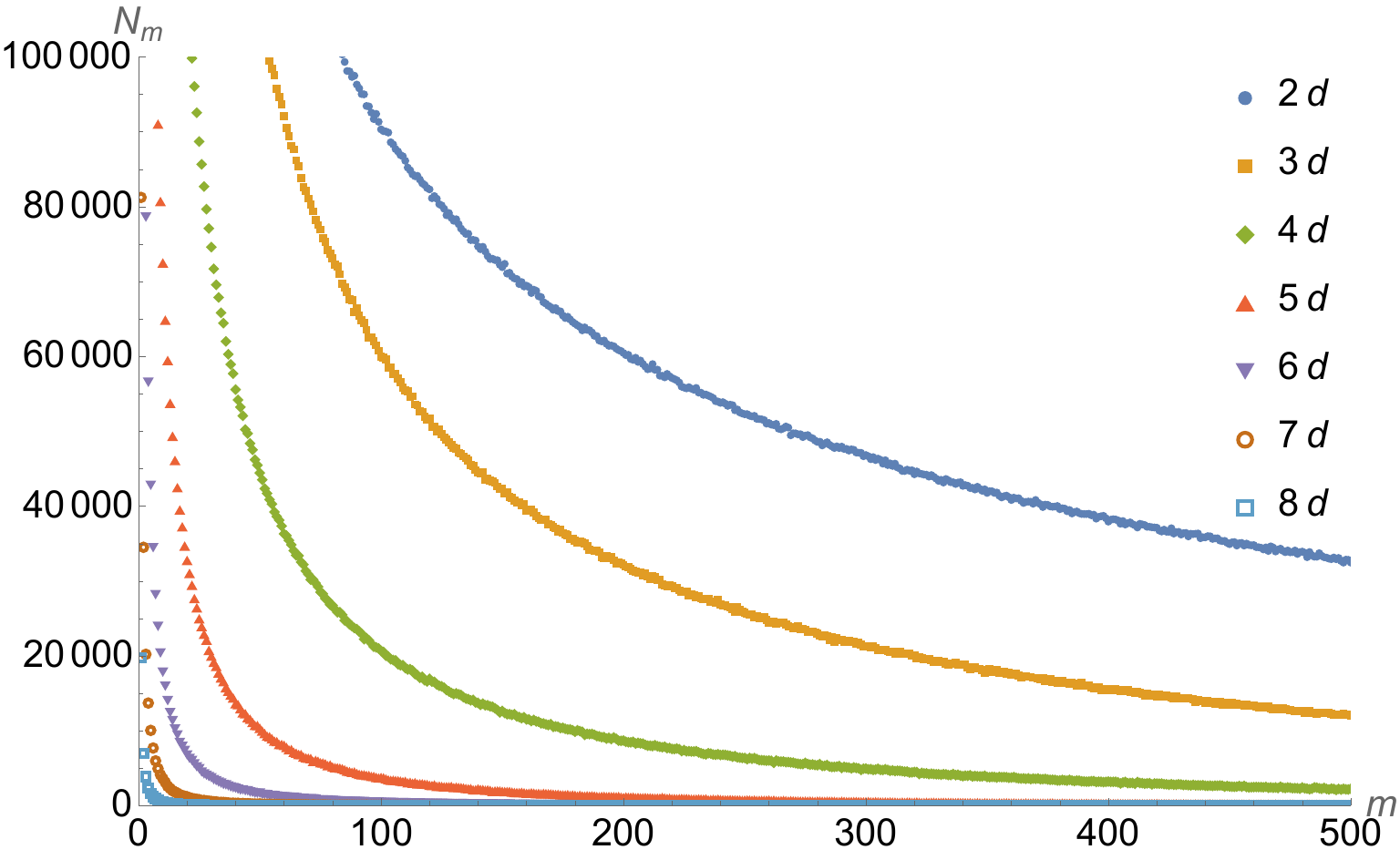}\caption{$\spec_n(\diam^d
     \times \mathbb I_t)$}
\end{subfigure} \begin{subfigure}{.5\textwidth}
   \includegraphics[height=4.9cm]{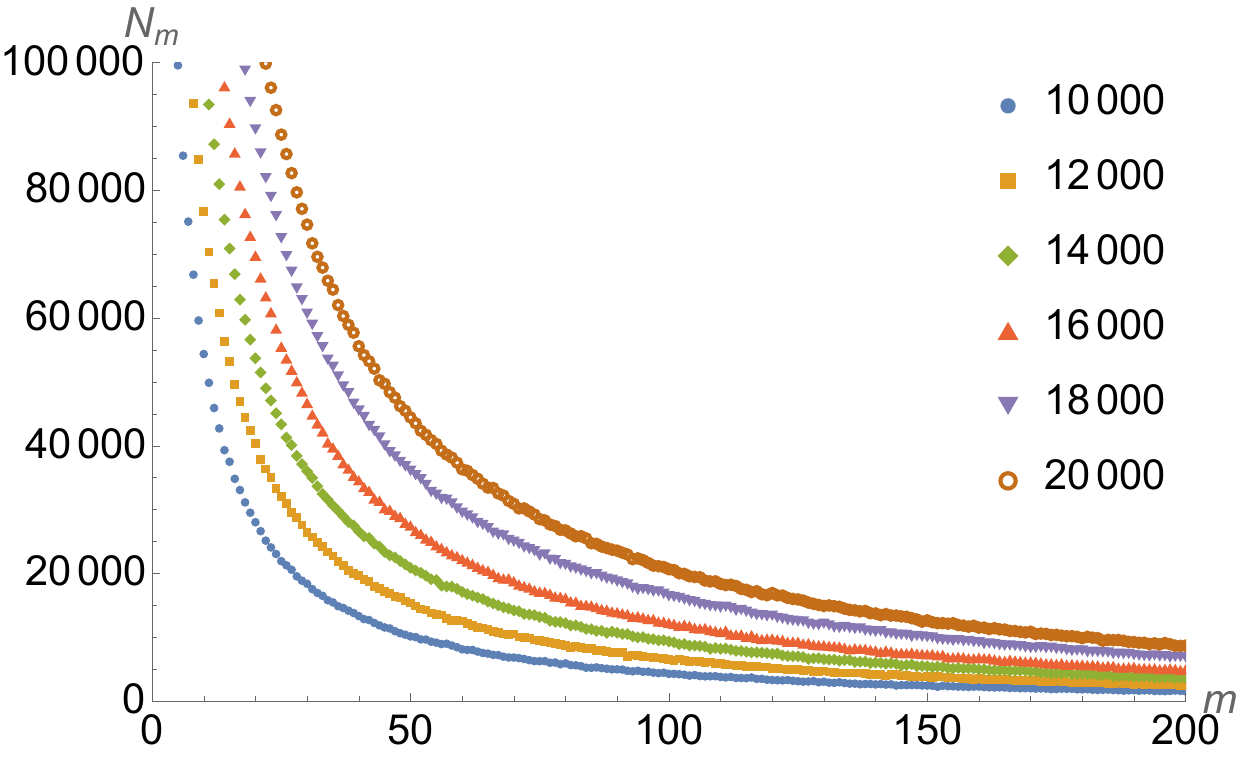} \caption{$\spec_n(\diam^4
     \times \mathbb I_t)$}\end{subfigure}
                \caption{
                  (a) $\spec_n(\diam^d\times \mathbb I_t)$ for $t=0.15$
                  and $d=2,\ldots, 8$. (b) $\spec_n(\diam^4\times
                  \mathbb I_t)$ for $n=10,000, \ldots 20,000$. The qualitative
                  behaviour is very similar to that of $\diam^d$. 
                               } \label{Fig4a.fig}
              \end{figure}

\begin{figure}[!htbp]
  \begin{subfigure}{.5\textwidth}
 \includegraphics[height=4.9cm]{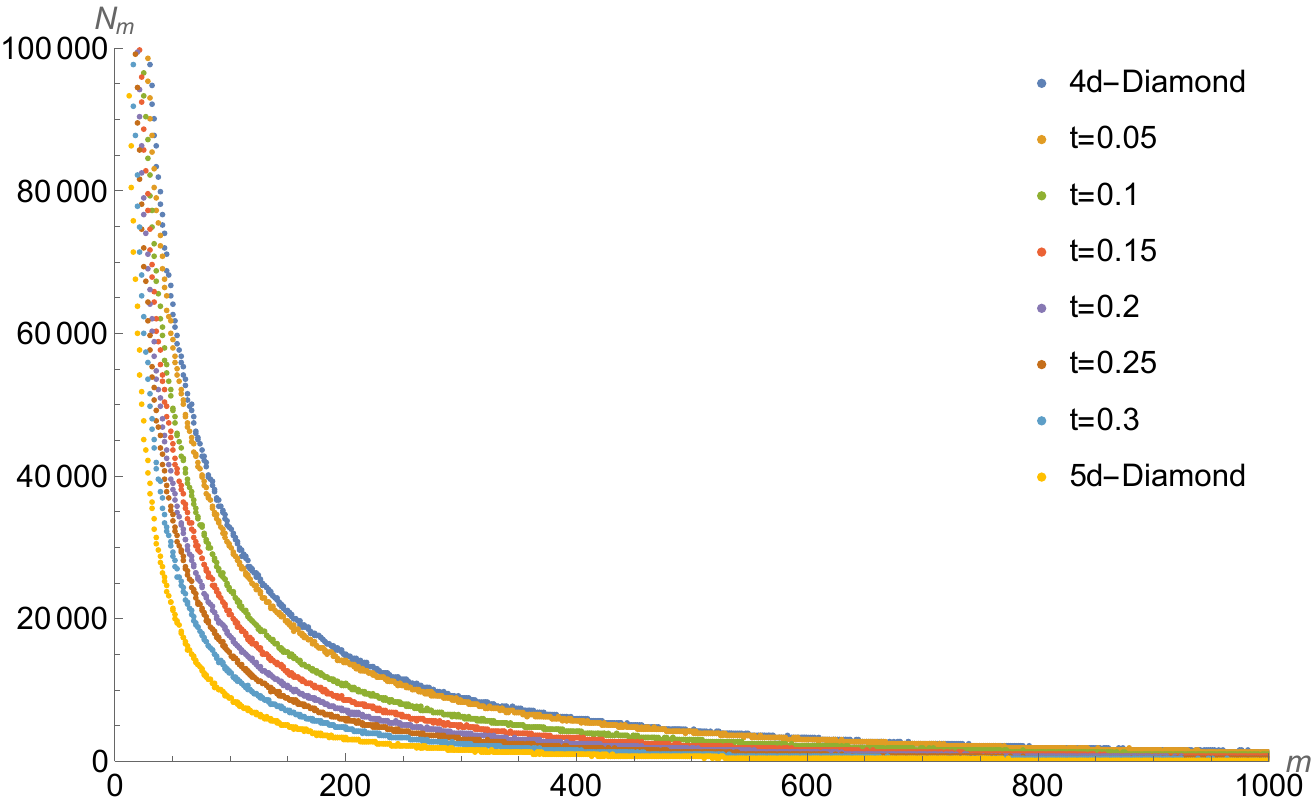} \caption{$\spec_n(\diam^4
   \times \mathbb I_t)$ along with $\spec_n(\diam^{4,5})$}\end{subfigure} \begin{subfigure}{.5\textwidth}
 \includegraphics[height=4.9cm]{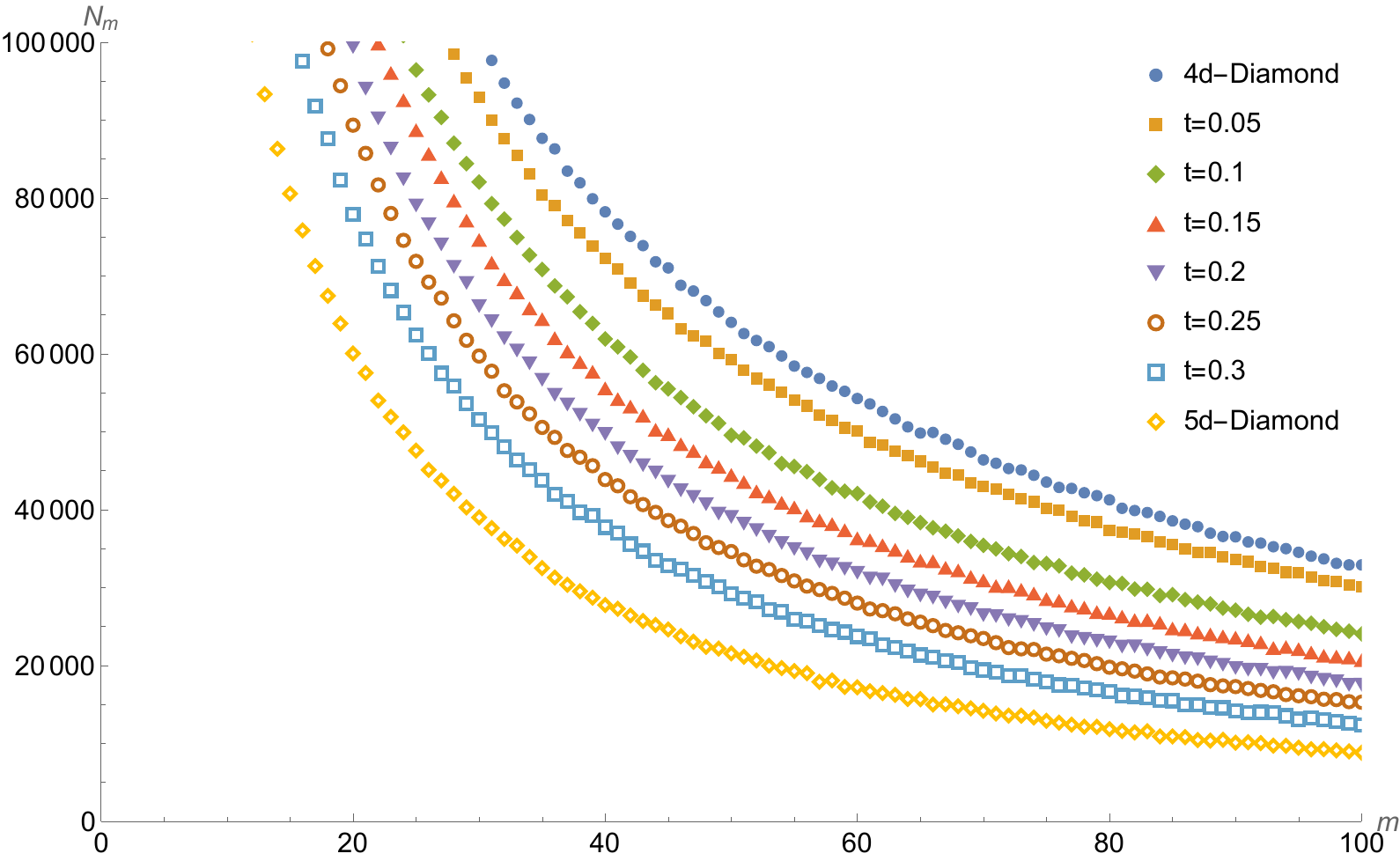} \caption{Zoomed
   in view} \end{subfigure}
                \caption{These figures contrast 
                  $\spec_n(\diam^4 \times \mathbb I_t)$  with  $\spec_n(\diam^4)$ and $\spec_n(\diam^5)$, as one changes the
                  relative size $t$ of the internal dimension. One notes the 
                  interpolation from $\spec_n(\diam^4)$ to $\spec_n(\diam^5)$  as $t$
                  increases. In particular, there is a convergence to $\spec_n(\diam^d)$ as $t$ decreases.} \label{Fig4b.fig}
 \end{figure}

\begin{figure}[!htbp]
  \begin{center}
 \includegraphics[height=8cm]{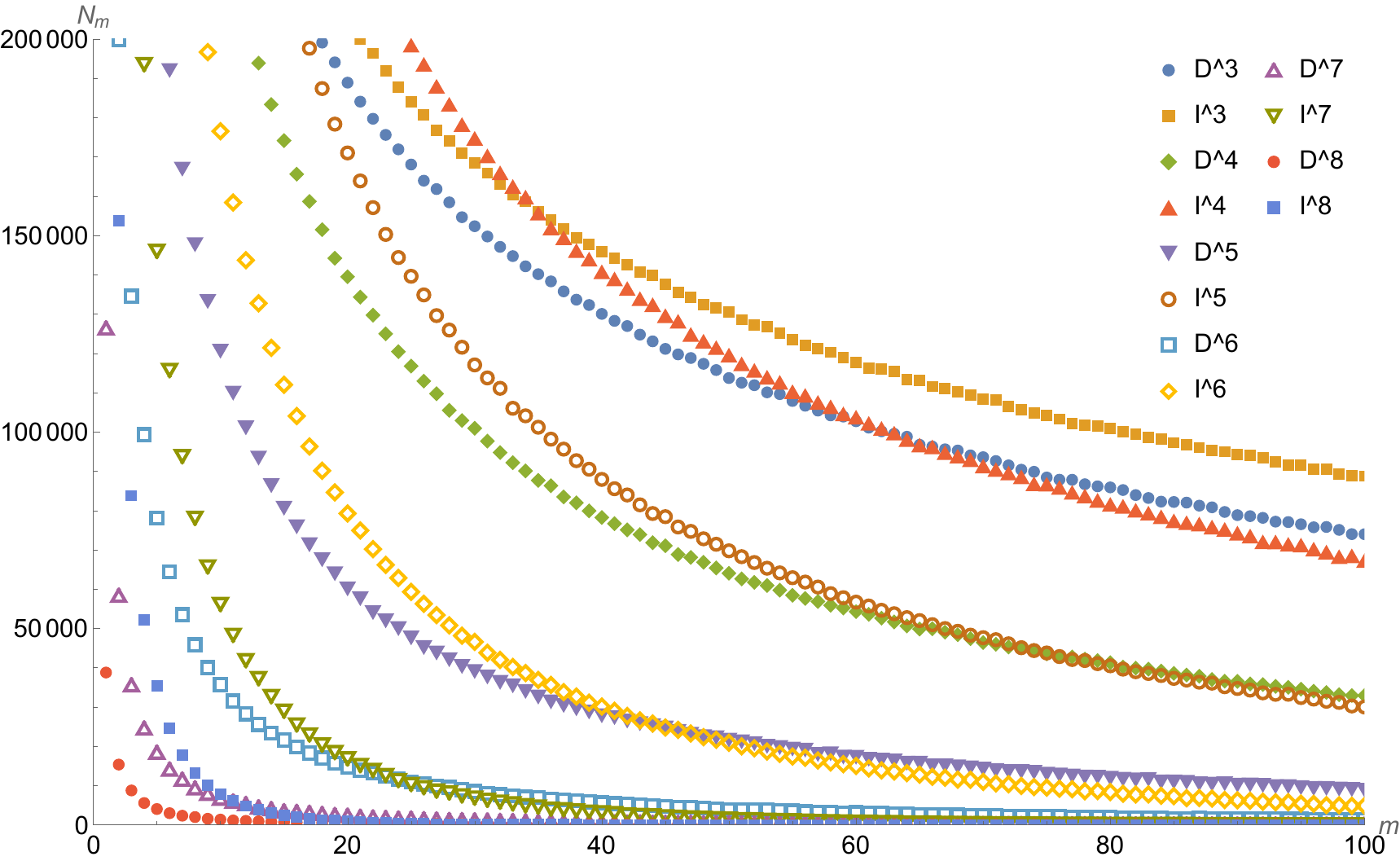} 
                \caption{ $\spec_n(\mathbb I^d)$ and $\spec_n(\diam^d)$, for 
                  $d=3,\ldots 8$, $n=20,000$. Even though the
                  spacetime is locally  Minkowski,  $\spec_n(.)$ can distinguish 
                  the global geometry.  
                               } \label{Fig5.fig}
                \end{center} 
              \end{figure}

\begin{figure}[!htbp]
  \begin{center}
    \includegraphics[height=8cm]{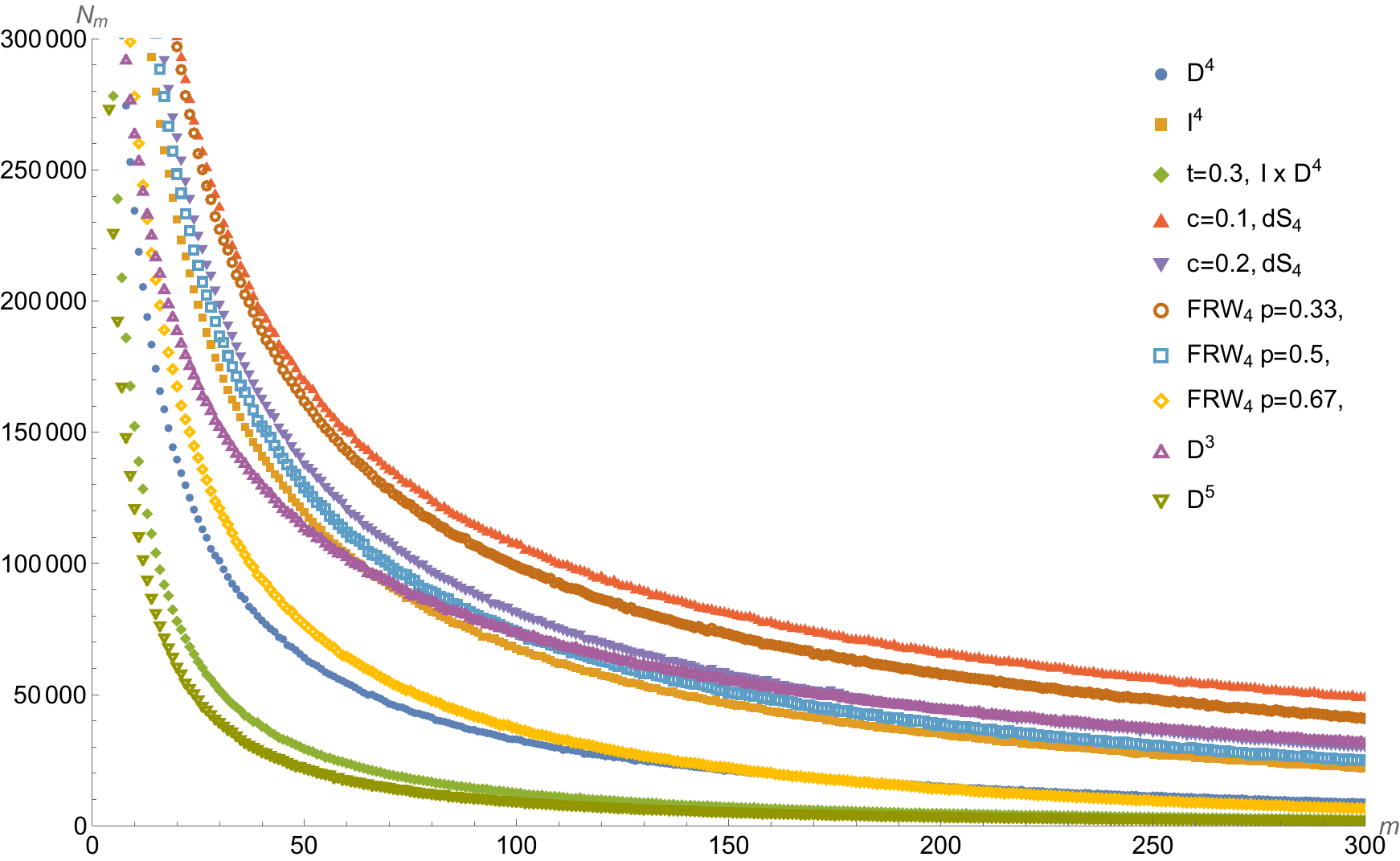} 
                \caption{ $\spec_n(.)$ for a range of
                  spacetimes with $n=20,000$ shows how  well it
                  is able to distinguish between them. 
                 } \label{Fig6.fig}
                \end{center} 
              \end{figure}

Figures \ref{Fig1.fig}-\ref{Fig6.fig} show examples 
of  $\spec_n\Mg$ from these simulations. What is striking about these
examples  is that
each spacetime is associated with a characteristic curve, its ``fingerprint''.
In most of these figures we have 
used  only a single realisation of the sprinkling, to avoid excess
computational burden.  In order to
justify this,  in Fig. \ref{Fig3.fig} we show the spectrum for $d=4$, $n=10,000$  over $30$
realisations  and also their  mean and
standard error.  While the $\bNmn\Mg$  in a given realisation can vary
significantly in absolute terms from the mean, 
there is a strong self-averaging which means that the deviation from
the full spectrum is relatively small. This suggests that even a single
realisation is sufficient to determine the spacetime, upto the
discreteness scale $\rho^{-1}=\vol\Mg/n$.

In Figure \ref{Fig1.fig} the $n$-interval spectrum $\spec_n(\diam^d)$ is shown for $d=2,\ldots
8$, for $n=20,000$ elements. The dimension dependence of the curves is
explicit and one can therefore use this as an alternative  to the flat spacetime
Myrheim-Myer dimension 
estimator \cite{myrheim,meyer} as suggested in \cite{ls}.   In Fig \ref{Fig2.fig} the
dimension fixed to $d=4$ while varying $n$. There is an obvious scaling
with $n$ which also suggests that even at around $n=10,000$, one is in
a fairly reliable large $n$-regime.  In Figure   \ref{Fig4a.fig}  we show the spectrum for a ``thickened''
diamond $\diam^d \times \mathbb I_t$, where $\mathbb I_t$ represents an
``internal'' dimension of relative thickness $t$. The local geometry
is Minkowski, but  the global features make the spectra 
distinct.  For fixed $t$, we
see a similar behaviour as in the case of  the $\diam^d$ for $d>2$, as
both $d$ and $n$ are varied.  In Fig.  \ref{Fig4b.fig} we fix $d=4,
n=20,000$  and change the $t$ values. As $t$ is made smaller, it
approaches $\diam^4$, and as it is made larger, it approaches
$\diam^5$, as one would expect.  In other words, for a fixed $n$,
there is a small enough $t$, such that $\diam^d$ and $\diam^d \times
\mathbf I$ are $n$-interval isospectral and cannot be distinguished
from each other.  
In Figure \ref{Fig5.fig} we  consider  the hypercube
$\mathbb I^d$ for $d=3, \ldots, 8$, again in Minkowski
spacetime and contrast it with $\diam^d$. The
relevance of the global features of the spacetime region are quite
explicit. Finally,  in Fig. \ref{Fig6.fig} we compare the spectra over a 
wider range of spacetimes including  non-flat, cosmological spacetime regions for fixed
$n$.

It is clear from these Figures and those in \cite{ls} that $\spec_n(M,g)$ differs significantly for
spacetimes of different dimensions as well as different geometries.
We have not made a comparison here  with non-continuum like
causal sets and refer the reader to the examples shown in  \cite{ls}.
While the $\spec_n(M,g)$ bring out qualitative differences and
similarities,  it is useful to have a quantitative measure of closeness
for both extrinsic and intrinsic comparisons in CST.


\section{An $n$-Interval Closeness Function}\label{distance.sec} 

We begin with the definition of Bombelli's closeness function using
$P_n(M,g)$.  Let $P_n(C|M,g)$ be the probability that causal set $C
\in \Omega_n$ is obtained in a sprinkling into $\Mg$.  
Since $1 \geq P_n(C|M,g) \geq 0$ and $\sum_{C \in
  \Omega_n}P_n(C|M,g)=1$, $P_n$ maps each spacetime $(M,g)$ onto the
 non-negative part of the unit sphere,  $\mathbb S^{|\Omega_n|-1,+}
\subset \re^{|\Omega|_n}$, i.e.,  $P_n: \ccL \rightarrow \mathbb
S^{|\Omega_n|-1,+}$. Using the angular separation between the points
on  $\mathbb S^{|\Omega_n|-1,+}$ as a  measure of
closeness in $\ccL$, the Bombelli closeness function is 
\begin{equation}
\bomd(g_1,g_2) \equiv \frac{2}{\pi} \arccos \sum_{C \in \Omega_n}
\sqrt{\PnCmgone}\sqrt{\PnCmgtwo}, \label{bom.eq} 
\end{equation}
where $\Mgi \in \ccL$.  For  finite $n$, $P_n$ is not injective:  for 
two spacetimes which are close with respect to $V_c$, the distribution
$P_n(M,g)$ is expected to be the same. Thus,   $\bomd(g_1,g_2)$ can
vanish  between non-isometric spacetimes and is hence not a distance
function. However, one expects that as $n$ increases,  $P_n$ can
resolve the difference between non-isometric
spacetimes. Assuming this, one can then construct  a distance function
on $\ccL$ \cite{bomclose}.  A recent  proof by   Braun proves this
assumption by showing that the 
(labelled) distributions $\PnCmgone$ and $\PnCmgtwo$ are equal for all $n$ iff $g_1$ and $g_2$ are strictly
isometric  \cite{braun}.  

We will now use the  $\Nmn(C_n)$  and $\bNmn\Mg$ for a similar
construction of a closeness function\footnote{For
  convenience of notation we henceforth replace $\av{.}$  with the
  random variable.}. These are non-negative integers and
real numbers, respectively, and having  no further
restrictions, lie in the non-negative hyperoctant $\oR \subset
\re^{n-1}$.  In what follows $\bZplus$ denotes the $n-1$ direct
product of $\overline{\bZ}^+$, the non-negative integers.

We begin by calling two spacetimes 
{\sl $n$-interval isospectral}, if $\forall \, \, n' \leq n$, $
\spec_{n'}\Mgone = \spec_{n'}\Mgtwo.$   We denote this equivalence by
$\Mgone \sim_{n} \Mgtwo$, and define the quotient space $\tcL_n \equiv
\ccL/\!\!\sim_n$.   A pair of non-isometric spacetimes will be said  to be   {\sl interval
   isospectral}  $\Mgone \sim \Mgtwo$ if $\spec_n\Mgone =
 \spec_n\Mgtwo $ $\forall \,\, n >0$.  
This defines the interval isospectral classes $[\Mg]$ in $\ccL$ and  the associated quotient space
   $\tcL \equiv \ccL/\!\!\sim$.
For example, distinct pairs of time reversed  causal sets or
spacetimes are interval isospectral. Other non-trivial discrete transformations on
a spacetime can also produce interval isospectral spacetimes, for
example the parity operator $x\rightarrow -x $ in $d=2$. A general 
classification of   interval  isospectral
spacetimes is of interest -- since these  have the same discrete
Einstein-Hilbert action, they are dynamically equally probable in
the quantum gravity path integral.

Define the map  $p_n: \Omega_n
\rightarrow \bZplus$ where  
\begin{equation}
p_n(C_n)\equiv \{ N_0(C_n), N_1(C_n), \ldots,  N_{n-2}(C_n) \} \in
\bZplus \subset \oR, 
\end{equation}
and $ N_m(C_n) \in \overline{\bZ}^+$.  Similarly, let  $\pi_n: \ccL_n \rightarrow \oR$, where 
\begin{equation}
\pi_n\Mg\equiv \{ \bN_0^{(n)}\Mg, \bN_1^{(n)}\Mg, \ldots, \bN_{n-2}^{(n)}\Mg  \} \in
\oR. 
\end{equation}
Note that (i) $p_n(\Omega_n)$ and $\pi_n(\ccL)$ are sets of measure zero
in  $\oR$,   (ii)  $p_n(C_n)$ is invariant under relabellings
of $C_n$,  and
similarly, $\pi_n\Mg$ is diffeomorphism invariant,   and  
(iii) it is possible for $p_n(C_n)=p_n(C_n')$, or  $\pi_n\Mg =
\pi_n\Mgprime$, i.e., $p_n$ and $\pi_n$  are  not
injective in general. This means that there cannot be an equivalent of  Braun's result for
$\pi_n(M,g)$  since non-isometric spaces can be interval
isospectral.  

We can use  any of the possible $L^r$ distance functions  on
$\oR$ for $r \geq 1 $ to define a  closeness function  between causal sets in
$\Omega_n$   
\begin{equation}
\nndr_n(C_n,C_n') = \biggl( \sum_{m=0}^{n-2} |N_m(C_n)-N_m(C_n')|^r \biggr)^{1/r}, \label{causetddn.eq}
\end{equation}
or  between spacetimes in $\ccL$
\begin{equation}
\nndr_n(g_1,g_2) = \biggl( \sum_{m=0}^{n-2} |\bNmn\Mgone-\bNmn\Mgtwo|^r
\biggr)^{1/r}.
\label{sptddn.eq}
\end{equation}
By definition,  $\nndr_n(.,.)$ satisfies symmetry and the triangle inequality, even
though it is possible for $\nndr_n(.,.)=0$, i.e., it is not positive.
Since our arguments do not depend on the choice of $r$,  we will
henceforth choose  the simplest, the $L^1$  ``taxi-cab''
distance and drop the $r$ label.  

\begin{figure}[!htbp]
  \begin{center}
     \includegraphics[height=4.5cm]{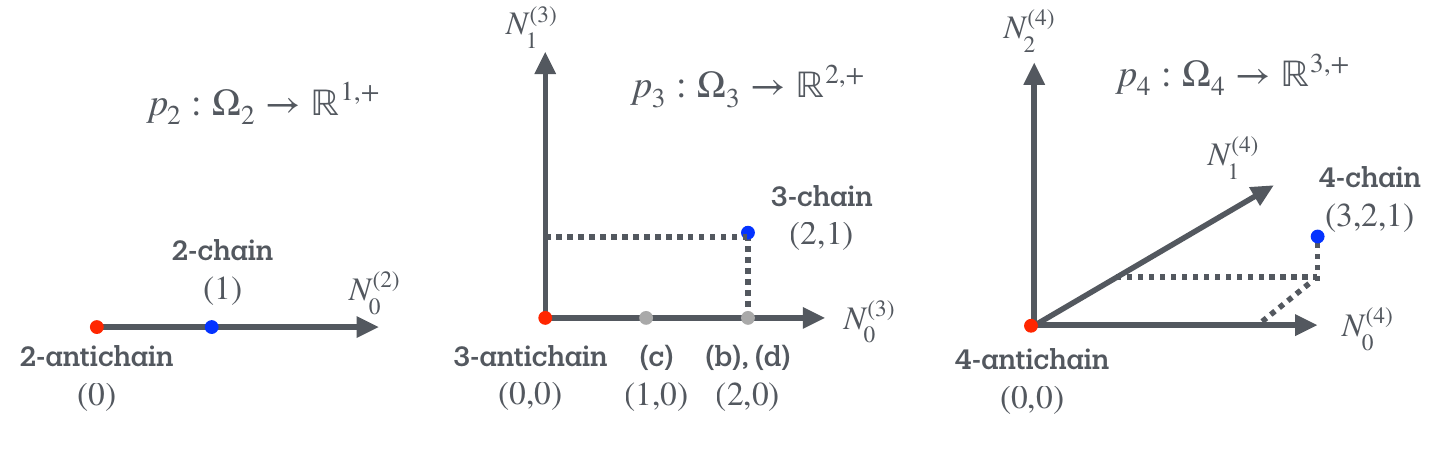} \caption{The
       maps $p_2,p_3$ and $ p_4$}\label{pttf.fig}.
  \end{center}
\end{figure}

We begin with examples from $\Omega_n$. Fig. \ref{pttf.fig} shows
examples of the $p_n$ maps for  $n=2,3$ and $4$.   
$\Omega_2$ consists of  the 2-element chain $\chain_2=\pcauset[small]{1,2}$ and the 2-element
antichain $\achain_2=\pcauset[small]{2,1}$. Hence $p_2:\Omega_2 \rightarrow
\re^1$, with   $p_2(\pcauset[small]{1,2})=\{1\}$, while
$p_2(\pcauset[small]{2,1})=\{0\}$. The latter
lies at the origin of $\oRone$, and the former at the integer
$1$.  Thus, $\nnd_2(\chain_2,\achain_2)=1$.

\begin{figure}[!htbp]
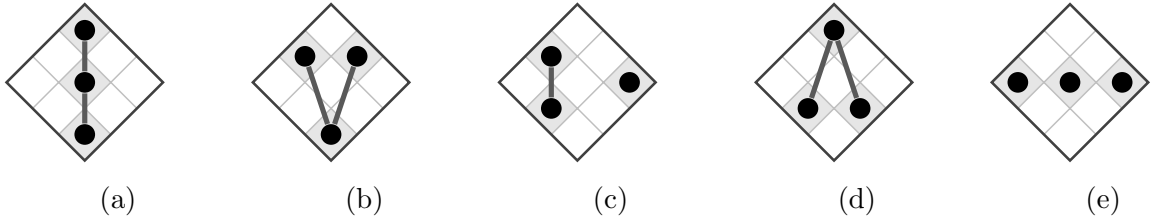

   \begin{subfigure}{.19\textwidth}
      {\pcausetP[huge,event color=black,unlabeled u, unlabeled
        v]{1,2,3}}  \caption{}\end{subfigure}      \begin{subfigure}{.19\textwidth}    {\pcausetP[huge,event color=black,unlabeled u, unlabeled v]{1,3,2}}  \caption{}\end{subfigure}      \begin{subfigure}{.19\textwidth} {\pcausetP[huge,event color=black,unlabeled u, unlabeled v]{3,1,2}} \caption{} \end{subfigure}      \begin{subfigure}{.19\textwidth} \ {\pcausetP[huge,event color=black,unlabeled u, unlabeled v]{2,1,3} } \caption{} \end{subfigure}      \begin{subfigure}{.19\textwidth}  {\pcausetP[huge,event color=black,unlabeled u, unlabeled v]{3,2,1}} \caption{} \end{subfigure} 
\caption{The set $\Omega_3$ of $n=3$-element  unlabelled
     posets.)}\label{omegathree.fig}
\end{figure}
The causal sets in $\Omega_3$ are shown in Fig. \ref{omegathree.fig}.
In order of their appearance in Fig. \ref{omegathree.fig}, these
causal sets  get
mapped to the following points in $\oRtwo$: $(2,1)$, $(2,0)$, $(1,0)$, $(2,0)$
and $(0,0)$ (see Fig. \ref{pttf.fig}). The degeneracy in the
spectrum is already evident  since the causal sets
$\pcauset[small]{1,3,2}$ and $\pcauset[small]{2,1,3}$ are time reversals of each other, and hence map
to the same point  in $\oRtwo$. The distances  between the interval
non-isospectral causal
sets are $\nnd_3(\chain_3=\pcauset[small]{1,2,3},\achain_3=\pcauset[small]{3,2,1})=3$, 
$\nnd_3(\pcauset[small]{1,2,3},
[\pcauset[small]{1,3,2}])=1$, $\nnd_3(\pcauset[small]{1,2,3},
\pcauset[small]{3,1,2})=2$,  $\nnd_3([\pcauset[small]{1,3,2}],
\pcauset[small]{3,1,2})=1$,  $\nnd_3([\pcauset[small]{1,3,2}],
\pcauset[small]{3,2,1})=2$, and $\nnd_3(\pcauset[small]{3,1,2},
\pcauset[small]{3,2,1})=1$, where $[\pcauset[small]{1,3,2}]$ refers to
either $\pcauset[small]{1,3,2}$ or $\pcauset[small]{2,1,3}$.

\begin{figure}[]
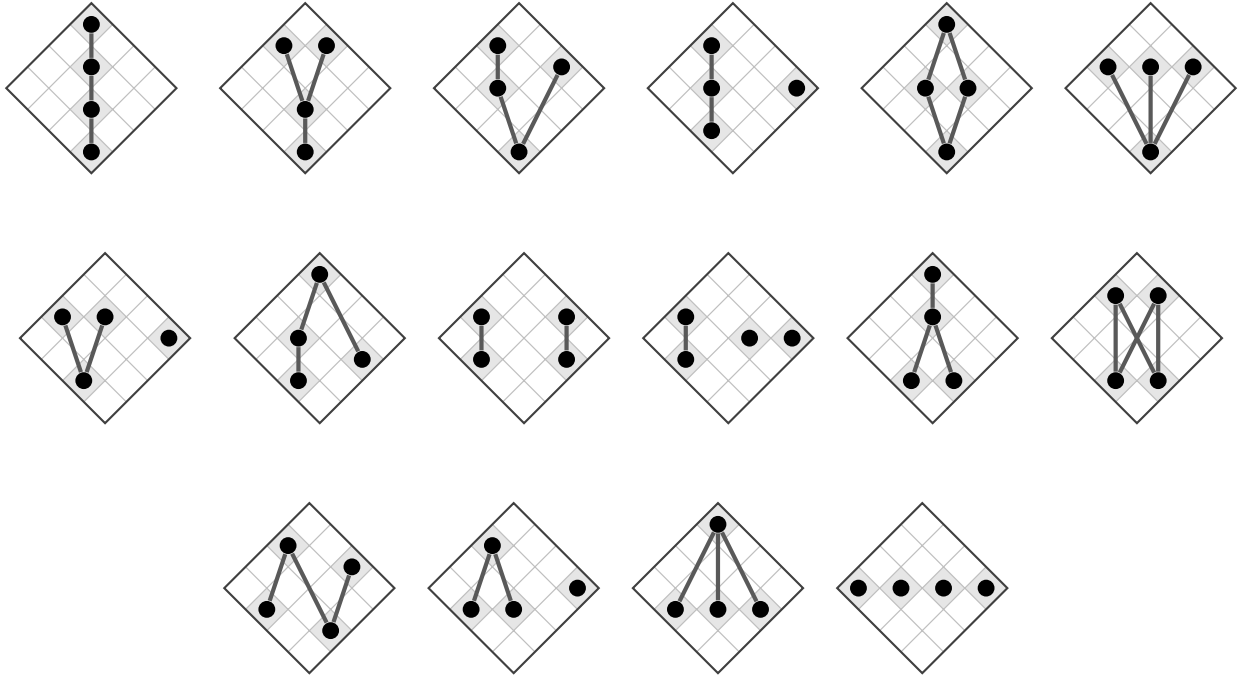

  \begin{center}
    \pcausetP[tile size=0.4cm,event color=black,unlabeled u, unlabeled v]{1,2,3,4} \hspace{0.15cm}
     \pcausetP[tile size=0.4cm,event color=black,unlabeled u, unlabeled v]{1,2,4,3} \hspace{0.15cm}
     \pcausetP[tile size=0.4cm,event color=black,unlabeled u, unlabeled v]{1,4,2,3}
     \hspace{0.15cm}
     \pcausetP[tile size=0.4cm,event color=black,unlabeled u, unlabeled v]{4,1,2,3} \hspace{0.15cm}
     \pcausetP[tile size=0.4cm,event color=black,unlabeled u, unlabeled v]{1,3,2,4} \hspace{0.15cm}
    \pcausetP[tile size=0.4cm,event color=black,unlabeled u, unlabeled v]{1,4,3,2} \\
    \vspace{1cm}
    \pcausetP[tile size=0.4cm,event color=black,unlabeled u, unlabeled v]{4,1,3,2}  \hspace{0.15cm} 
    \pcausetP[tile size=0.4cm,event color=black,unlabeled u, unlabeled v]{3,1,2,4}\hspace{0.15cm}
    \pcausetP[tile size=0.4cm,event color=black,unlabeled u, unlabeled v]{3,4,1,2} \hspace{0.15cm}
    \pcausetP[tile size=0.4cm,event color=black,unlabeled u, unlabeled v]{4,3,1,2}\hspace{0.15cm}
    \pcausetP[tile size=0.4cm,event color=black,unlabeled u, unlabeled v]{2,1,3,4}\hspace{0.15cm}
    \pcausetP[tile size=0.4cm,event color=black,unlabeled u, unlabeled v]{2,1,4,3} \\
     \vspace{1cm}
    \pcausetP[tile size=0.4cm,event color=black,unlabeled u, unlabeled v]{2,4,1,3}\hspace{0.15cm}
    \pcausetP[tile size=0.4cm,event color=black,unlabeled u, unlabeled v]{4,2,1,3}\hspace{0.15cm}
    \pcausetP[tile size=0.4cm,event color=black,unlabeled u, unlabeled v]{3,2,1,4} \hspace{0.15cm}
    \pcausetP[tile size=0.4cm,event color=black,unlabeled u, unlabeled v]{4,3,2,1} 
\vspace{0.6cm}\caption{The set $\Omega_4$ of $n=4$-element unlabelled posets}\label{omegafour.fig} 
 \end{center} 
\end{figure}

The non-trivial degeneracies become evident as one moves to
$\Omega_4$ as depicted in  Fig \ref{omegafour.fig}.
$p_n(\Omega_4)$ is given by the list (from left to right):
\begin{eqnarray} 
&\{ \{3,2,1\}, \{3,2,0\}, \{3,1,0 \}, \{2,1,0\}, \{4,0,1\}, \{3,0,0\},
  \{2,0,0\}, \{3,1,0 \}, \{2,0,0 \}, &\nonumber  \\
  &\{1,0,0 \}, \{3,2,0 \}, \{4,0,0\},  \{3,0,0\}, \{2,0,0\}, \{3,0,0\},
  \{0,0,0\} \}. &
 \end{eqnarray} 
Apart from pairs of time-reversed causal sets, one finds two other non-trivial
degeneracies in the spectra: (i)  $\pcauset[small]{1,4,3,2}$ and
$\pcauset[small]{2,4,1,3}$ and (ii)  $\pcauset[small]{3,4,1,2}$ and $\pcauset[small]{4,1,3,2}$
 
One way to break the degeneracy in the spectrum is to include new  coordinates. For example, to distinguish between some of
the  time-reflected causal sets, one could include the sizes of the
initial  and the final antichains. For $\Omega_3$ this is enough to lift the
degeneracy. For $\Omega_4$, this lifts the first non-trivial
degeneracy but not necessarily the second, if all unrelated elements are considered to be
in the initial antichain. In this specific instance, we can
distinguish the causal sets  by also counting the number $F_m$ of elements with $m$ elements
in their future,  and similarly the number $P_m$ of elements with $m$ elements in their past. 

It becomes rapidly more difficult to repeat this calculation for
larger $n$, since   $|\Omega_n| \sim
2^{\frac{n^2}{4}+o(n^2)}$ for large $n$.  However the simplicity of the
$n$-element chain $\achain_n$ and $\chain_n$ allows us an easy
analytic calculation. For these causal sets, the coordinates in $\oR$
are         
\begin{equation}
p_n(\achain_n) = (0,0, \ldots, 0), \, \,  p_n(\chain_n) =(n-1,n-2,n-3, \ldots,1), 
  \end{equation} 
and hence 
\begin{equation}
\nnd_n(\achain_n,\chain_n)=\frac{(n-1)(n)}{2} \sim n^2/2, 
\end{equation}
for large $n$. We can also calculate $p_n$ for slightly more
complicated causal sets, like the 
$n$-element ``L'' poset $\mathbf L_n$  obtained by inserting a single
link between two elements of $\achain_n$  as well as the $n$-element ``Y'' poset $\mathbf
Y_n$ obtained by removing the maximal element of $\chain_n$ and linking
it to the $(n-2)^{\mathrm{th}}$ (or  last but one) future most element
of $\chain_{n-1}\subset \chain_n$. For these
\begin{equation}
 p_n(\mathbf L_n)=
(1,0, \ldots, 0), \, \,p_n(\mathbf Y_n)=(n-1,n-2,n-3, \ldots,2,0), 
\end{equation} 
and 
\begin{eqnarray} \nnd_n(\achain, \mathbf L_n)=\nnd_n(\chain, \mathbf Y_n) =1, &\, \,& \nnd_n(\achain, \mathbf
                                                         Y_n)=\frac{(n-1)(n)}{2}, \nonumber \\
  \nnd_n(\chain, \mathbf
L_n)=\frac{(n-1)(n)}{2}-1 &\, \,& \nnd_n(\mathbf
L_n,\mathbf Y_n)=\frac{(n-1)(n)}{2}-2. 
\end{eqnarray} 
Thus, as $n$ increases, the closeness function between
$\achain_n$ and $\chain_n$ increases, as does that between $\achain_n$
and $\mathbf Y_n$, $\chain_n$
and $\mathbf L_n$ and $\mathbf L_n$ and $\mathbf Y_n$. All these
functions grow as $n^2/2$.  However, the closeness function   between
$\achain_n$  and $\mathbf L_n$, as well as that between $\chain_n$
and $\mathbf Y_n$ is independent of $n$ and hence relatively speaking,
they get closer to each other as $n$ increases, while remaining spectrally 
distinct from each other for all $n$. This ties in with our intuition that for
large $n$, the differences between $\achain_n$ and $\mathbf L_n$ wash
out, as do those between  $\chain_n$ and $\mathbf Y_n$.  

While expanding the  repertoire of causal sets for which $\nnd_n(.,.)$  is
analytically calculable is an interesting exercise in combinatorics,
our main goal here is to use it on $\ccL$, the space of
all finite 
volume spacetimes. By fixing $n$ but not the spacetime volumes,  we can 
use $\nnd_n(.,.)$ to compare  spacetimes of different volumes.  For
example, consider the  product spacetime
$\Mg=(M_1\times M_2, g_1 \otimes g_2)  $, where $M_2$ is an
``internal''  compact space, with $\rho^{-1} = \vol(M)/n
\gg \vol(M_2)$. In this case we expect $\bC_n\Mg $ and $\bC_n\Mgone$
to be very similar, since the sprinkling is  too coarse
to see the internal manifold $\Mgtwo$ and  thus only captures aspects of  $\Mgone$. As $n$
increases, however, the differences in  $\bC_n\Mg $ and $\bC_n\Mgone$
will begin to manifest themselves.  While  $\nnd_n(g, g_1) \approx
0$ for $n\ll \vol(M)/\vol(M_2)$, we would expect  it to increases with
$n$, thus  pushing the 
spacetimes further apart.  In our simulations we
have illustrated  this in the spectral comparisons of  $\diam^d$ with its  ``thickened'' version $\diam^d \times \mathbb I_t$,
in  Figs. \ref{Fig4a.fig} and \ref{Fig4b.fig}
and the closeness function in Fig. \ref{distmdtc.fig}.

We would like to use $\nnd_n(., .)$ as a  measure of an approximate, {\sl $n$-isometry} between
spacetimes, but because of interval isospectrality, we will need to be careful.  If  two spacetimes  are $n$-interval isospectral and further, there exists an $n_0>n$ such that $\forall \, n'>n_0$
$\Mgone \nsim_{n'} \Mgtwo $, then from
Braun's result \cite{braun} we know that $P_{n'}\Mgone \neq P_{n'}\Mgtwo$ and hence
$\Mgone$ and $\Mgtwo$ are strictly non-isometric. Hence,  in a well defined
sense, $\Mgone $ and $\Mgtwo$ can be said to be   {\sl approximately  $n$-interval
  isometric} in $\ccL$. Thus, even if we cannot hope to extend this
definition to all non-isometric pairs of spacetimes, 
we can still give a precise notion of approximate
isometry for a  sub-class of spacetimes.   We define the following  convergence condition on
$\tcL$: 
\begin{definition} 
Consider a sequence of equivalence classes of finite volume spacetimes
$\{[\Mgi]\}$, $[\Mgi] \in \tcL$. We 
will say that it  {\sl interval-converges}  to $[\Mg]\in \tcL$ if for
every $\epsilon>0$ and $n>0$, 
there exists an  $i_0>0$ such that for all 
$i >i_0$, $\nnd_n([\Mgi],[\Mg])<\epsilon$.  We will say that $\{[\Mgi]\}$ are
$(\epsilon,n)$-close to $[\Mg]$ for $i>i_0$. 
\end{definition} 
As $n$ increases, the closeness function between non-interval
isospectral spacetimes should typically increase. Hence if
$\{[\Mgi]\}$ is $(\epsilon, n)$- close to $[\Mg]$, then it need not be
$(\epsilon, n')$-close to $[\Mg]$ for $n'>n$.  One can also tie $n$ to
$\epsilon$ by defining $\epsilon=1/F(n)$, where $F(n)$ is a
monotonically increasing function of $n$, to define a stricter
$1/F(n)$-closeness, with $\nnd_n([\Mgi],[\Mg])<1/f(n)$. 

Despite being strictly defined only on $\tcL$,  this convergence
criterion can be usefully  applied to special classes of spacetimes in 
$\ccL$. 
Consider for example a one parameter family $\{ g(s) \} $ of
cosmological FRW  spacetimes where the  scale
factor is given by $a_{s}(t)=t^{f(s)}$ or, in the case of de Sitter,
$a_s(t)=e^{f(s) t}$.  Let  $f(s) >0$ be a monotonically increasing
function of  $s$.  Consider a sequence of spacetimes  $\{
a_{s_i}(t)\}$, or equivalently a sequence  of functions $\{f(s_i)\}$
such that $\lim_{s_i \rightarrow \infty} f(s_i) =  f_0 <\infty$,
equivalently,  $\lim_{s_i \rightarrow \infty}  a_{s_i}(t)  = 
a_0(t)$.

This sequence of (non-isometric) montonically expanding FRW and de Sitter
are  also not interval isospectral as we now argue.

Consider two spacetimes $(M,g(s_1))$  and $(M,g(s_2))$,
with $s_1 < s_2$.  Let us assume that $\rho$ is a constant, and compare the interval
spectrum of equal volume  regions of the spacetimes, i.e., $\vol(M,g(s_1))=\vol(M,g(s_2))$.  Since the $g(s_i)$ are  conformally related, the
continuum Alexandrov intervals $(p,q)_{s_i}$ remain the same, but
their volumes do not, i.e., $\vol((p,q)_{s_1})
<\vol((p,q)_{s_2})$. Hence on average, every $m_1$-element interval in
$(M,g(s_1))$ is an $m_2$-element interval in $(M,g(s_2))$ where $m_2 > m_1$, i.e.,
$\bN_{m_2}(s_2)=\bN_{m_1}(s_1)$. Since 
$\bN_m$ is a monotonically decreasing function of $m$,
 $\bN_{m_1}(s_2) >
\bN_{m_2}(s_2)=\bN_{m_1}(s_1)$. Thus, these spacetimes are not
interval 
isospectral.  It is important to note that there are no smoothness
assumptions on $a_{c,p}(t)$ as a function of $c,p$ or even $t$.

\begin{figure}[!htbp]  \begin{subfigure} {.5\textwidth}
   \includegraphics[height=4.9cm]{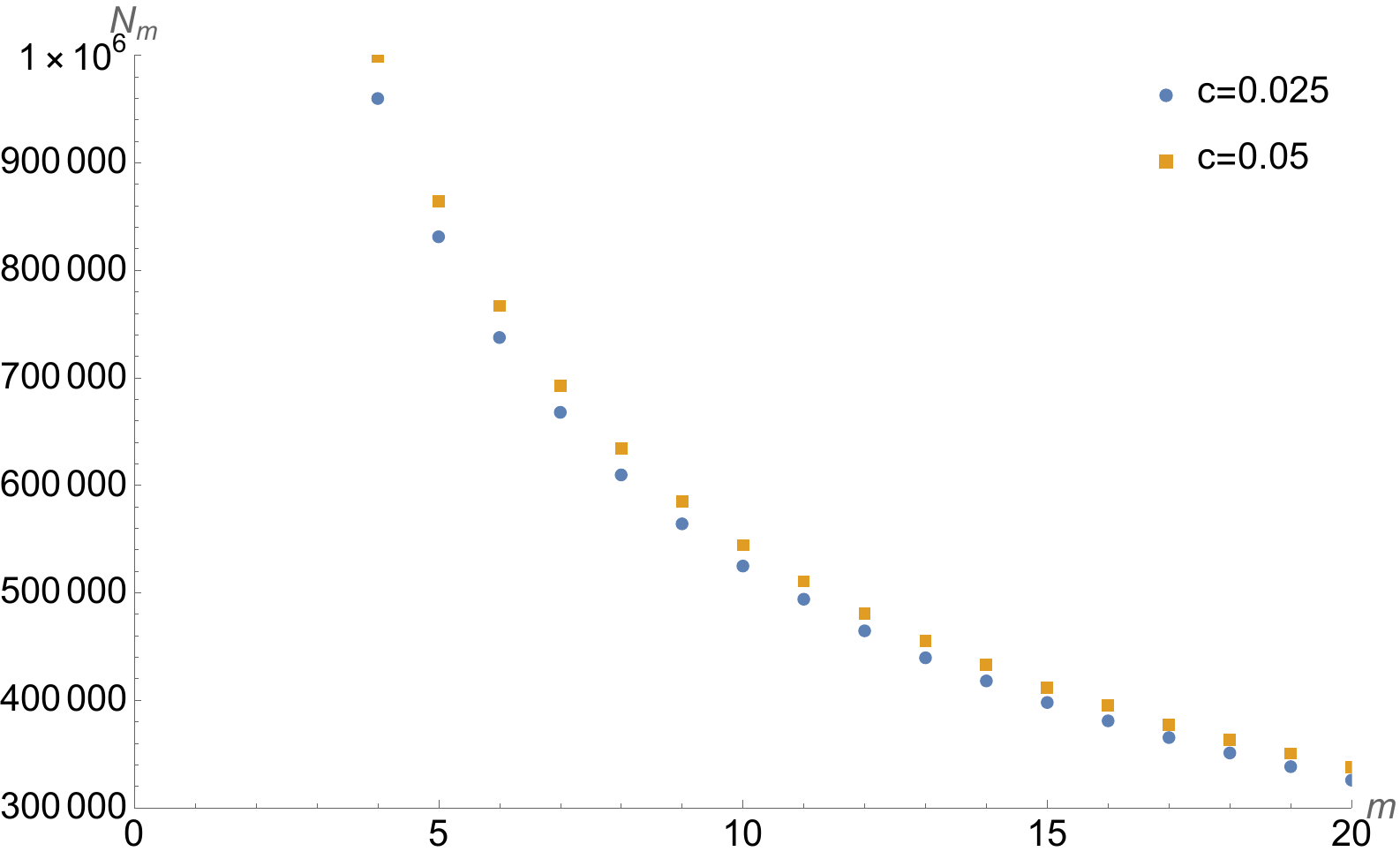} 
                \caption{$\dS_5(e^{c t})$, small $m$} \label{CloserLook.fig}\end{subfigure}
\begin{subfigure} {.5\textwidth}
  \centering  \includegraphics[height=4.9cm]{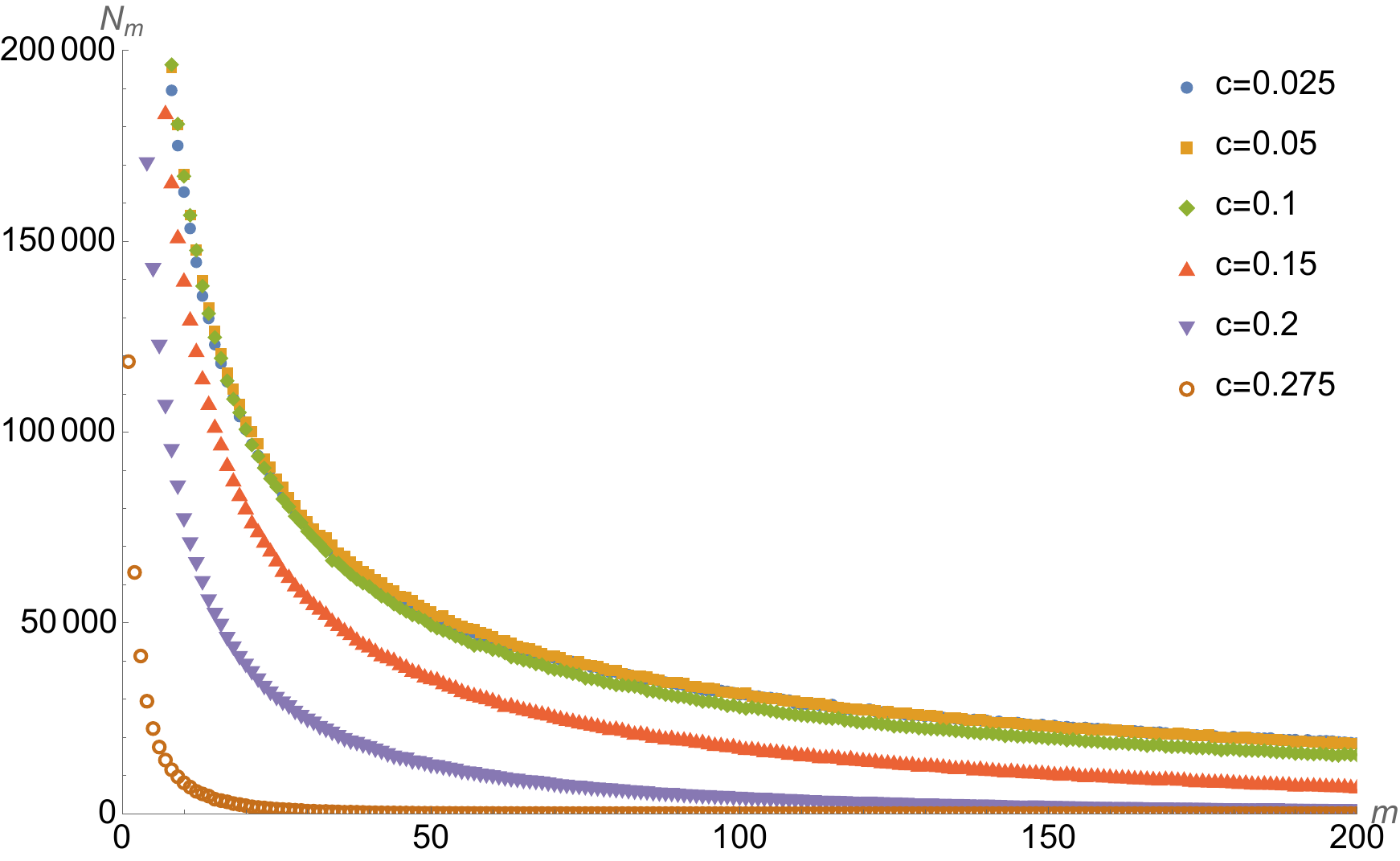} \caption{$\dS_5(c)$,
      $n=10,000$}  \label{intfrwfrwdSa.fig}\end{subfigure}  \begin{subfigure} 
    {.5\textwidth}  \centering \includegraphics[height=4.9cm]{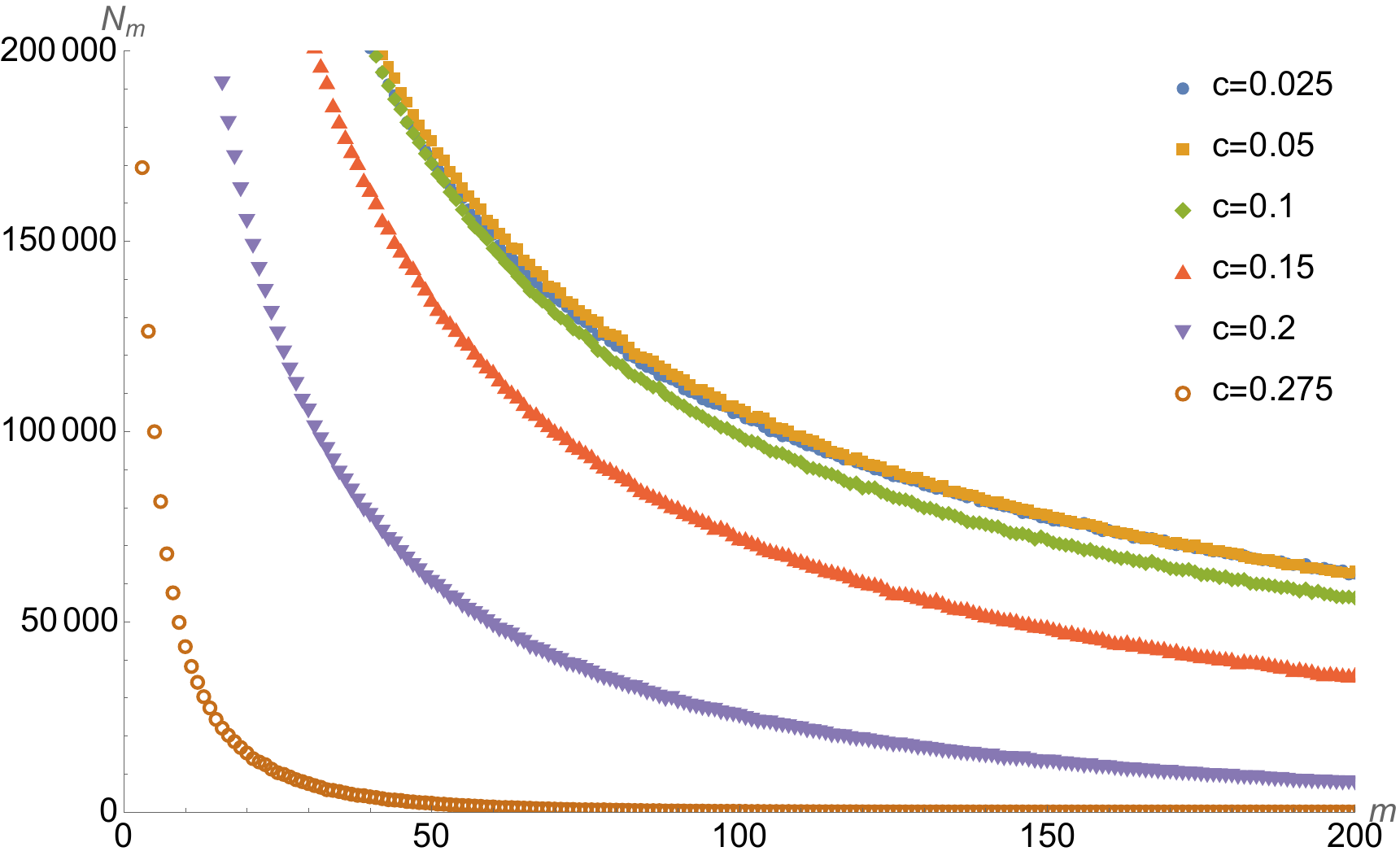} \caption{$\dS_5(c))$,
      $n=20,000$}  \label{intfrwfrwdSb.fig}\end{subfigure}
\begin{subfigure} {.5\textwidth}       \centering 
    \includegraphics[height=4.9cm]{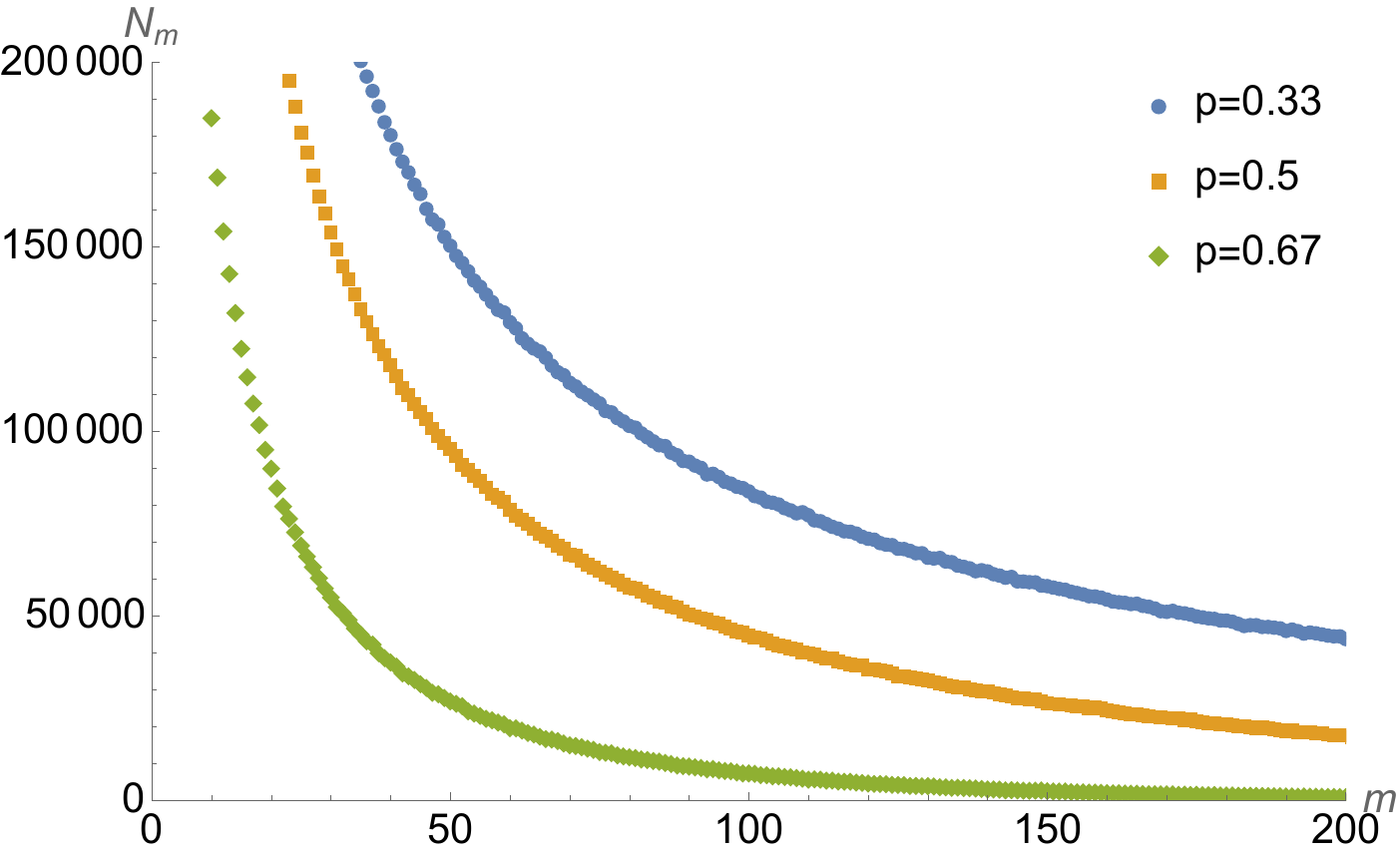} 
               \caption{$\bbF_5(p)$, $n=20,000$.}  \label{intfrwfrwdSc.fig}\end{subfigure}
               \caption{Fig. (a) shows $\bNmn$ 
                  for smaller values of $m$ 
                  for $\dS_5(e^{c t})$ with two different scale
                  factors $c=0.025, 0.05$ for $n=20,000$. 
              Figs. (b) and (c) show $\dS_5(e^{c t})$ for larger $m$
              values for a range of $c$ values for  $n=10,000$ and
              $20,000$. One can see how the 
                 $n$-interval spectral degeneracy is lifted as $n$
                 increases. Fig. (d) shows the 
               $n$-interval spectra
                  for FRW spacetimes $\bbF_5(t^p)$ for  $p=0.33,0.5,0.67$. 
                } 
                 \label{intfrwfrwdS.fig}
              \end{figure}

How does this square with our simulations? An example of this expected 
behaviour is shown in Fig. \ref{CloserLook.fig} for small values of $m$.   
However, because we fix $n$, and consider different 
volumes of the spacetime regions,  the analytic argument is too
simplistic (or equivalently our simulations are too limited) to
explain the behaviour for all $m$.  In particular, in our
simulations of  $\dS_d$ and $\bbF_d$ we have fixed the initial and final
coordinate times, for different scale factors.  While the small $m$
behaviour is as we expect analytically, there is a cross over at
larger $m$ values with $\bN_m(s_1) > \bN_n(s_2)$ for $s_1<s_2$. This
is because the small $m$ values are less affected by boundary effects,
while for large $m$, they do matter. In particular, the spacetime
region grows ``wider'' with increasing $s_i$, and thus there are fewer large
$m$ intervals. In either case, the simulations support the
overall argument that these  spacetimes are not interval isospectral.  Figures   
\ref{intfrwfrwdSa.fig} and \ref{intfrwfrwdSb.fig} show  simulations 
comparing  the interval spectra for $\dS_5(c)$, $a_c(t)=e^{c t}$ for different values of
$c$. For $n=10,000$ the spectra for $c=0.025,0.05$ and $0.1$ lie
almost on top of each other, but for  $n=20,000$, the $c=0.1$ spectrum
is distinguishable from the others.  In Fig. \ref{intfrwfrwdSc.fig} the
interval spectrum $a_p(t)=t^{p}$ is shown for
$p=0.33,0.5,0.67$.
In principle we could include both monotonically increasing and decreasing
functions of $t$ to our set of FRW-type spacetimes,  i.e., expanding and 
contracting spacetimes, and lift the degeneracy in the $n$-interval
spectrum  by including the sizes of the initial and
final antichains as an additional pair of order invariants, with the closeness function now
calculated in  $\bar{\re}^{n+1,+}$.  The interval-convergence
condition would 
then trivially  extended to this larger space, with  
contracting  and expanding spacetimes being strictly {\it not}
$(\epsilon,n)$ close to each other for arbitrarily large $n$.   


Since we have an analytic expression Eqn. \eqref{nmid.eq} for the
$\bNmn(\diam^d)$,  we can begin by testing the closeness  function
 by varying the  dimension. For 
small $n$, $\nnd_n(\diam^{d_1}, \diam^{d_2})$ varies
exactly as one might wish, growing as the difference in
dimension grows.  However, for large 
$n$, from  Eqn. \eqref{asympnmid.eq} we see that  $\nnd_n(\diam^{d_1}, \diam^{d_2})$ grows like 
$n^{2-2/d}$ , where $d=maximum(d_1,d_2)$. Hence,  the contribution
to the closeness function  is dominated by $|\bNmn(\diam^d)|$. This in turn means
that as $n$ becomes larger,  $\diam^d$ becomes equally far away from all 
$\diam^{d'}$ for $d'<d$.

It is possible to replace  the  
$\bNmn\Mg$ by the normalised $\nbNmn\Mg$ in Eqn. \eqref{normNs.eq},  which is
independent of $n$ to leading order.  The ``normalised'' closeness
function is then 
\begin{equation}
\nd_n(\Mgone,\Mgtwo) = \sum_{m=0}^{n-2} |\nbNmn\Mgone-\nbNmn\Mgtwo|. 
\end{equation}
While this can  lead to better convergence conditions for large $n$ , it
 doesn't satisfy the triangle inequality which makes it less desirable
 to work with. 

\subsubsection{Simulations}

We show examples of  $\nnd_n(.,.)$ using the $n$-interval
spectra $\spec_n(.)$ of the various types of spacetime regions shown in Fig.\ref{regions.fig}.  These simulations show
that the closeness function behaves in a manner one expects in simple
cases, but also uncovers new features. For
example, it is not always the case that the closeness function between
spacetimes of different dimensions increases with the difference in
dimension -- other geometric features can make them closer to lower or
higher dimensional spacetimes with different geometries.

\begin{figure}[!htbp]
  \begin{center}
    \includegraphics[height=6cm]{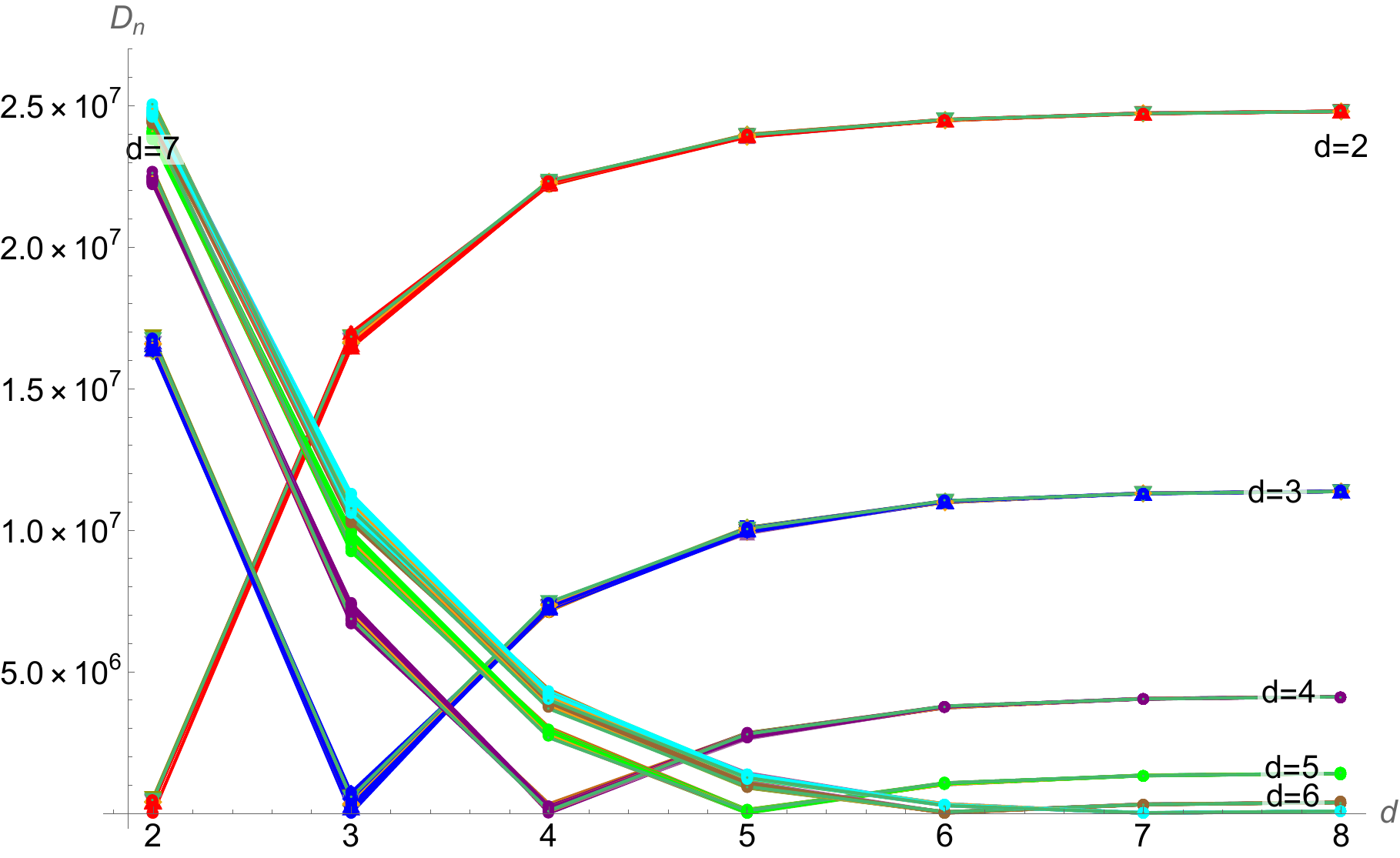} 
                \caption{The figure shows $\nnd_n(\diam^{d},\diam^{d'})$ for $n=10,000$ 
              and  $d=2,\ldots 7$ and $d'=2,\ldots, 8$, over 30 trials.} \label{distmdmdtrials.fig}
                \end{center} 
              \end{figure}

Figure \ref{distmdmdtrials.fig} shows the closeness function 
$\nnd_n(\diam^{d_1},\diam^{d_2})$ as one varies over $d_1,d_2$ for
$d_i=2, \ldots 8$ for $n=10,000$, and over 30 trials. The closeness function increases
monotonically with $|\Delta d|$ where $ \Delta d \equiv (d_1-d_2)$,  but the rate of change 
depends on $\Delta d$. For fixed $d_1$, and $d_2<d_1$ the rate of
change increases with  $|\Delta d|$ while for $d_2>d_1$, it decreases with  $|\Delta d|$.
Because there is no significant difference between trials,
as before  we use a single realisation of the sprinkled causal
set.

\begin{figure}[!htbp]
  \begin{center}
    \includegraphics[height=6cm]{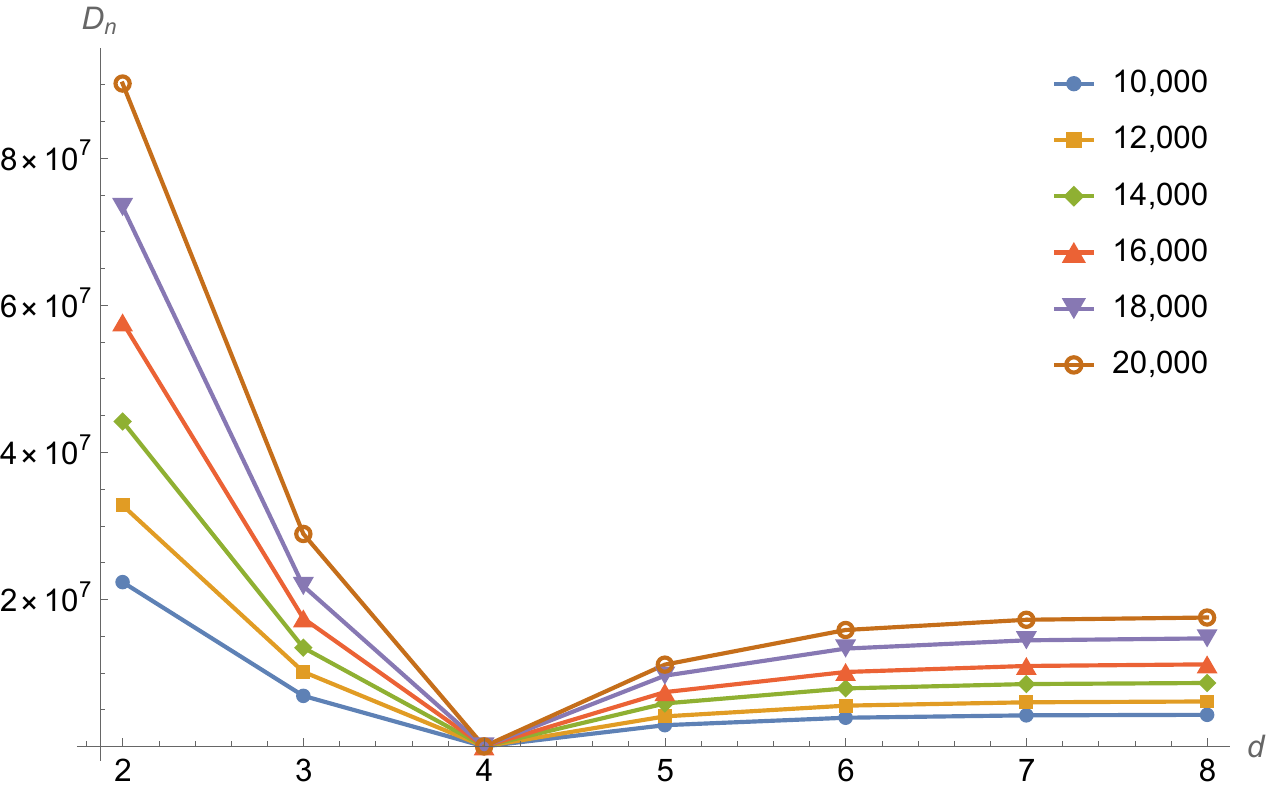} 
                \caption{The figure shows $\nnd_n(\diam^{4} ,\diam^{d}
                  )$ for a 
                  range of $n$ values as  
            $d=2,\ldots, 8$. } \label{distmdmdn.fig}
                \end{center} 
              \end{figure}

\begin{figure}[!htbp]
\begin{subfigure} {.5\textwidth}
 \centering{ \includegraphics[height=4.9cm]{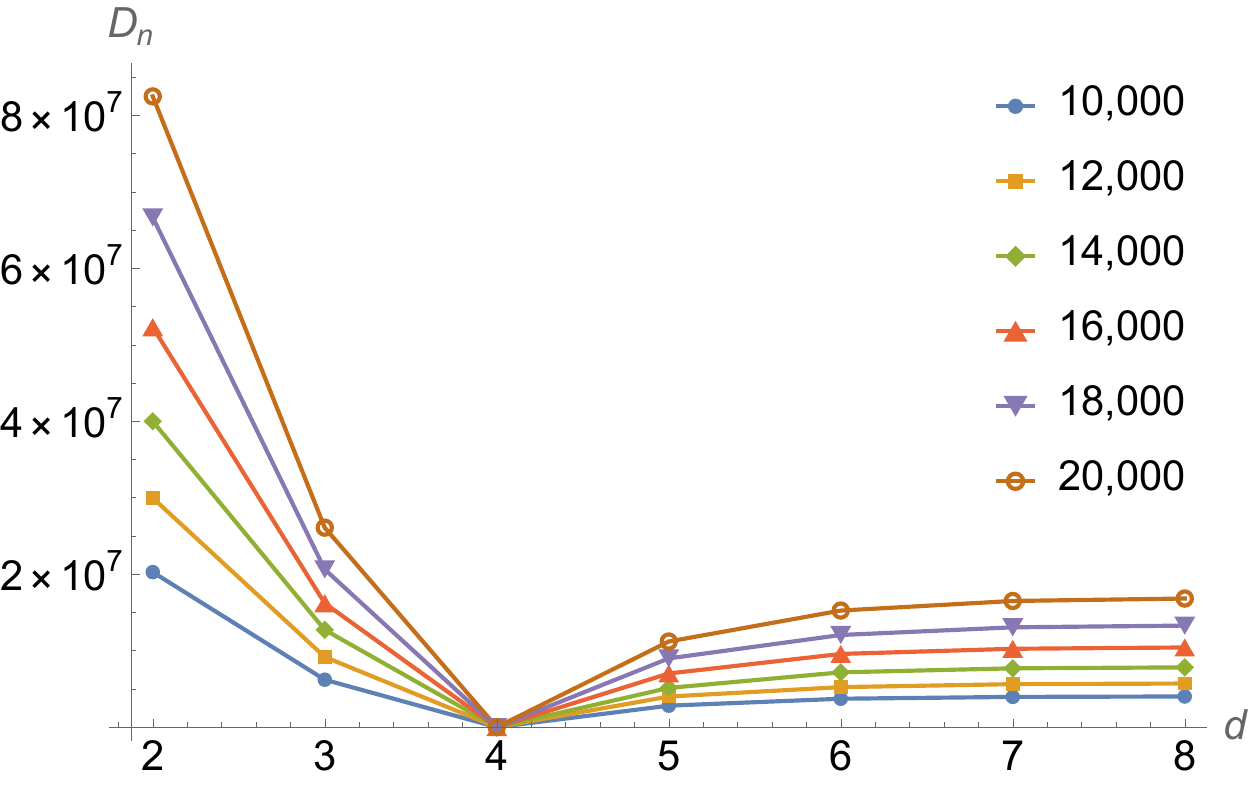} \caption{$\nnd_n(\diam^{4} \times
                  \mathbb I_{0.05} ,\diam^{d} \times \mathbb I_{0.05} )$.} \label{tctcfixtd.fig} }
\end{subfigure} \begin{subfigure} {.5\textwidth}
\centering{  \includegraphics[height=4.9cm]{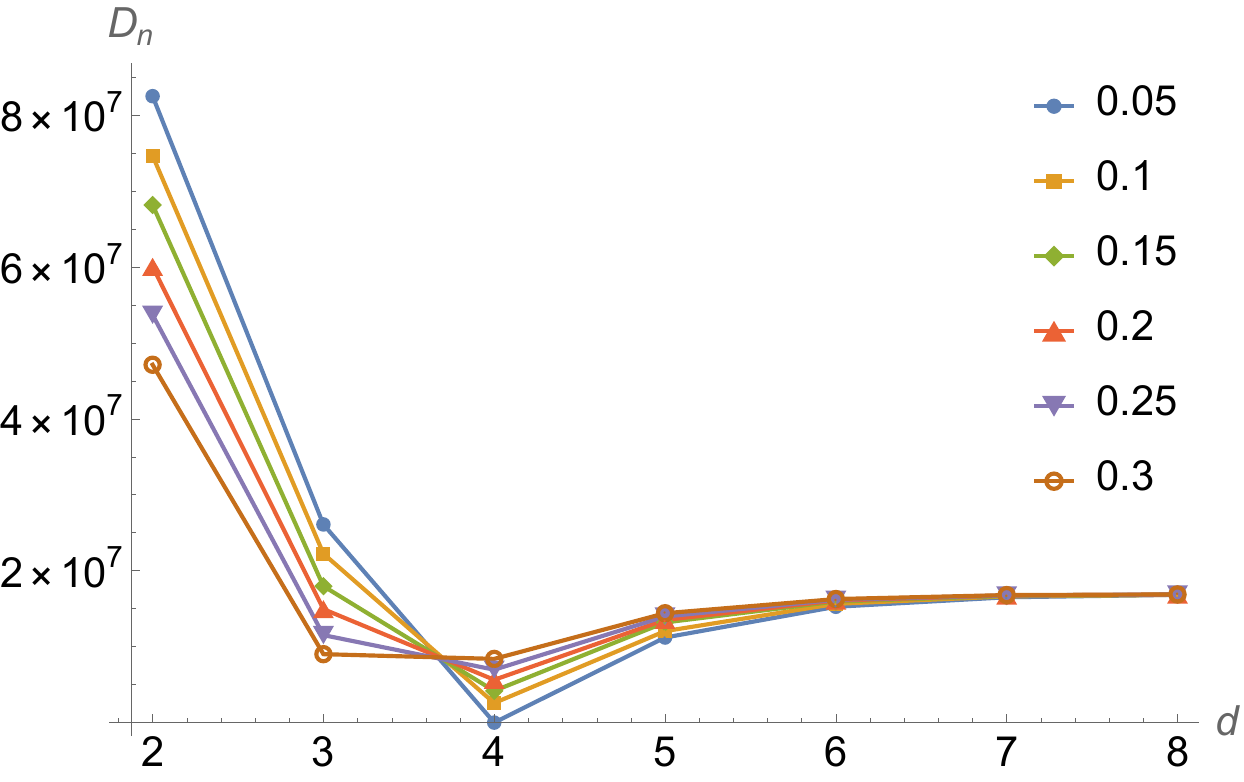}\caption{$\nnd_n(\diam^{4} \times
                  \mathbb I_{0.05} ,\diam^{d} \times \mathbb I_t)$.} \label{tctctt.fig} } \end{subfigure} 
                \caption{$\nnd_n(\diam^{4} \times
                  \mathbb I_{0.05} ,\diam^{d} \times \mathbb I_{t} )$
                  as a function of dimension  $d=2,\ldots, 8$. (a) $t=0.05$ is fixed and $n$ is varied.  
     (b)  $n=20,000$ and  $t$ is varied. } \label{disttctc.fig}
\end{figure}

Figure \ref{distmdmdn.fig} shows how the distance
$\nnd_n(\diam^{4},\diam^{d})$ changes as a function of $n$ as one
varies $d$.
Figure \ref{tctcfixtd.fig} shows the closeness function $\nnd_n(\diam^4
\times \mathbb I_t, \diam^d \times  \mathbb I_t)$ for thickness $t=0.05$ for both
spacetimes.  The behaviour in this case 
            is similar to that of the $\diam^d$ shown in
            Fig. \ref{distmdmdn.fig}.    Fig.  \ref{tctctt.fig}
            compares  $\nnd_n(\diam^4
\times \mathbb I_t, \diam^d \times  \mathbb I_t)$ between $t=0.05$ and
different thicknesses.   As $t$ increases, there is an interesting
cross over where the closeness  function  between $d=4, t=0.05$ and
$d=2,3$ decreases as $t$ increases. One can see that this is
reasonable: as the thickness of $\diam^3\times \mathbb I_t$ increases, it
becomes closer to $\diam^4 \times \mathbb I_{0.05}$.    

\begin{figure}[!htbp] \centering 
                \begin{subfigure}{.3\textwidth}
 \centering 
    \includegraphics[height=3cm]{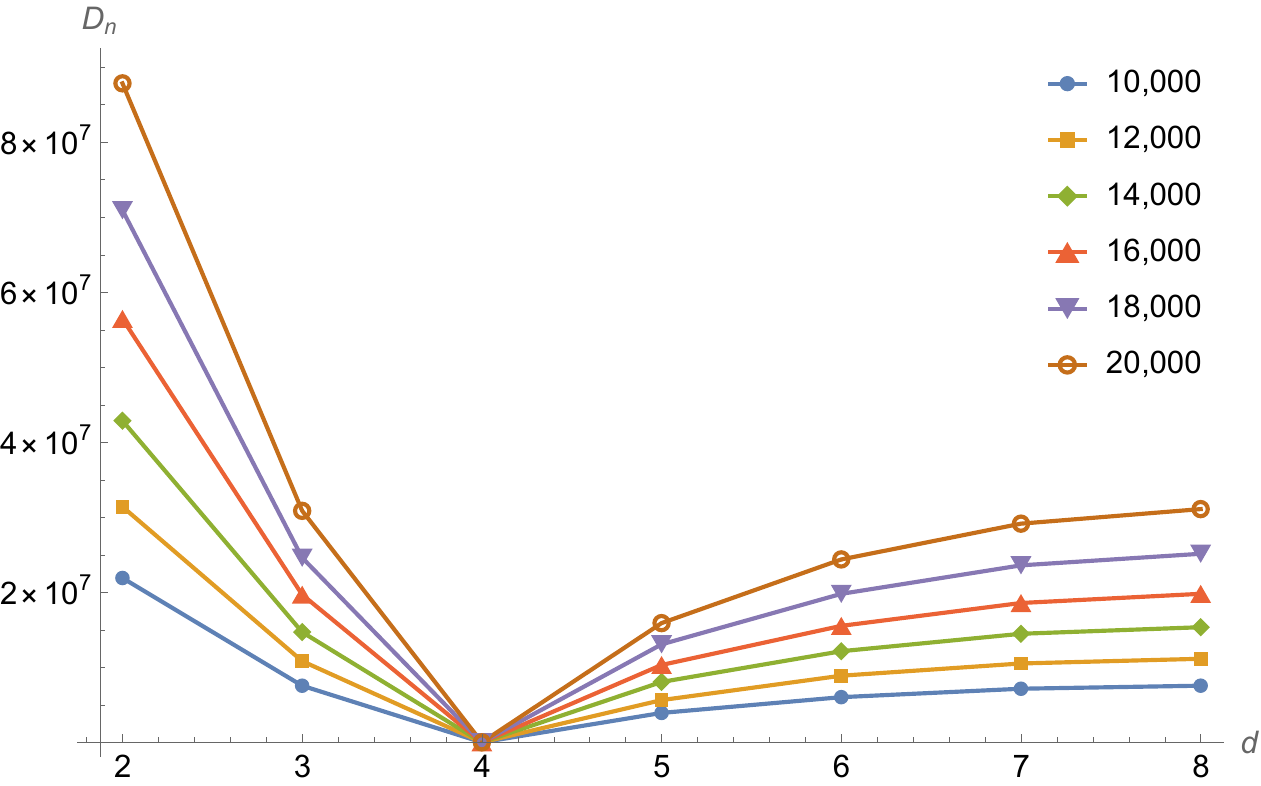} \caption{$\nnd_n(\mathbb I^4,\mathbb I^d)$}  \label{distccccfour.fig}
            \end{subfigure}
                \begin{subfigure}{.3\textwidth}
 \centering 
    \includegraphics[height=3cm]{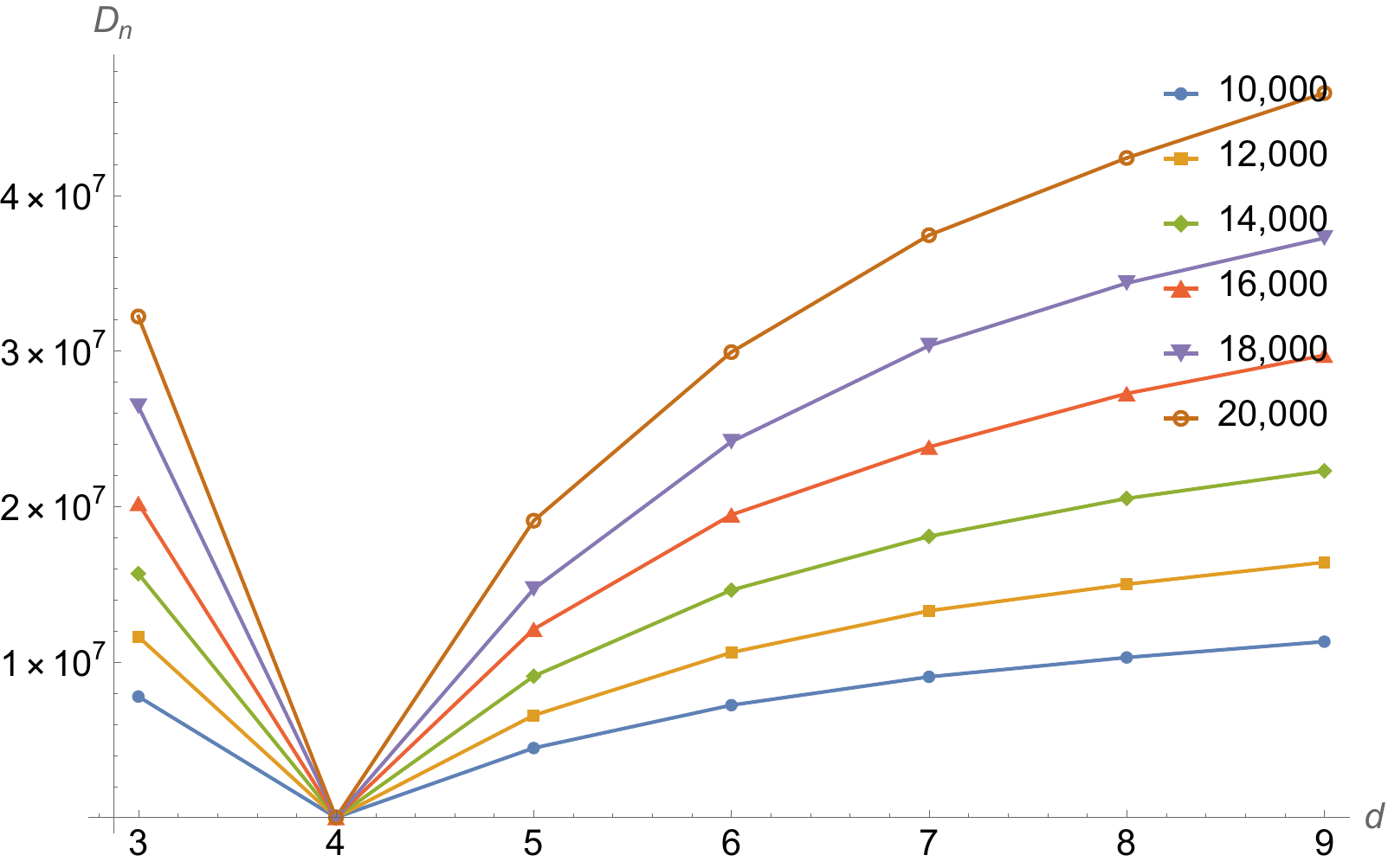} \caption{$\nnd_n(\bbF_4(t^{1/3}), \bbF_d(t^{1/3}))$ } 
  \end{subfigure}
  \begin{subfigure}{.3\textwidth}
 \centering 
    \includegraphics[height=3cm]{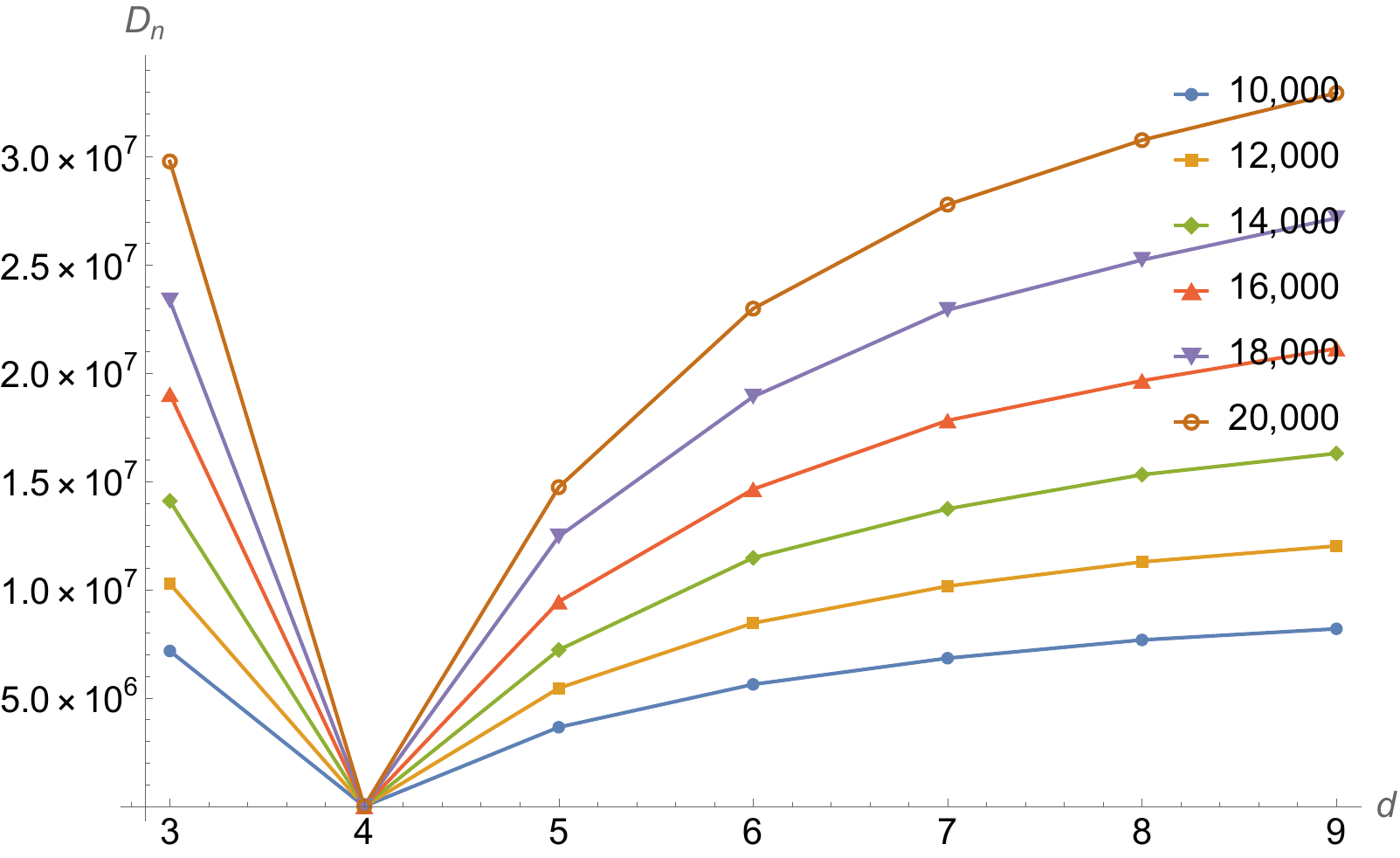} \caption{$\nnd_n(\bbF_4(t^{1/2}), \bbF_d(t^{1/2}))$ } 
  \end{subfigure}
  \begin{subfigure}{.3\textwidth}
 \centering 
    \includegraphics[height=3cm]{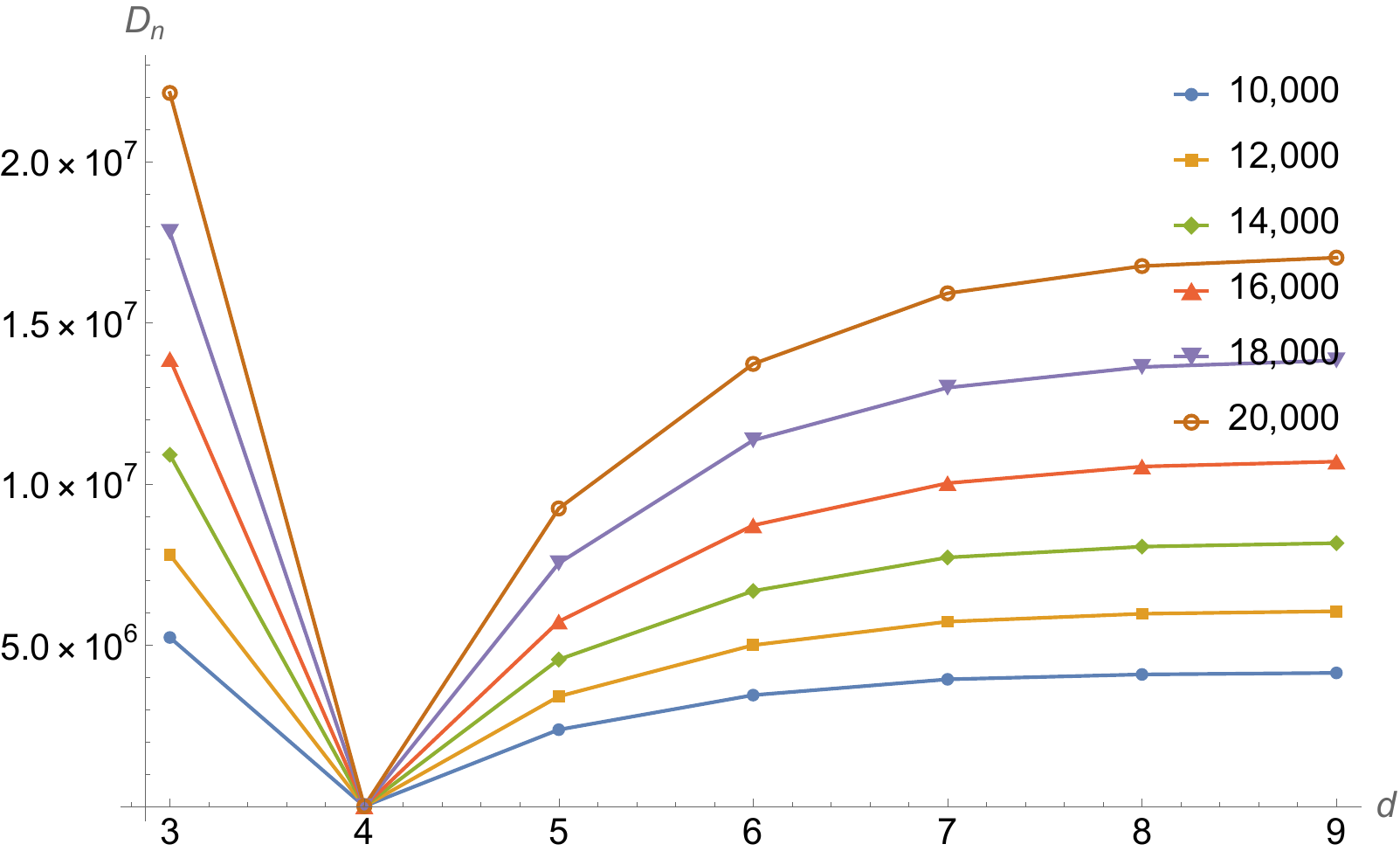} \caption{$\nnd_n(\bbF_4(t^{2/3}), \bbF_d(t^{2/3}))$ } 
  \end{subfigure}
    \begin{subfigure}{.3\textwidth}
 \centering 
    \includegraphics[height=3cm]{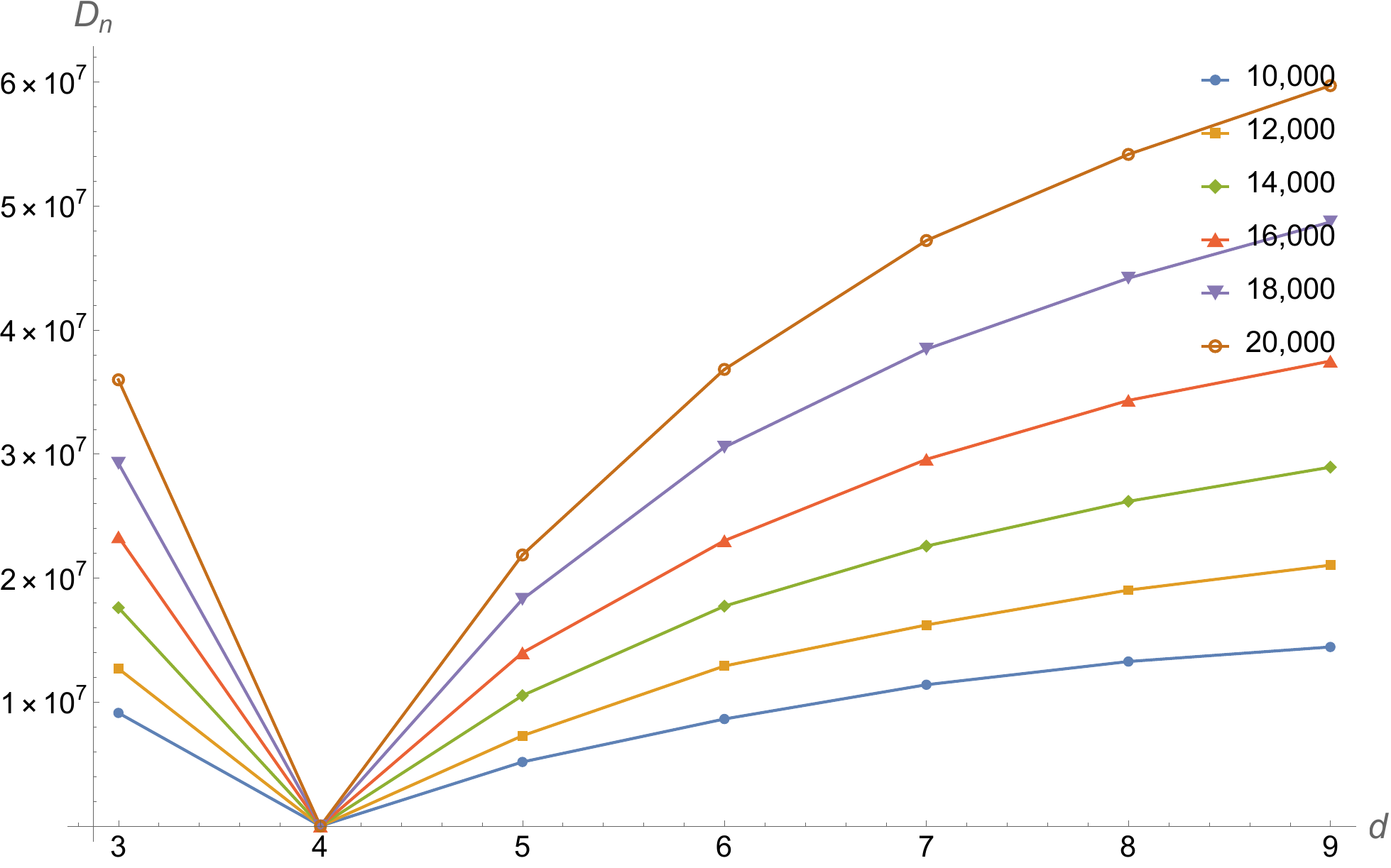} \caption{$\nnd_n(\dS_4(e^{0.05\,t}), \dS_d(e^{0.05\, t}))$ } 
  \end{subfigure}
  \begin{subfigure}{.3\textwidth}
 \centering 
    \includegraphics[height=3cm]{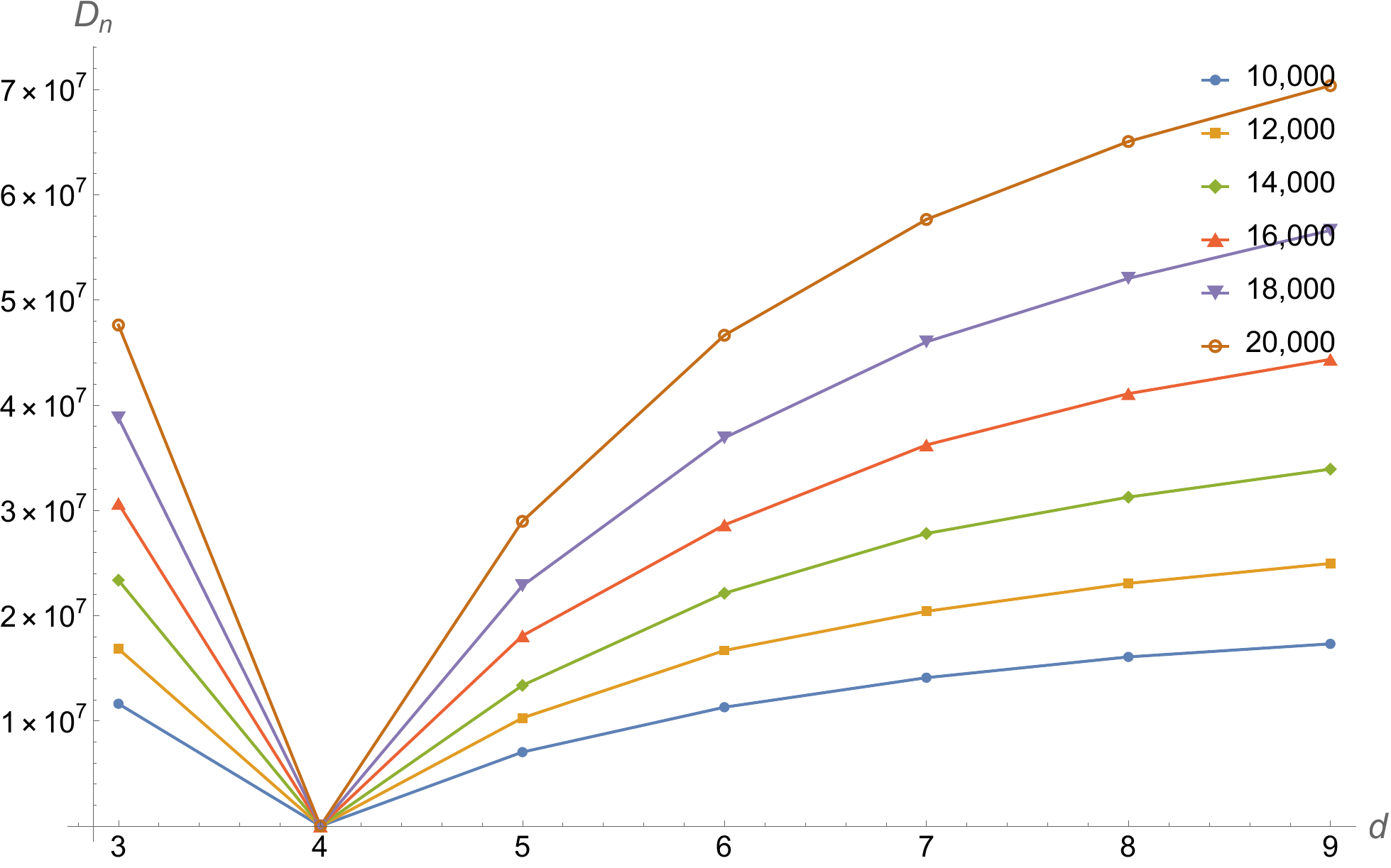} \caption{$\nnd_n(\dS_4(e^{0.1\,t}), \dS_d(e^{0.1, t}))$ } 
  \end{subfigure}
  \begin{subfigure}{.3\textwidth}
 \centering 
    \includegraphics[height=3cm]{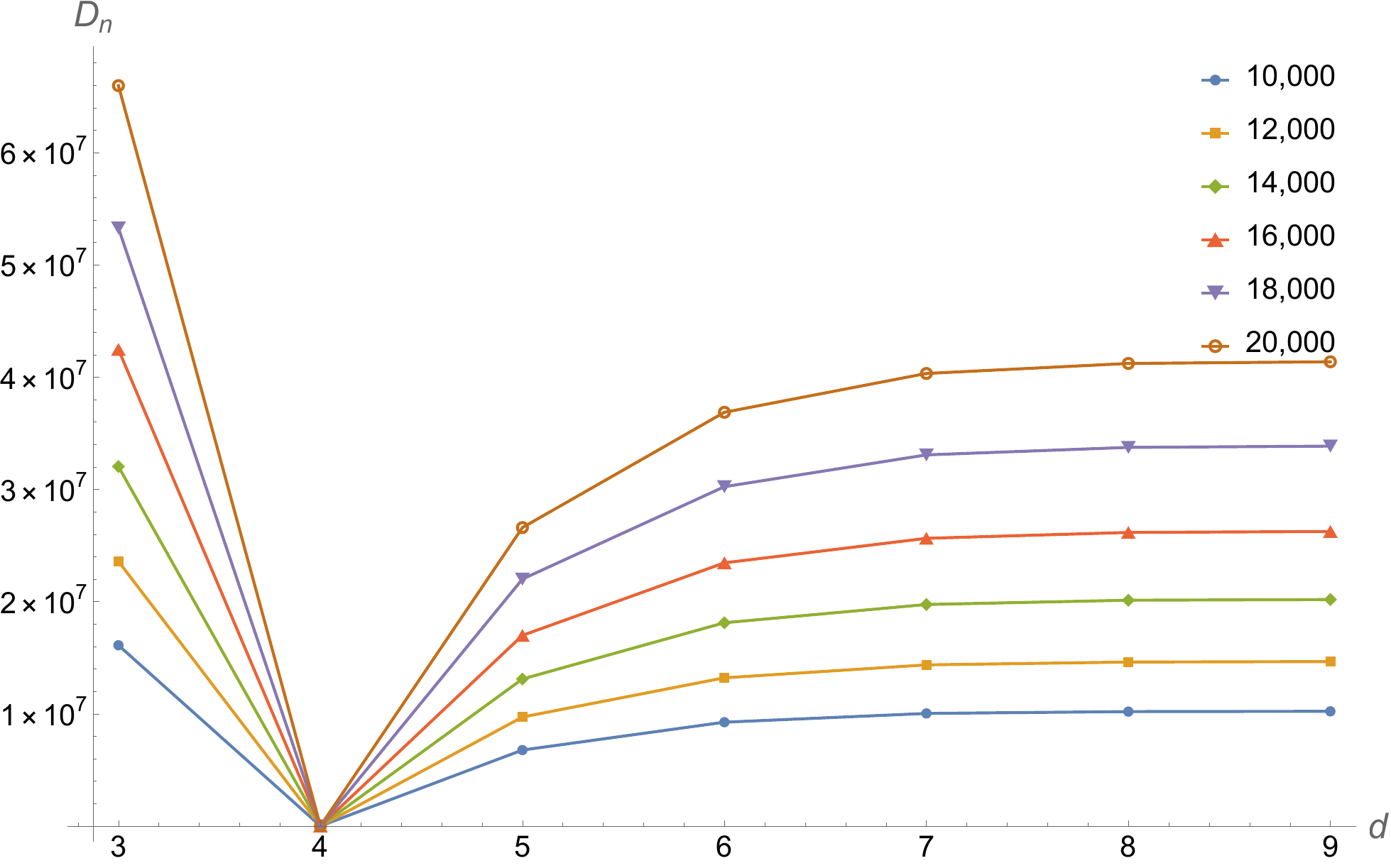} \caption{$\nnd_n(\dS_4(e^{0.2\,t}), \dS_d(e^{0.2\, t}))$ } 
  \end{subfigure}
  \caption{The $\nnd_n(4,d)$ between a spacetimes of dimension $4$
and $d$, within the same geometric class of spacetimes. The behaviour
is similar to that of $\nnd_n(\diam^4,\diam^d)$ in Fig. \ref{distmdmdn.fig}.}  \label{samespt.fig}
\end{figure}

Next, in Fig. \ref{samespt.fig} we illustrate the behaviour of $\nnd_n(.,.)$ between spacetimes
of the same kind, but with varying dimensions. Again, the  qualitative
behaviour is very similar to Fig. \ref{distmdmdn.fig}. We see that
$\nnd_n(.,.)$  between spacetimes of the same kind, but of different
dimensions grows with $|\Delta d|$, but that the rate of change of
depends on the sign of $\Delta d$.    

\begin{figure} \centering 
\begin{subfigure}{.3\textwidth}
 \centering 
    \includegraphics[height=3cm]{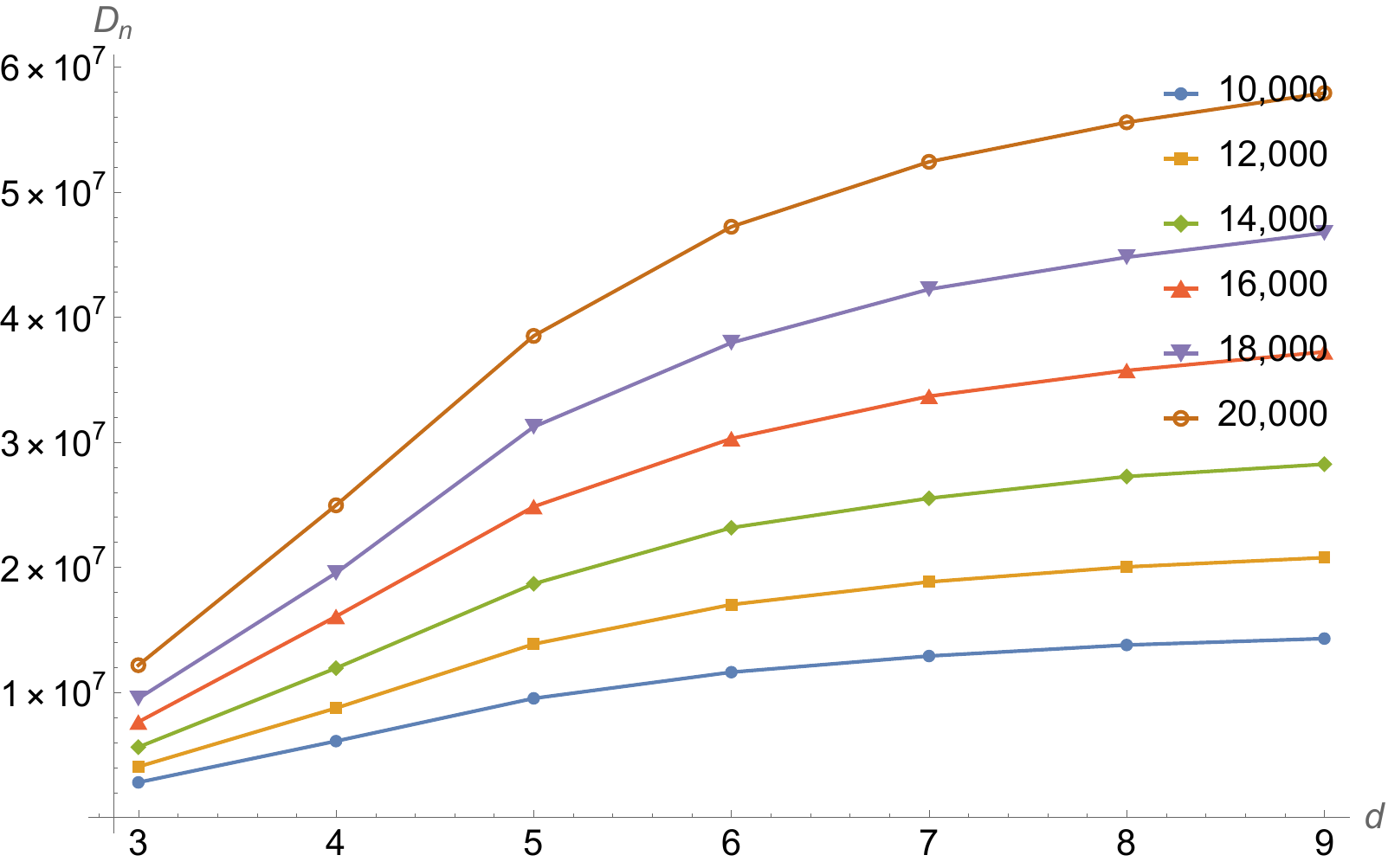} \caption{$\nnd_n(\bbF_4(t^{1/3}), \bbF_d(t^{1/2}))$ } \label{distfrwfrwfourthirdhalf.fig}
  \end{subfigure}
\begin{subfigure}{.3\textwidth}
 \centering 
    \includegraphics[height=3cm]{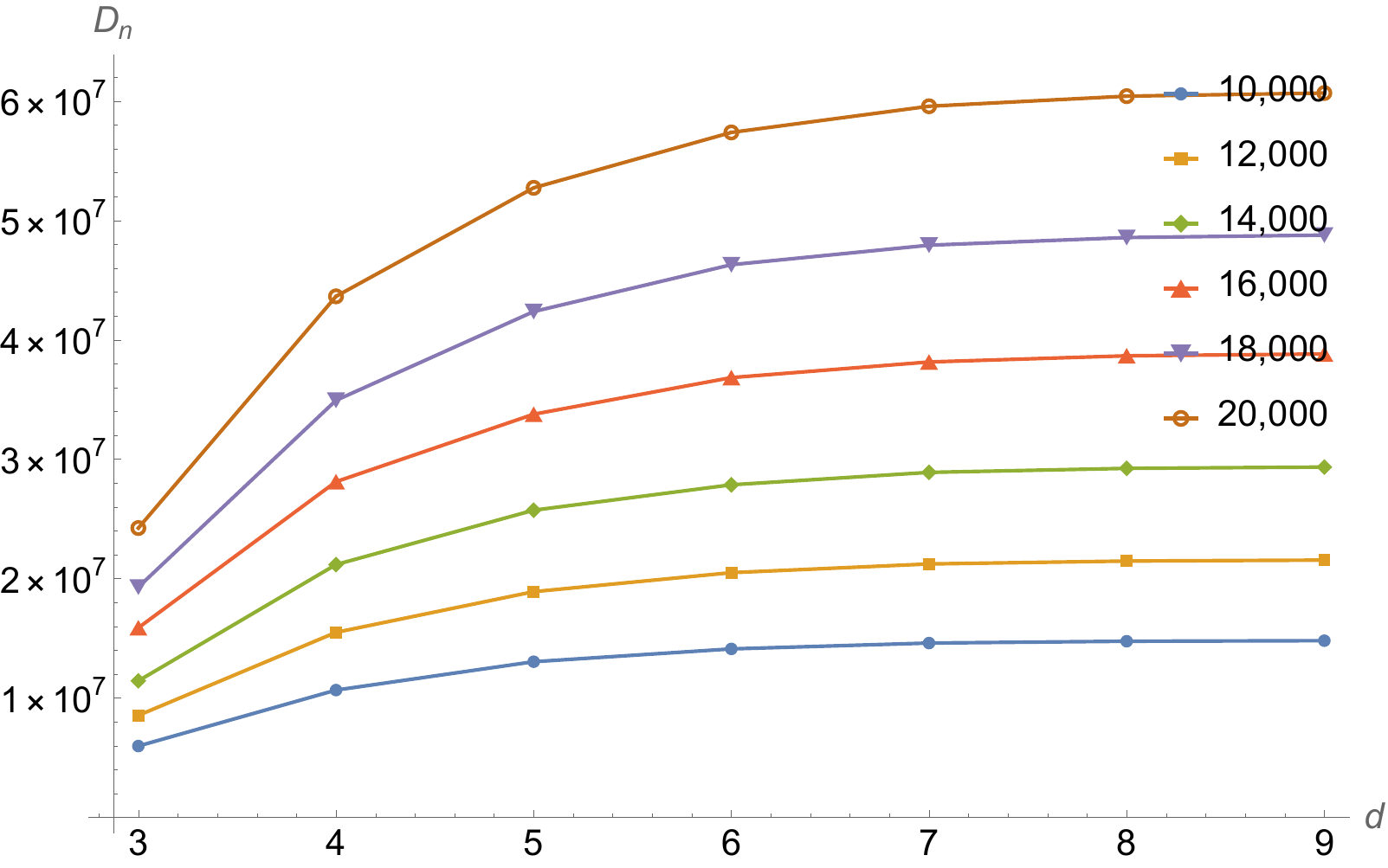} \caption{$\nnd_n(\bbF_4(t^{1/3}), \bbF_d(t^{2/3}))$ } \label{distfrwfrwfourthirdtwothird.fig}
  \end{subfigure}
  \begin{subfigure}{.3\textwidth}
 \centering 
    \includegraphics[height=3cm]{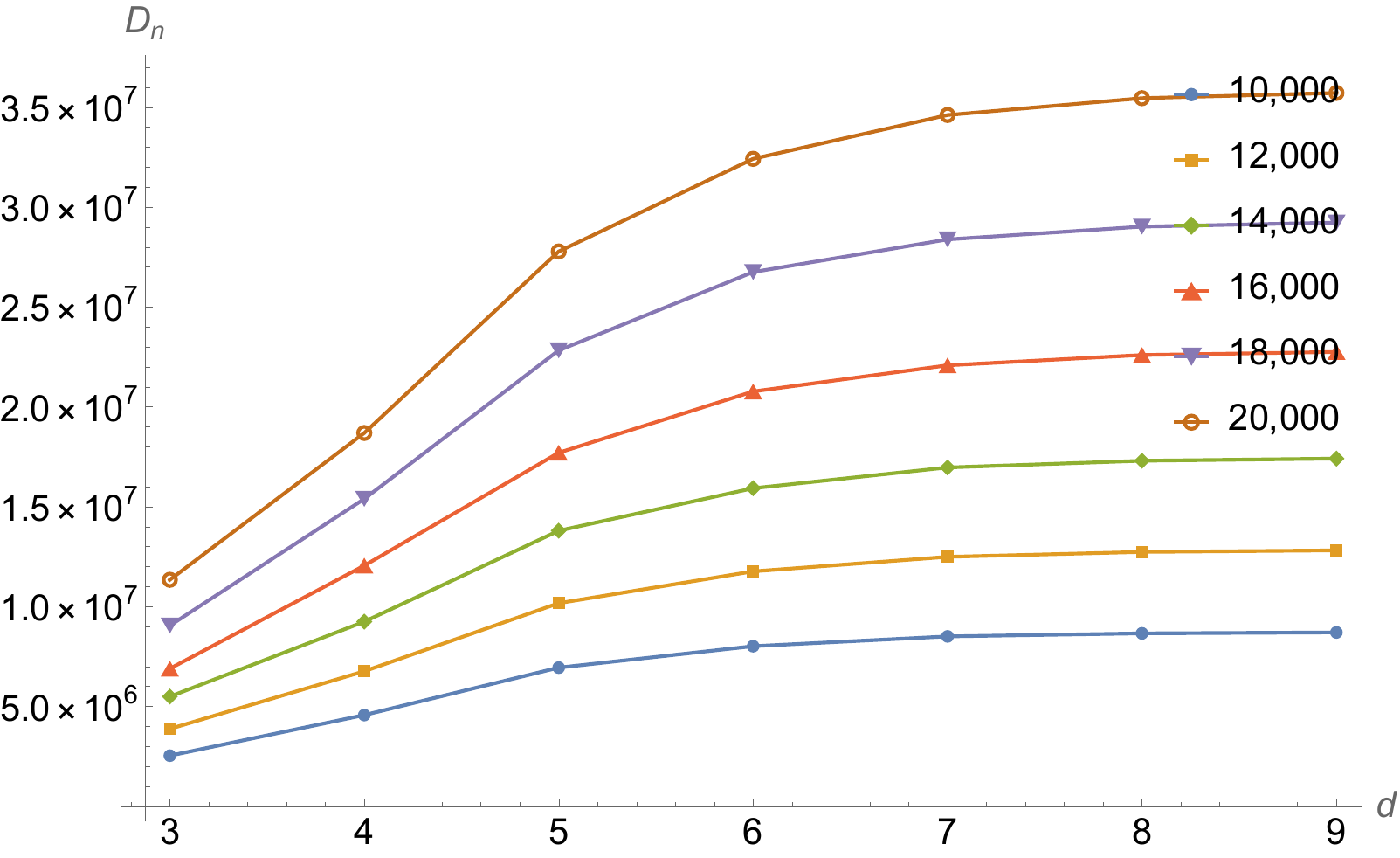} \caption{$\nnd_n(\bbF_4(t^{1/2}), \bbF_d(t^{2/3}))$ } \label{distfrwfrwfourhalftwothird.fig}
  \end{subfigure}
  \caption{$\nnd_n(\bbF_4(p_1), \bbF_d(p_2))$ does not just depend on the difference in the
    dimension but also crucially on the different scale factors. Thus
    $\bbF_4(t^{1/3})$ is closer to $ \bbF_{3}(t^{1/2})$ than to $
    \bbF_{4}(t^{1/2})$, and similarly,  $\bbF_4(t^{1/3})$ is closer to
    both $ \bbF_{2,3}(t^{2/3})$ than to $
    \bbF_{4}(t^{2/3})$. 
  }\label{difffrws.fig} 
\end{figure}

In Fig. \ref{difffrws.fig} we compare  different FRW $\bbF_d(t^p)$
spacetimes for different $d$ and $p$ values. In these plots we see
that the geometry can play a more dominant role than the topological
dimension. In all these cases, and for $p_1 \neq p_2$, 
$\nnd_n(\bbF_4(t^{p_1}),\bbF_d(t^{p_2}))$ is smaller for $d=3$ than
for $d=4$.

\begin{figure}
  \begin{subfigure}{.5\textwidth}
 \centering 
    \includegraphics[height=4.8cm]{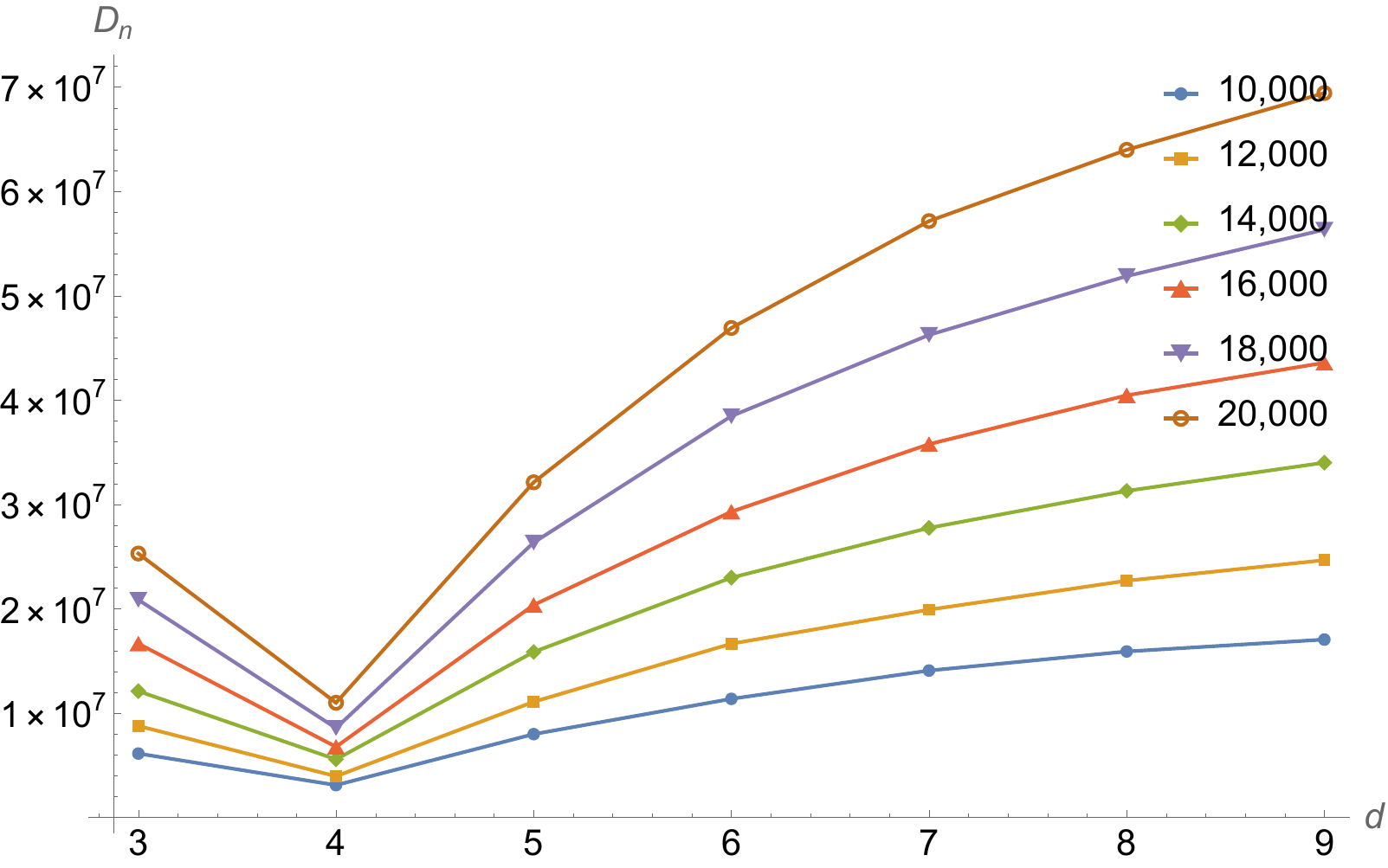} \caption{$\nnd_n(\dS_4(e^{0.025 \,t}), \dS_d(e^{0.05\,t}))$ } \label{distdsfrwdsfrwone.fig}
  \end{subfigure}
\begin{subfigure}{.5\textwidth}
 \centering 
    \includegraphics[height=4.8cm]{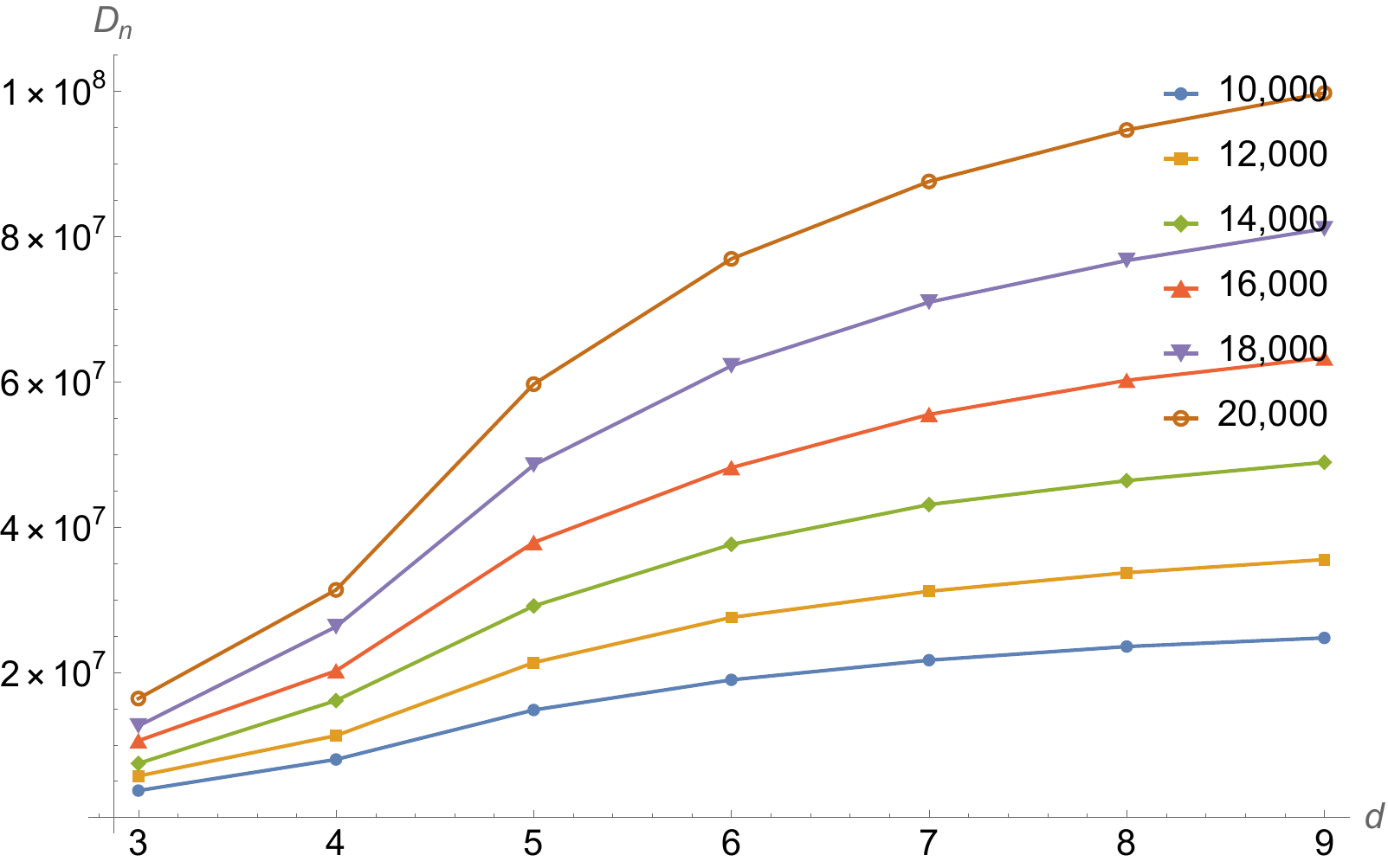} \caption{ $\nnd_n(\dS_4(e^{0.025 \,t}), \dS_d(e^{0.1\,t}))$}  \label{distdsfrwdsfrwtwo.fig}
  \end{subfigure}
\begin{subfigure}{.5\textwidth}
 \centering 
    \includegraphics[height=4.8cm]{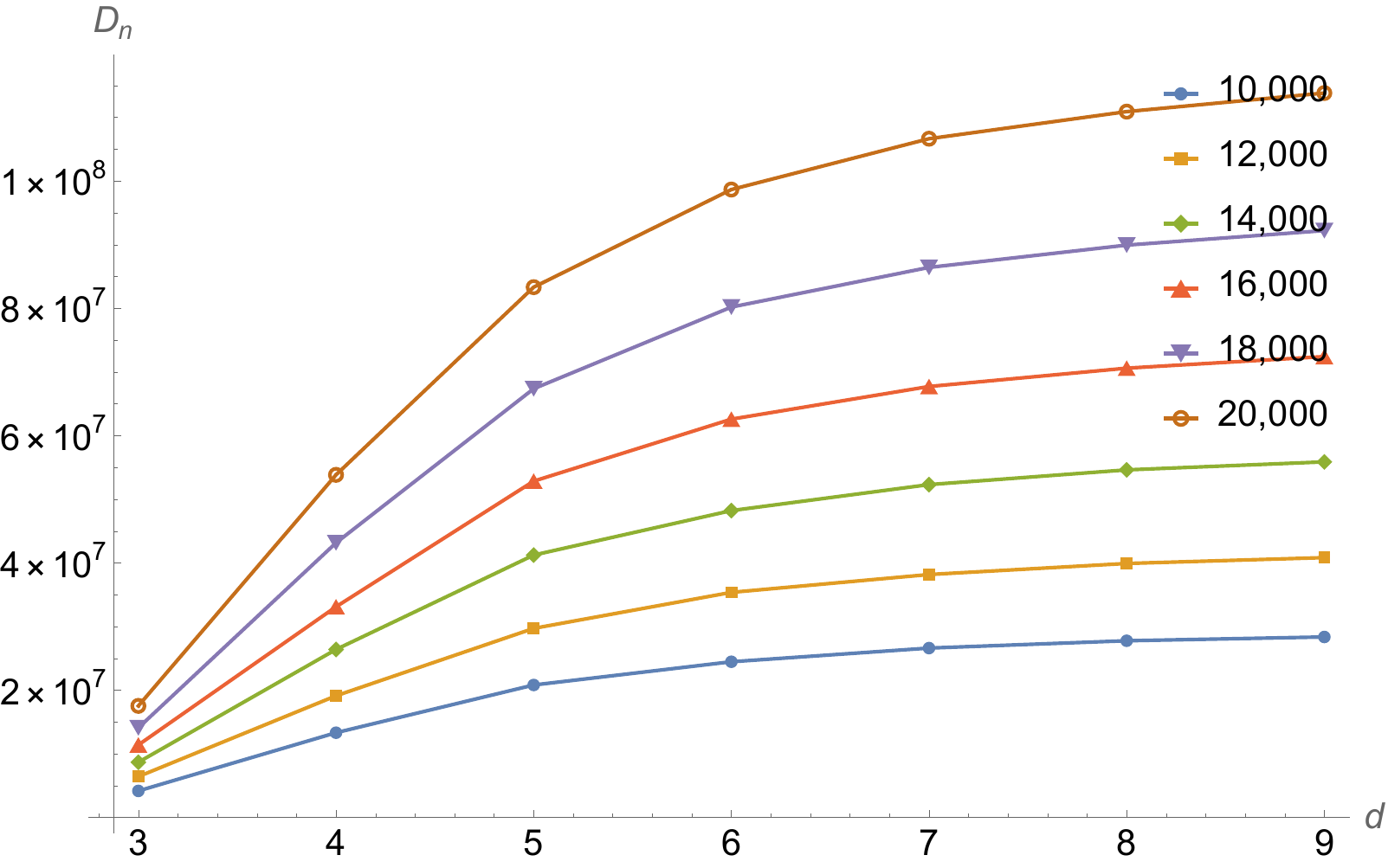} \caption{ $\nnd_n(\dS_4(e^{0.025 \,t}), \dS_d(e^{0.15\,t}))$} \label{distdsfrwdsfrwthree.fig}
  \end{subfigure}
  \begin{subfigure}{.5\textwidth}
 \centering 
    \includegraphics[height=4.8cm]{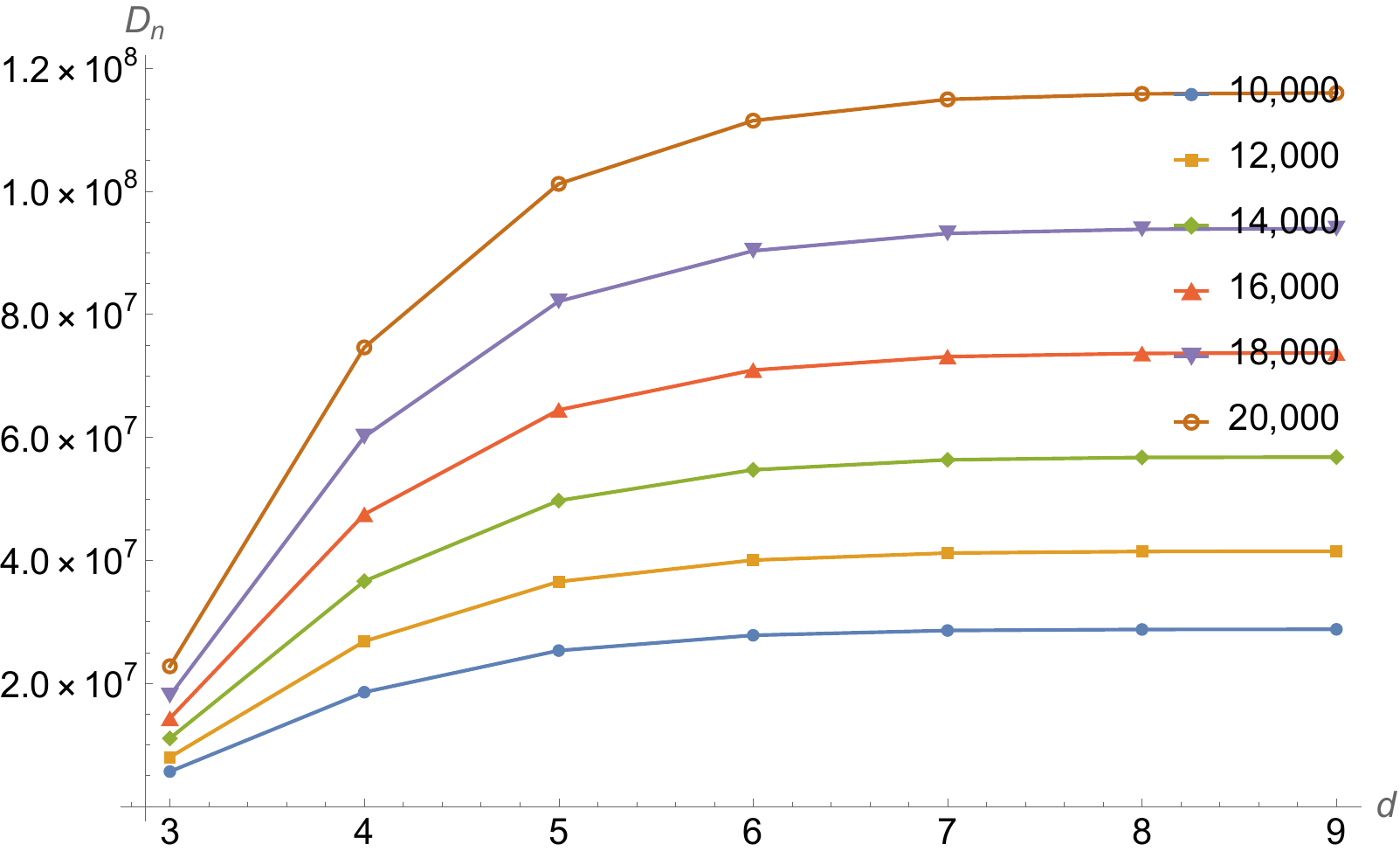} \caption{$\nnd_n(\dS_4(e^{0.025 \,t}), \dS_d(e^{0.2\,t}))$ } \label{distdsfrwdsfrwfour.fig}
  \end{subfigure}
  \caption{$\nnd_n(\dS_4(c_1), \dS_d(c_2))$ again does 
not just depend on the difference in the
    dimension but also on the different scale factors. 
  }\label{diffdsfrws.fig}
\end{figure}

In Fig. \ref{diffdsfrws.fig} we compare the different de Sitter
spacetimes $\dS_d(e^{ct})$. Here, the interplay between geometry and
topological dimension is more evident. Between  smaller  $c$ values,
topological dimension remains more important than the geometry, but as
the difference in $c$ values increases, this reverses.

\begin{figure}
  \begin{subfigure}{.5\textwidth}
  \centering
  \includegraphics[height=4.8cm]{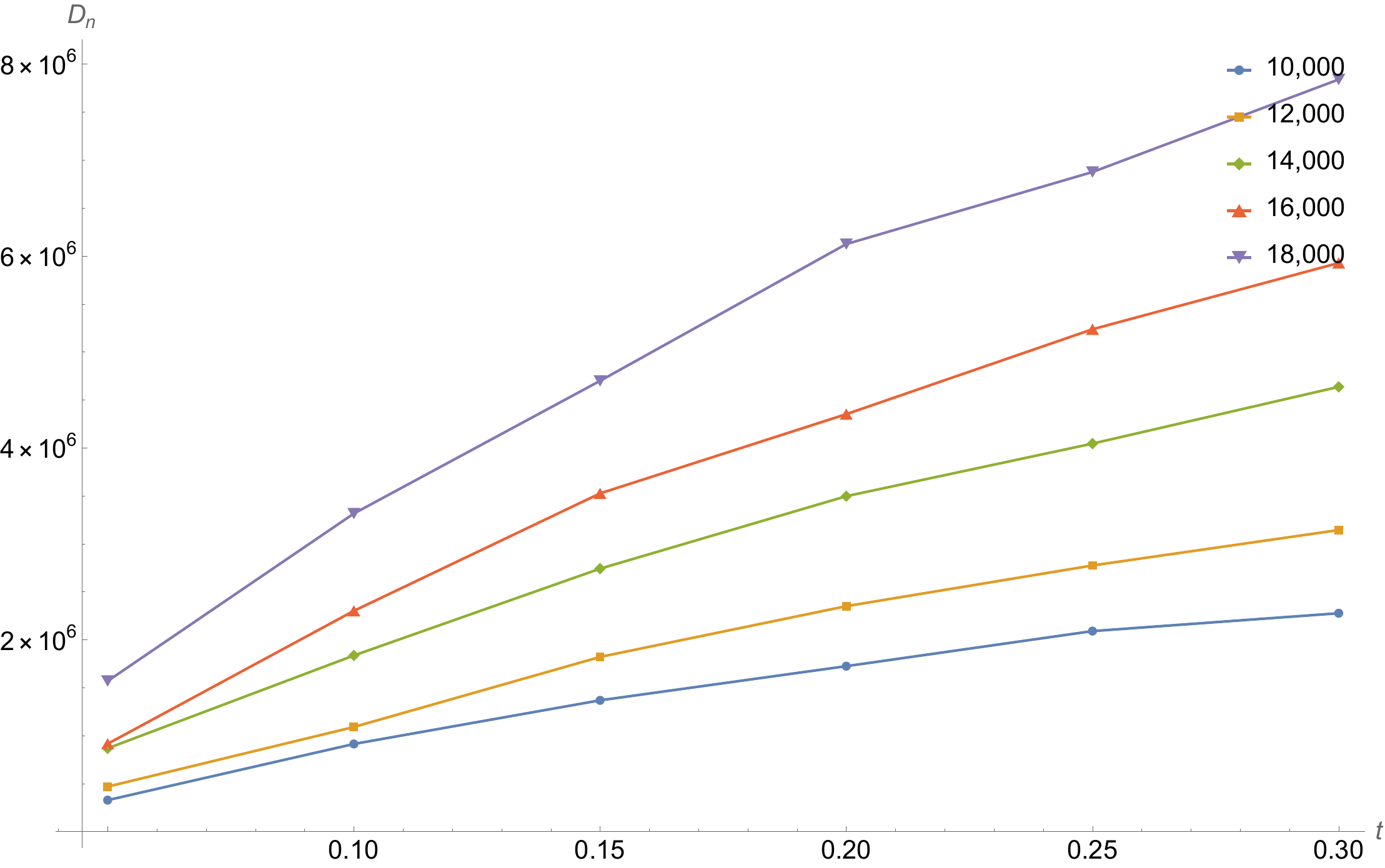}
\caption{}
  \label{distmdccthree.fig}
\end{subfigure}
  \begin{subfigure}{.5\textwidth}
\centering
 \includegraphics[height=4.8cm]{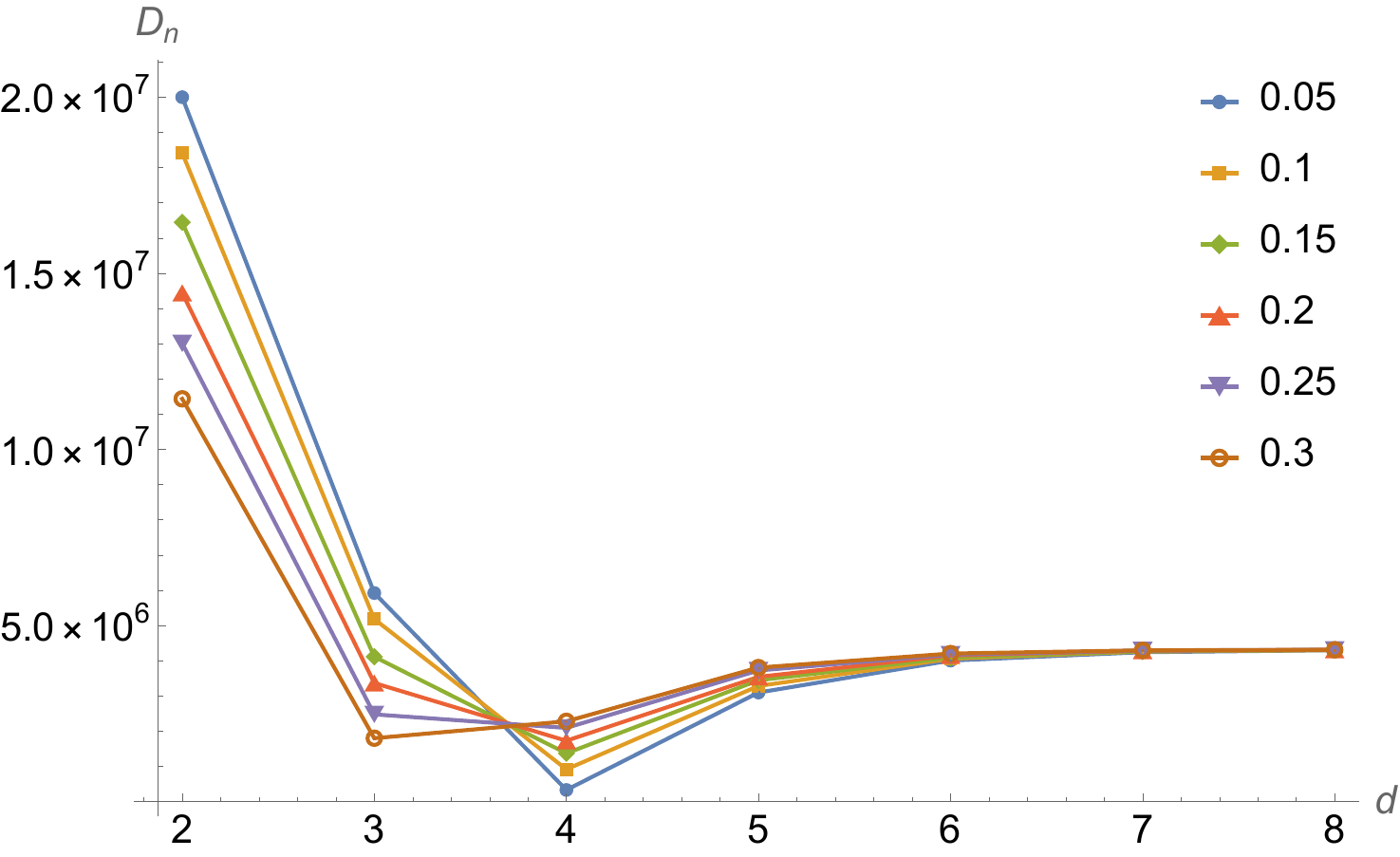} \caption{} \end{subfigure} 
  \caption{$\nnd_n(\diam^{4},\diam^{d}\times \mathbb I_t)$ is shown as
    a function of $d=2,\ldots 8$ and for various values of the
    thickness $t$. In (a) $d=4$ and one varies over $n$ and $t$, while in (b) $n=10,000$ and one varies over  $d$ and
    $t$. Note that  the minimum value $\nnd_n(\diam^{4},\diam^{4}\times
    \mathbb I_{0.05}) \sim 3 \times 10^5$ is not  small! }
  \label{distmdtc.fig}
\end{figure}

In Fig. \ref{distmdtc.fig} we find $\nnd_n(.,.)$ between  $\diam^4$
and $\diam^d \times \mathbb I_t$ as $t$ varies. Here for smaller $t$
values, the geometry is more important, for example  with $\diam^4 \times \mathbb I_{0.05}$
being closer to $\diam^4$ than to $\diam^3 \times \mathbb I_{0.05}$,
as one might expect. There is a cross over as $t$ increases. For $t=0.3$ $\diam^4$ gets closer to
$\diam^3 \times \mathbb I_{0.3}$ than $\diam^3 \times \mathbb
I_{0.3}$ which again not surprising.  

\begin{figure}\centering{
\begin{subfigure}{.3\textwidth}
  \centering
  \includegraphics[height=3cm]{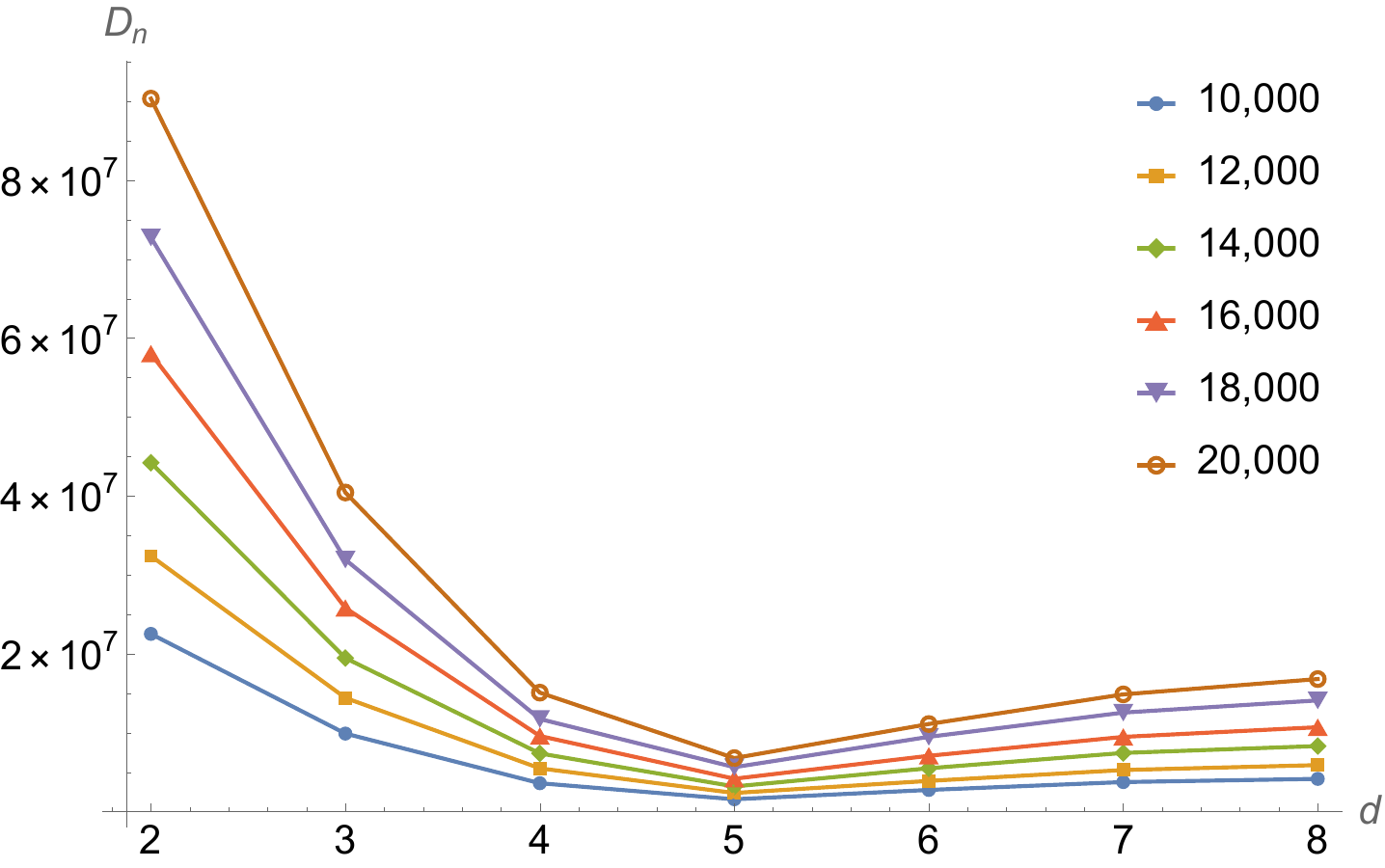}
  \caption{$\nnd_n(\diam^{4},\mathbb I^{d})$}
  \label{distmdccfour.fig}
\end{subfigure}
\begin{subfigure}{.3\textwidth}
  \centering
  \includegraphics[height=3cm]{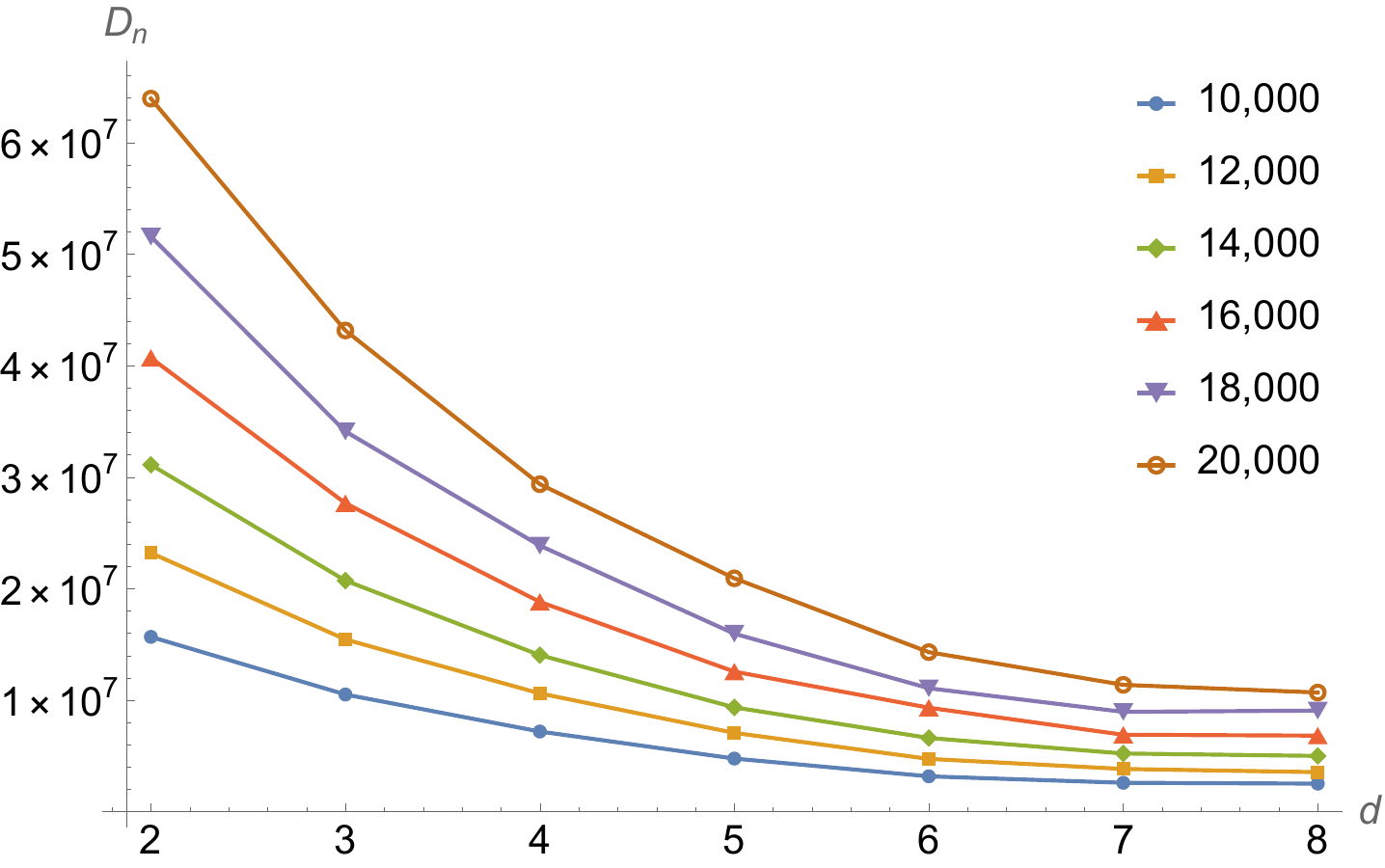}
  \caption{$\nnd_n(\diam^4,\bbF(t^{1/3}))$}
  \label{distmdfrwonethird.fig}
\end{subfigure}
\begin{subfigure}{.3\textwidth}
  \centering
  \includegraphics[height=3cm]{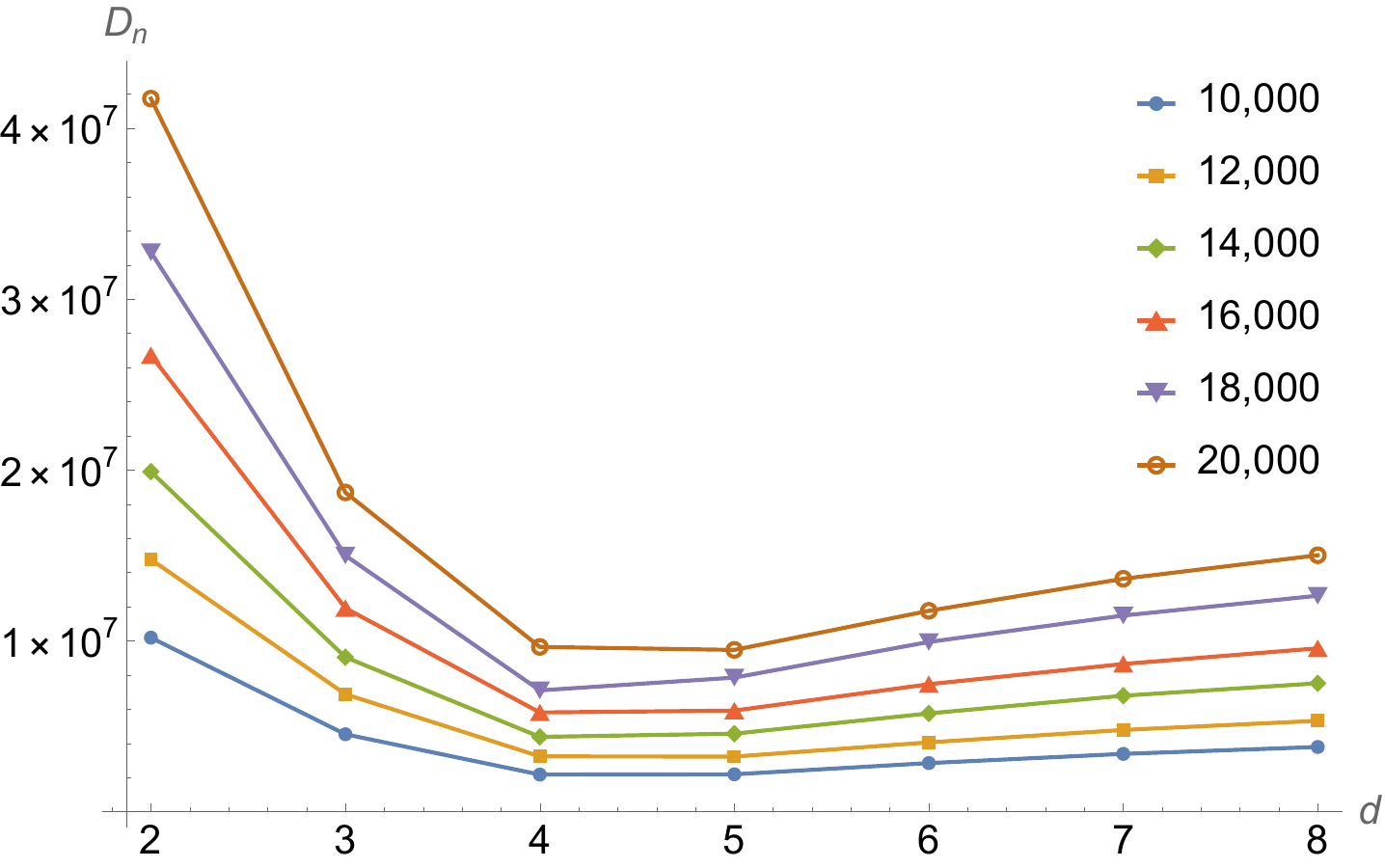}
  \caption{$\nnd_n(\diam^4,\bbF(t^{1/2}))$}
  \label{distmdfrwhalf.fig}
\end{subfigure}
\begin{subfigure}{.3\textwidth}
  \centering
  \includegraphics[height=3cm]{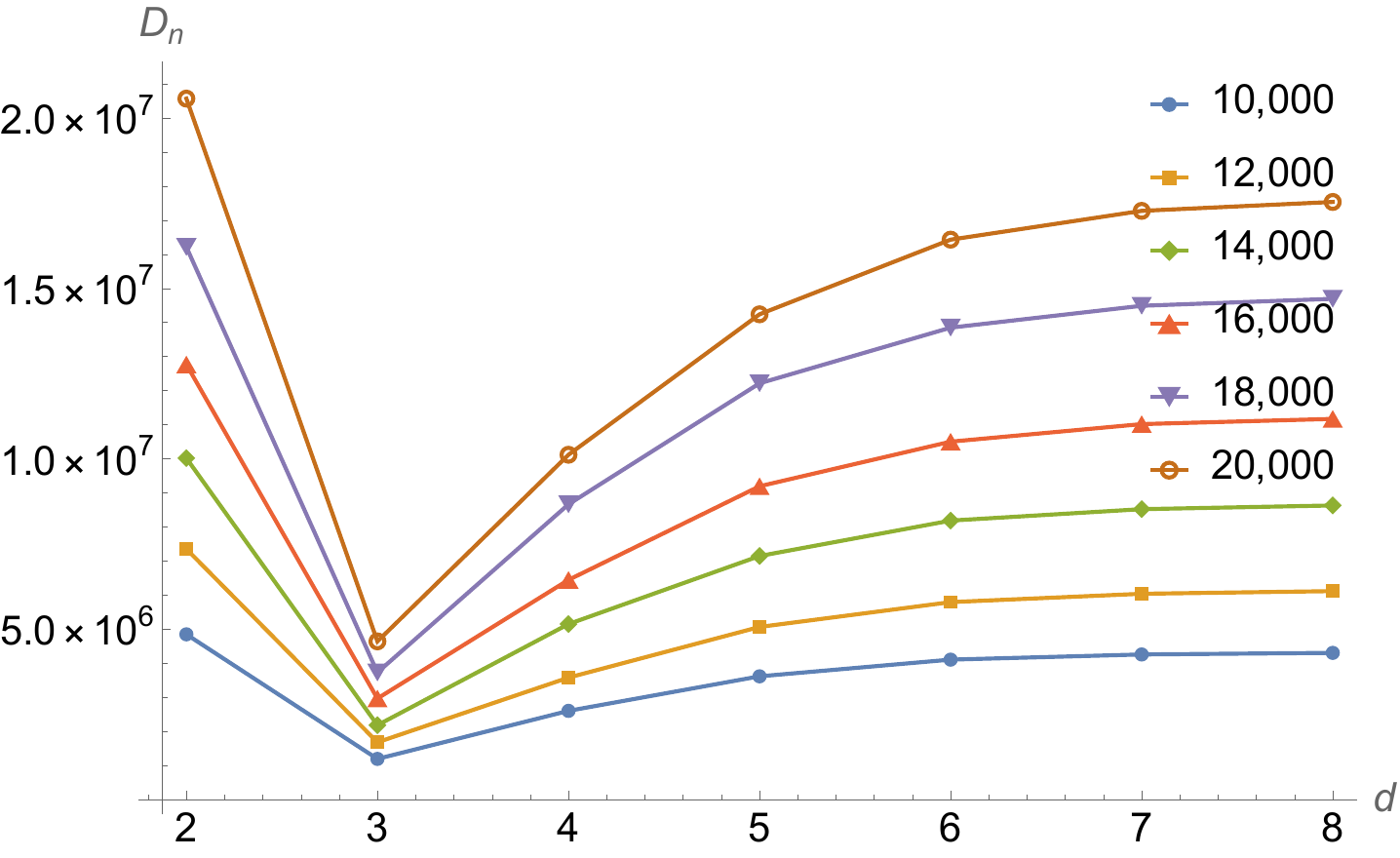}
  \caption{$\nnd_n(\diam^4,\bbF_d(t^{2/3}))$}
  \label{distmdfrwtwothird.fig}
\end{subfigure}
\begin{subfigure}{.3\textwidth}
  \centering
  \includegraphics[height=3cm]{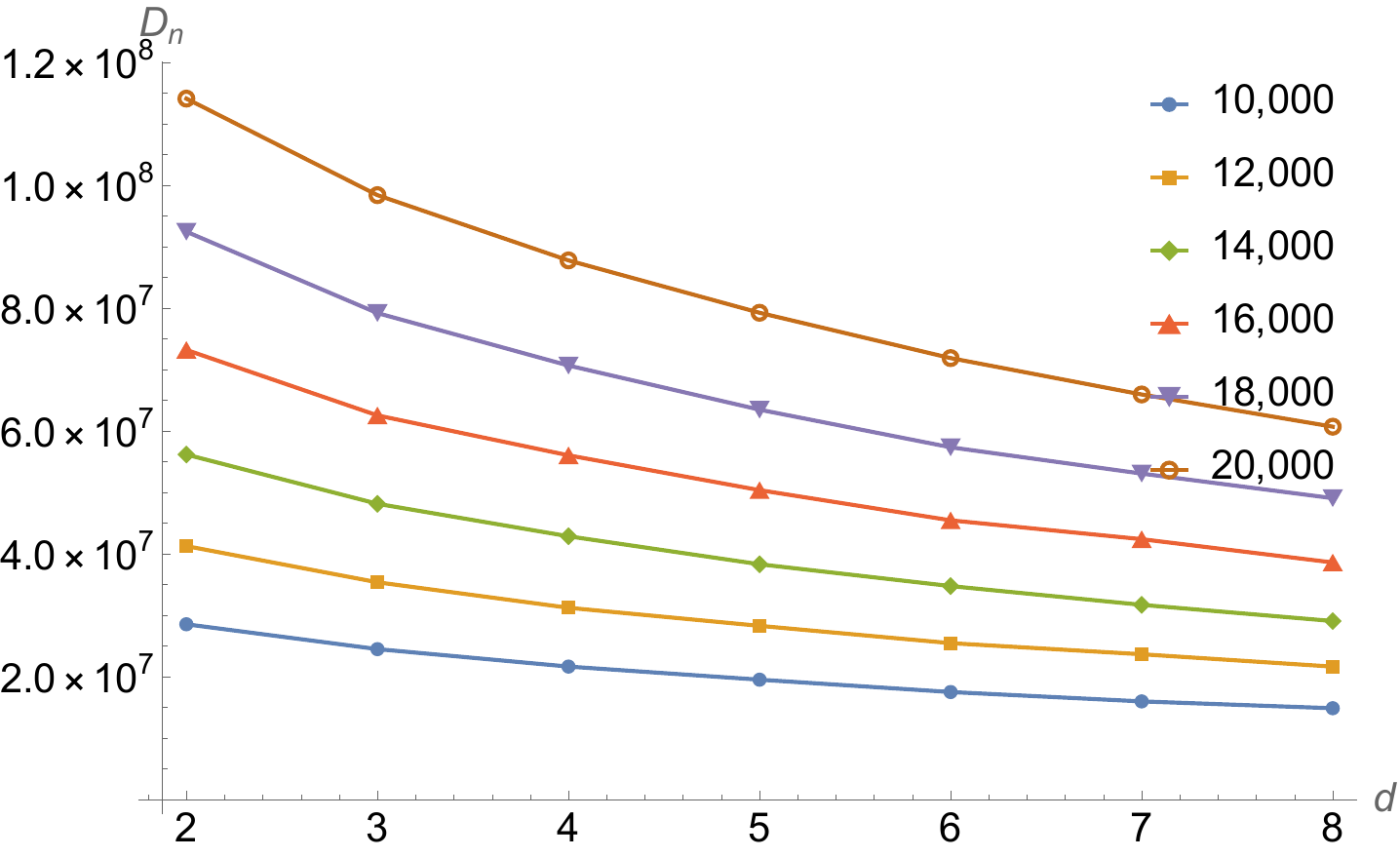}
  \caption{$\nnd_n(\diam^4,\dS_d(e^{0.025\, t}))$}
  \label{distmdfrwdsone.fig}
\end{subfigure}
\begin{subfigure}{.3\textwidth}
  \centering
  \includegraphics[height=3cm]{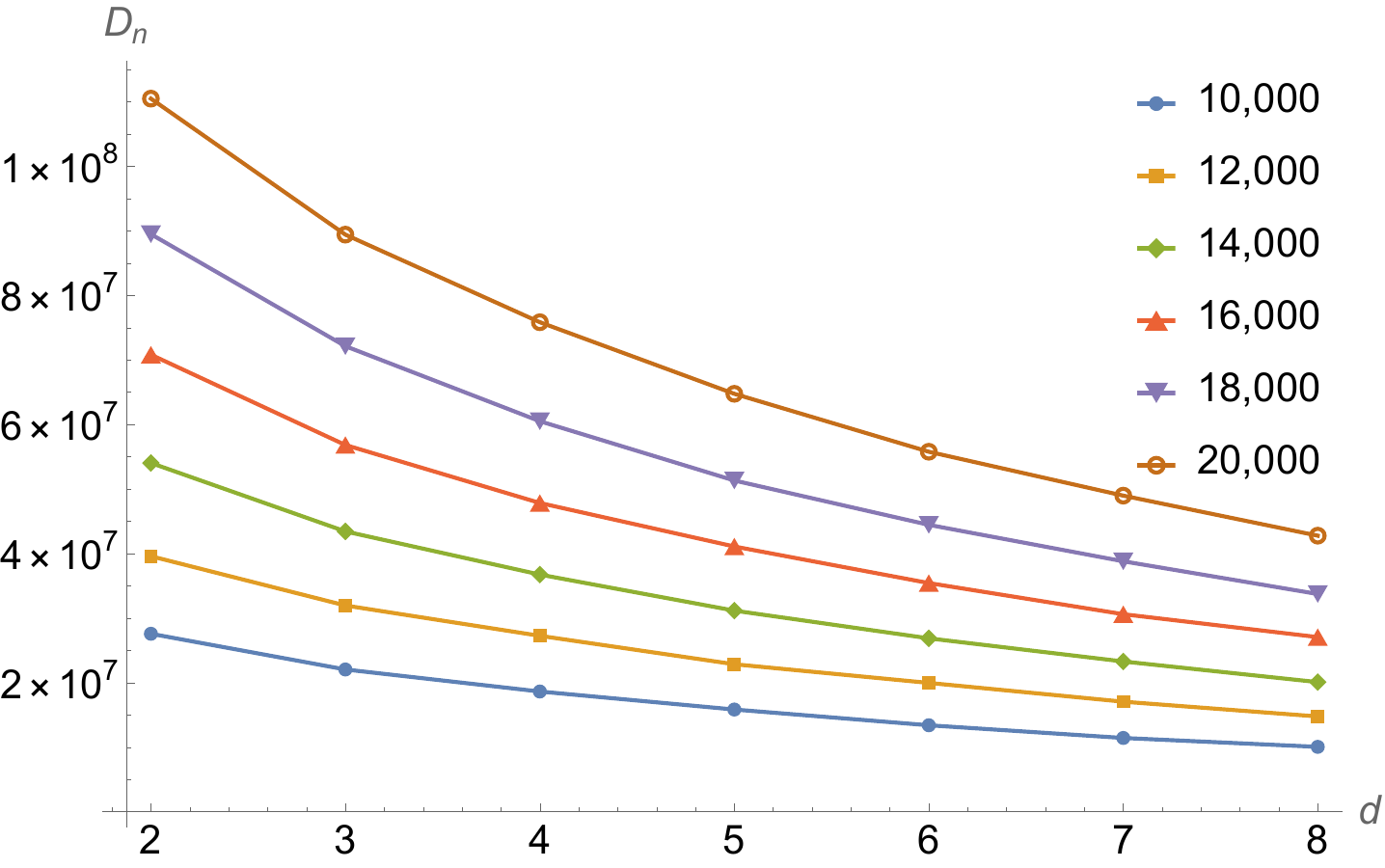}
  \caption{$\nnd_n(\diam^4,\dS_d(e^{0.05\, t}))$}
  \label{distmdfrwdstwo.fig}
\end{subfigure}
\begin{subfigure}{.3\textwidth}
  \centering
  \includegraphics[height=3cm]{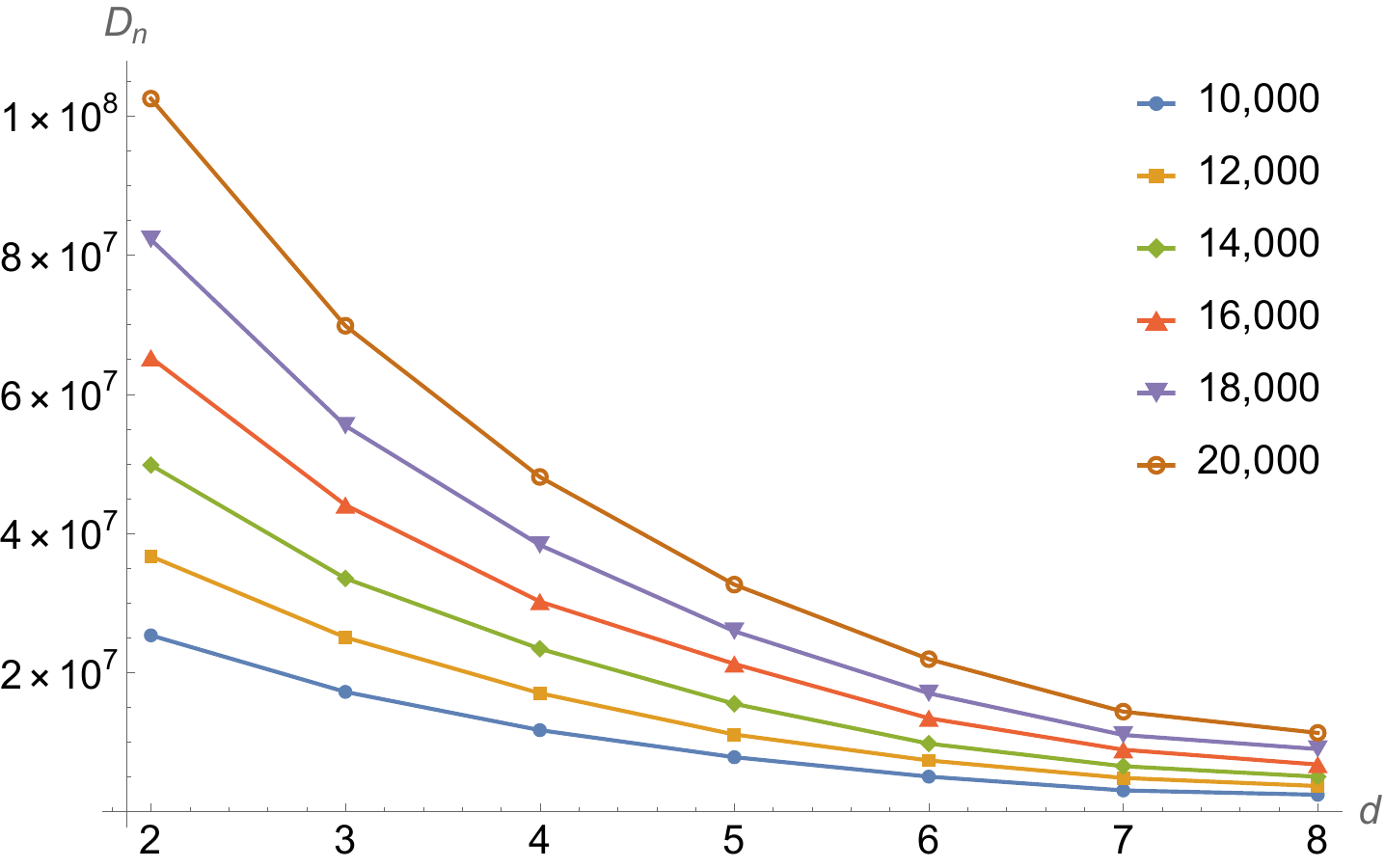}
  \caption{$\nnd_n(\diam^4,\dS_d(e^{0.1\, t}))$}
  \label{distmdfrwdSthree.fig}
\end{subfigure}
\begin{subfigure}{.3\textwidth}
  \centering
  \includegraphics[height=3cm]{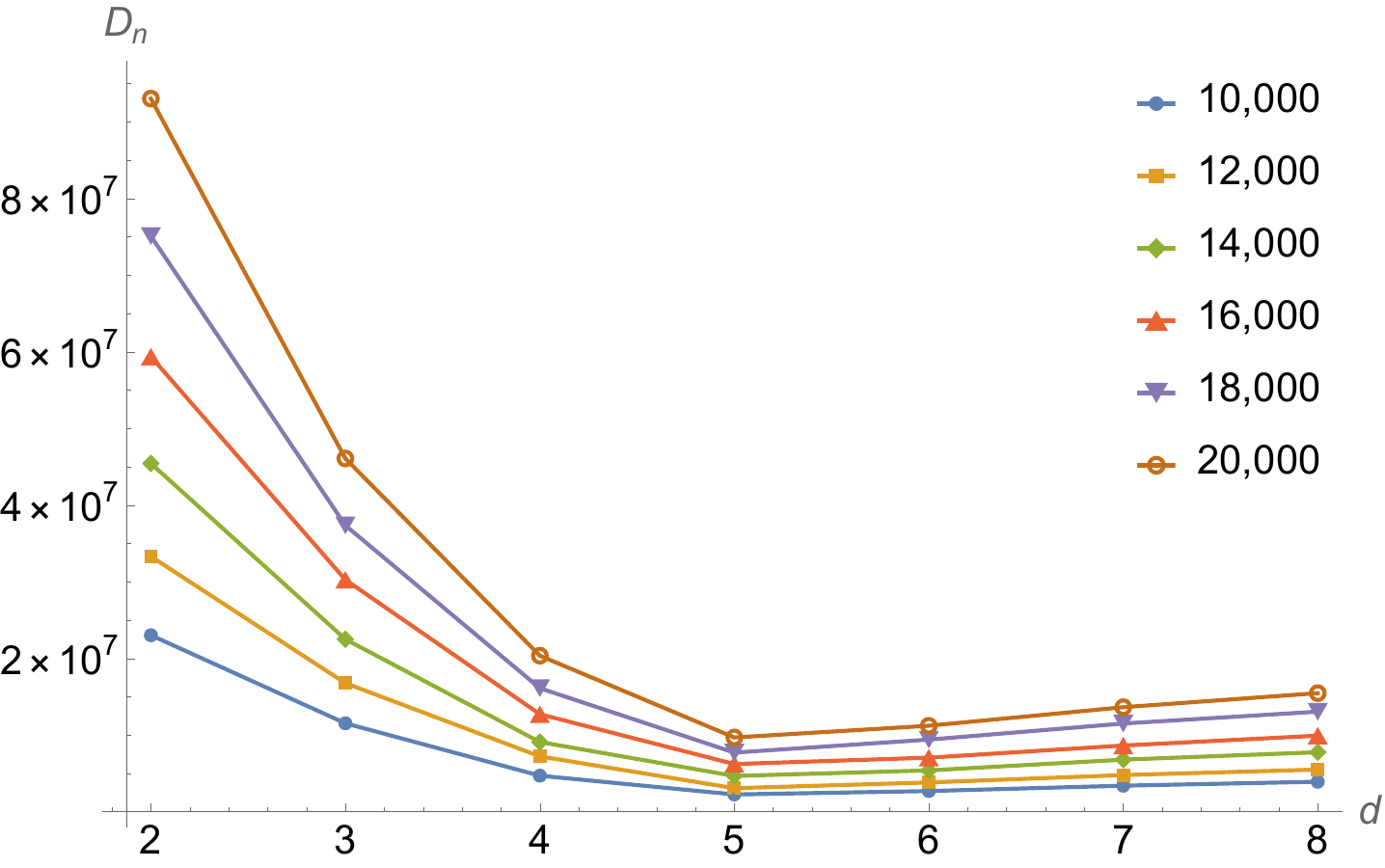}
  \caption{$\nnd_n(\diam^4,\dS_d(e^{0.15\, t}))$}
  \label{distmdfrwdsfour.fig}
\end{subfigure}
\begin{subfigure}{.3\textwidth}
  \centering
  \includegraphics[height=3cm]{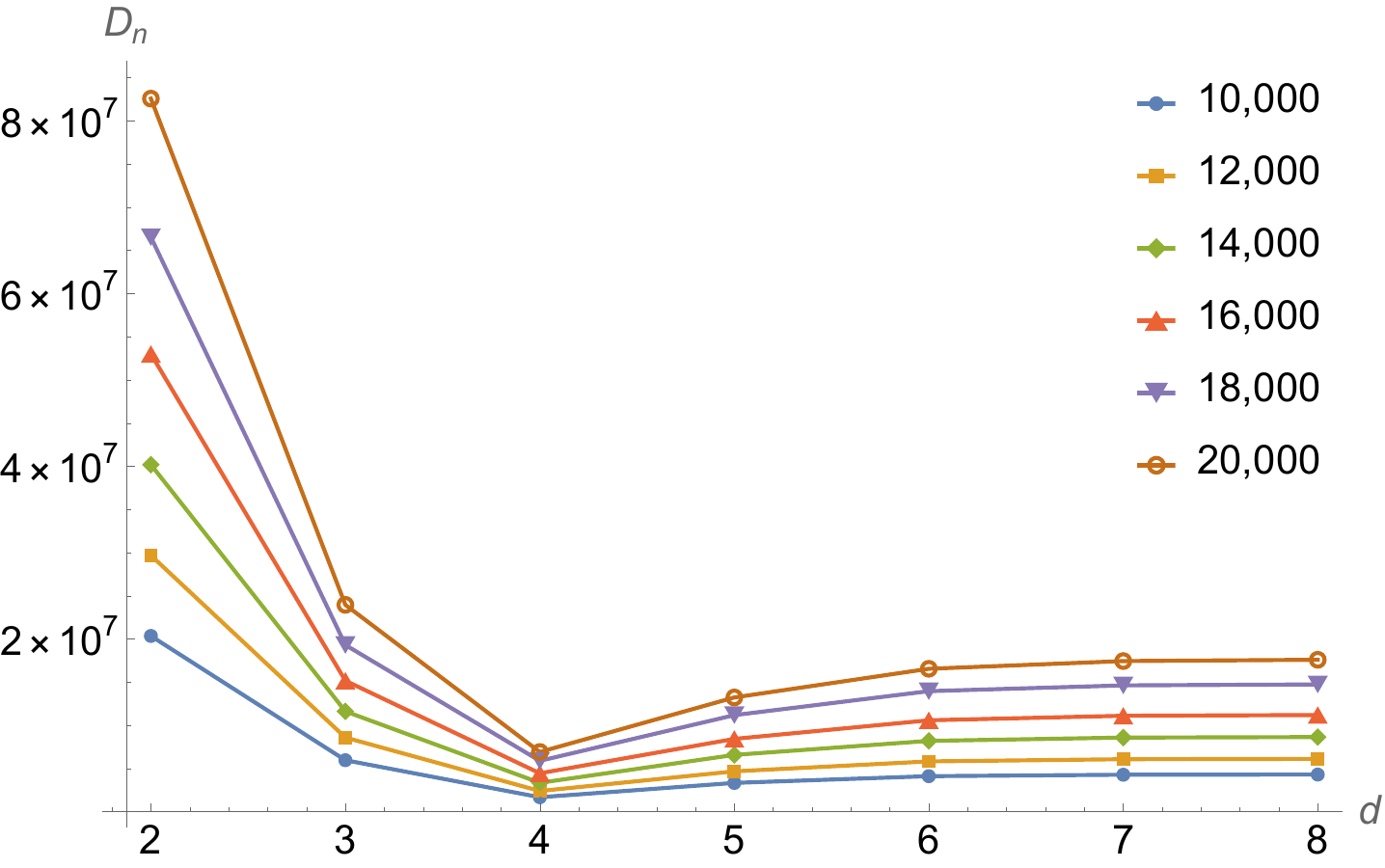}
  \caption{$\nnd_n(\diam^4,\dS_d(e^{0.2\, t}))$}
  \label{distmdfrwdsfive.fig}
\end{subfigure}
\caption{ Plots of $\nnd_n(\diam^{4},.)$ for different spacetimes.}
\label{MDothers.fig}}
\end{figure}

Finally, in Fig. \ref{MDothers.fig} we show $\nnd_n(\diam^d, .)$ for
different spacetimes. Again we see cases where the topological
dimension dominates the geometry and vice-versa.

An important feature of $\nnd_n(.,.)$ is that it is strictly
global in character. Hence even if the spacetime regions  are locally isometric, their
global ``shapes''  determine closeness. For example, the 
causal diamond and hypercube, both in Minkowski spacetime, 
differ in their  $n$-interval-spectrum and are hence are not
arbitrarily close. This global character  is also a feature of  Bombelli's distance
function \cite{bomclose}. A local comparison can be made by extracting a causal
diamond from the hypercube and comparing with an equal volume causal
diamond.

These figures are illustrative -- we have not shown our extensive
data, but they  corroborate  the general features shown here. The
simulations of $\spec_n(.)$ themselves were intensive,  with some parameter values
taking several days to compute on a high performance cluster. Since
we are not (at least at this stage)  interested in the actual value of $\nnd_n(.,.)$ but only  its qualitative
features,  one might want to simplify simulations in the future by
considering only a subset of the $m$'s,  for example all $m \leq  n/2$. 
For the range of $n$ values we have considered, perhaps one
could even stop at $m=100$.   However, this makes the definition
somewhat ad-hoc, and while this might be justified a posteriori in
specific cases, it is more natural to use all the $n-1$ values of $m$.    

\section{Summary and Discussion} \label{conclusion.sec} 

In this work we have proposed  a new closeness function
$\nnd_n^{(r)}(.,.)$ on the space of finite volume causal Lorentzian
geometries $\ccL$.  $\nnd_n^{(r)}(.,.)$  satisfies  the triangle inequality, but is 
not a true distance function since it  can vanish 
between non-isometric  spacetimes.  It is 
constructed from a family of order invariants $\bNmn$ in the underlying $n$-element
random causal sets associated with each spacetime. 
The $\bNmn$ are the abundances of order intervals of
size $m$ whose expectation values define  the $n$-interval spectrum
$\spec_n\Mg$ of a spacetime $\Mg$ \cite{ls}. 
Using the map 
$\spec_n\Mg \rightarrow \re^{n-1}$,  we define 
$\nnd_n^{(r)}(.,.)$  to be the $L^r$ distance function on $\re^{n-1}$,
which in turn provides a closeness function on $\ccL$.   Without loss of
generality we consider the $r=1$ case, i.e., the taxi-cab distance
$\nnd_n(.,.) \equiv \nnd_n^{(1)}(.,.)$ on $\re^{n-1}$.  We quotient $\ccL$
by the equivalence class of  
interval isospectral spacetimes to define a notion of
convergence on $\tcL=\ccL/\!\!\sim$. Because of interval  isospectrality, $\nnd_n(.,.)$
is strictly weaker than Bombelli's closeness function but has 
the advantage that it can be numerically calculated for a  
wide range of spacetimes.

Despite the obvious drawback of interval
isospectrality, our closeness function $\nnd_n(.,.)$ has its uses in 
Lorentzian geometry.  In this work we have used it to define converge
for monotonically increasing (or decreasing)  scale factors in
FRW-type spacetimes. When supplemented by a small set of additional
order invariants, the convergence can be defined more generally.  
At present $\av{\bNmn\Mg}$ has been calculated 
analytically for the $d$-dimensional Minkowski causal diamond
$\diam^d$.  Extending this calculation  to other spacetimes -- for
example,  de Sitter or general Riemann Normal Neighbourhoods -- would provide
a better analytic understanding of the strengths and weaknesses of the
current proposal.

The simulations in this work, though fairly extensive are largely for
illustrative purposes.  Constructing $\nnd_n(.,.)$ for a much wider
trial set, as a catalogue of sorts, would expand the scope of the
present work.  There are of course other types of order invariant
spectra like the abundance of $m$-element chains or the abundance of
$m$-element past or future sets \cite{PF}, from which closeness
functions can be constructed on $\ccL$.  Exploring these possibilities
and comparing them with our closeness function would be a useful
future direction to pursue.

In summary, our closeness function on $\tcL$ serves two purposes. The
first is to quantify when an abstract causal set is
continuumlike. The second is to define an approximate isometry between
interval non-isospectral spacetimes, which in turn gives rise to a
convergence condition on $\ccL$ for certain sequences of spacetimes. The
family of order invariants we have chosen in this work can be
supplemented by other sets of order invariants, and in special cases
yields stronger convergence conditions.

\section{Acknowledgements} 
The author was supported in part by the ANRF MATRICS grant,
MTR/2023/000831. She attended the  2nd Workshop on  “A new geometry for Einstein’s
theory of relativity and beyond” at the University of Vienna thanks to
the Austrian Science Fund (FWF) [Grant DOI 10.55776/EFP6]. She
would  like to thank the 
organisers of this meeting and the  stimulating discussions that
followed with the participants. She would also like to thank  Dr. Emma
Albertini for discussions.
\bibliography{Submission-Final.tex}
\bibliographystyle{ieeetr}

\end{document}